\documentclass[useAMS,usenatbib]{mnras}

\usepackage{graphicx}
\usepackage{rotating}
\usepackage{longtable}
\usepackage{tablefootnote}

\newif\ifAMStwofonts                        
\newcommand{\lsimeq}{{_<\atop^{\sim}}}
\newcommand{\gsimeq}{{_>\atop^{\sim}}}

\title[Mid-IR Line Luminosity as a tracer of BHAR]{Tracing black hole accretion with SED decomposition and IR lines: from local galaxies to the high-$z$ Universe} 
\author[C. Gruppioni, S. Berta, L. Spinoglio et al.]
{\parbox{\textwidth}{\raggedright C. Gruppioni$^{1}$\thanks{E-mail: carlotta.gruppioni@oabo.inaf.it}, 
S. Berta$^{2}$, 
L. Spinoglio$^{3}$, 
M. Pereira-Santaella$^{4}$,  
F. Pozzi$^{5}$, 
P. Andreani$^{6}$,
M. Bonato$^{7}$,
G. De Zotti$^{8}$,
M. Malkan$^{9}$, 
M. Negrello$^{10}$,
L. Vallini$^{1,5}$,
C. Vignali$^{5}$
}
\vspace{0.4cm}\\
$^{1}$Istituto Nazionale di Astrofisica - Osservatorio Astronomico di Bologna, via Ranzani 1, I--40127 Bologna, Italy.\\
$^{2}$Max-Planck-Institut f\"{u}r Extraterrestrische Physik (MPE), Postfach 1312, D-85741 Garching, Germany.\\
$^{3}$Istituto Nazionale di Astrofisica - Istituto di Astrofisica e Planetologia Spaziali, via Fosso del Cavaliere 100, 00133 Roma.\\
$^{4}$Centro de Astrobiolog\`ia (CSIC/INTA), Ctra de Torrej\`on a Ajalvir, km 4, 28850, Torrej\`on de Ardoz, Madrid, Spain.   \\
$^{5}$Dipartimento di Fisica e Astronomia, Universit\`a di Bologna, viale Berti Pichat 6, I--40127 Bologna, Italy.\\
$^{6}$European Southern Observatory, Karl-Schwarzschild-Stra\"se 2, 85748 Garching, Germany.\\
$^{7}$Department of Physics and Astronomy, Tufts University, 574 Boston Ave., Medford, MA 02155, USA.\\
$^{8}$Istituto Nazionale di Astrofisica - Osservatorio Astronomico di Padova, vicolo del'Osservatorio 5, I--35122 Padova, Italy.\\
$^{9}$Dept. of Physics \& Astronomy, UCLA, Los Angeles, CA 90095-1547, USA.\\
$^{10}$School of Physics and Astronomy, Cardiff University, Queen's Buildings, The Parade, Cardiff CF24 3AA, UK.\\
}
\begin{document}

\date{Accepted 2016 March 7. Received 2016 March 4; in original form 2015 November 30.}

\pagerange{\pageref{firstpage}--\pageref{lastpage}} \pubyear{2015}

\maketitle

\label{firstpage}

\begin{abstract}
We present new estimates of AGN accretion and star-formation luminosity in galaxies obtained for the local 12 $\mu$m sample of Seyfert galaxies
(12MGS), by performing a detailed broad-band spectral energy distribution (SED) decomposition including the emission of stars,
dust heated by star formation and a possible AGN dusty torus. 
Thanks to the availability of data from the X-rays to the sub-millimetre, we constrain and test the contribution of the stellar, 
AGN and star-formation components to the SEDs. The availability of {\em Spitzer-IRS} low resolution mid-infrared (mid-IR) spectra is crucial to constrain 
the dusty torus component at its peak wavelengths. The results of SED-fitting are also tested against the 
available information in other bands: the reconstructed AGN bolometric luminosity is compared to those derived from X-rays and from 
the high excitation IR lines tracing AGN activity like [Ne~V] and [O~IV].
The IR luminosity due to star-formation (SF) and the intrinsic AGN bolometric luminosity 
are shown to be strongly related to the IR line luminosity. 
Variations of these relations with different AGN fractions are investigated, showing that the relation dispersions
are mainly due to different AGN relative contribution within the galaxy.
Extrapolating these local relations between line and SF or AGN luminosities to higher redshifts, by means of recent {\em Herschel} galaxy 
evolution results, we then obtain mid- and far-IR line luminosity functions useful to estimate how many star-forming galaxies and
AGN we expect to detect in the different lines at different redshifts and luminosities with future IR facilities (e.g., {\em JWST, SPICA}).

\end{abstract}


\section{Introduction}
\label{sec_intro}
The comoving rates of star formation and central black hole accretion follow a similar rise and fall, offering evidence for coevolution of black holes and their host galaxies (i.e., \citealt{shankar09, kormendy13} and references therein).
The data show that the star formation activity peaked at around 8--11 Gyr ago (at redshifts between 1 and 3; e.g. \citealt{burgarella13}). In this phase most of the energy
emitted in the star formation process was absorbed by dust and converted into infrared (IR), with less than 10\% remaining visible at ultraviolet (UV) to optical wavelengths (e.g., \citealt{madau14}). 
While the large majority of present-day stars have been produced in such obscured galaxies, we have very limited knowledge about the nature of infrared galaxies at those high redshifts, 
on what controlled and changed the star formation along the evolution of these galaxies and of their metal content at early epochs.
Dusty galaxies are the only ones that can tell us the whole story of galaxy and AGN evolution (whose activity peaks at about the same epoch) during that obscured era. 
So far, mid- and far-IR facilities like {\em ISO}, {\em Spitzer} and - more recently - {\em Herschel} have allowed us to observe the ``obscured'' side of the Universe at non local redshifts mostly in 
photometry, by measuring total fluxes of galaxies, but not peering into their internal physics. 
This could be done through mid- and far-IR spectroscopic observations, only for local galaxies or, at high redshifts, for
few lensed and/or ultra-luminous sources not representative of the bulk of the galaxy population. 

Emission-line intensities and emission-line ratios in the mid- and far-IR domain, not suffering from dust extinction as the optical and UV emission lines, provide unique information on the physical 
conditions (i.e., electron density and temperature, degree of ionisation and excitation, chemical composition) of the gas responsible for emitting the lines within the dust-obscured regions of galaxies with intense star formation activity or surrounding an AGN (\citealt{spinoglio92}; \citealt{rubin94}; \citealt{panuzzo03}). 
The typical electron density of the emitting gas can be identified through IR fine structure lines, since different IR fine structure transitions of the same ion have different critical density to collisional de-excitation (e.g.,
[O~III] at 52 and 88 $\mu$m). Moreover, the relative strengths of the fine-structure lines in different ionisation stages can probe the primary spectrum of the ionising source, with these line ratios
providing information on the age of the ionising stellar population and on the ionisation parameter in single H II regions or in starbursts of short duration. 
In addition, these line ratios are sensitive to the presence of non-stellar sources of ionising photons, like the AGN, with some of these lines with higher ionisation potential ([Ne~V] at 14.32 and 24.32 $\mu$m, and [O~IV] at 25.89 $\mu$m) excited only -- or primarily -- by an AGN (\citealt{genzel98}; \citealt{armus07}).

In the analysis performed in this work, we have considered the two PAH features at 6.2 and 11.25 $\mu$m and several mid- and far-IR fine-structure lines, including four emission lines from 
photodissociation regions, PDRs ([Si~II] 34.8$\mu$m, [O~I] 63$\mu$m, [O~I] 145.5$\mu$m and [C~II] 157.7$\mu$m), seven stellar/H~II region lines ([Ne~II] 12.8$\mu$m, [S~III] 18.7$\mu$m, [S~III] 33.5$\mu$m, 
[O~III] 51.8$\mu$m, [N~III] 57.3$\mu$m, [O~III] 88$\mu$m and [N~II] 122$\mu$m) and four AGN lines ([Ne~V] at 14.3 and 24.3$\mu$m, [Ne~III] 15.5$\mu$m and [O~IV] 25.9$\mu$m).

In \citet{spinoglio12} ``global'' relations between the luminosity of the above lines and the total IR
luminosity were derived for all the sources  of the 12-$\mu$m selected sample of Seyfert galaxies 
(12MGS; \citealt{rush93}), regardless the AGN type and the fraction of AGN
contributing to the IR luminosity, while in \citet{bonato14b} an attempt was made to
correlate the line luminosity with the SF IR luminosity and the AGN bolometric
luminosity, although the latter was derived using just a few representative SED templates,
normalised to the observed 12$\mu$m flux density. 

What is new in our approach with respect to the previous works is that we use the broad-band photometry, from UV to millimeter (mm), to constrain the stellar and SF components, and 
the {\em Spitzer--IRS} spectra to constrain the torus. Moreover, we derive new relations between the line and the total IR/AGN accretion luminosities by accounting for the 
different AGN contributions, finding different relations for AGN dominated sources with respect to star-formation dominated ones. 

The total IR, SF and AGN bolometric luminosities have been derived by performing a detailed 3-components (emission from stars, dust heated by star formation and a possible AGN dusty torus)
spectral energy distribution (SED) decomposition (\citealt{berta13}) of the well studied and extensively observed 12MGS local sample.
We note that, thanks to the very detailed data (IR spectrum and photometry along the whole SED), available for this local sample, we are able to constrain the crucial 
physical quantities characterising the AGN and its host galaxy like, i.e., star-formation rate (SFR), AGN luminosity, stellar mass, AGN fraction. 
On the contrary, similar works performed on higher-$z$ samples cannot reach the same kind of precision on the physical parameter derivation, 
because the constraints are not as many, as good and as detailed as here. 

The resulting SF and the reconstructed AGN bolometric luminosities are then 
related to the available mid- and far-IR line luminosities and used to derive new estimates of the relationships between the luminosity of PAH features and fine structure lines and 
AGN bolometric or SF luminosity. Variations of these relations with different AGN fractions are investigated.
By extrapolating these local line vs. AGN (or SF) luminosity relations to higher redshifts, by means of recent {\em Herschel} galaxy 
evolution results (e.g., \citealt{gruppioni13}), we then obtain mid- and far-IR line luminosity functions (LFs). Thanks to the line LFs we can then estimate how many 
SF galaxies and AGNs we expect to detect in the different IR lines at different redshifts with future IR facilities, like, e.g., the {\em James Webb Space Telescope} ({\em JWST}; \citealt{gardner06}) and the 
{\em SPace IR telescope for Cosmology and Astrophysics} ({\em SPICA}; \citealt{nakagawa12}).

The paper is organised as follows.  We present the 12MGS Seyfert sample in Section 2. In Sections 3 and 4 we present the SED decomposition method and discuss the main results obtained. 
In Sections 5 and 6 we derive the AGN/SB diagnostics and the relations between the line and the SF or accretion luminosity. In Section 7 we derive the line luminosity functions up to high redshifts
and the estimates for future IR facilities. Finally, in Section 8 we present our conclusions. 

\noindent Throughout this paper, we use a Chabrier (2003) initial mass function (IMF) and we adopt a $\Lambda$CDM cosmology with $H_{\rm 0}$\,=\,71~km~s$^{-1}$\,Mpc$^{-1}$, $\Omega_{\rm m}$\,=\,0.27, and $\Omega_{\rm \Lambda}\,=\,0.73$. 

\section{The sample}
\label{sec_sample}
The parent reference sample in our work is the extended 12-$\mu$m galaxy sample (12MGS) selected by \citet{rush93} (hereafter RMS) from the 
{\em Infrared Astronomical Satellite (IRAS)} Faint Source Catalogue Version-2 (FSC-2). The original RMS sample consists of 893 galaxies with {\em IRAS} 12-$\mu$m flux $>$0.22 Jy, 
118 of which were preliminary classified as AGN (51 Seyfert 1s or quasars, 63 Seyfert 2s and 2 blazars) from existing catalogues of active galaxies (\citealt{spinoglio89}; \citealt{veron91}; \citealt{hewitt91}). 
Although the selection of active galaxies based on their rest-frame mid-IR flux is commonly considered one of the best tools to provide almost unbiased AGN samples  
(e.g., \citealt{spinoglio89} argue that since all galaxies emit a similar fraction of their bolometric luminosity at 12 $\mu$m, the AGN selection based on their rest-frame 
12-$\mu$m flux is almost unbiased), a small anisotropy has been noted by some authors through the comparison with radio (e.g., \citealt{buchanan06,honig11}) or X-ray (e.g., \citealt{gandhi09,asmus14}) selected samples. 
With this cautiousness in mind, we can nonetheless consider the 118 mid-IR selected, optically classified AGN to constitute one of the largest IR-selected almost unbiased/mildly biased samples of AGN, 
likely to be almost representative of the true number and fractions of different active galaxy types in relation to each other, and thus well suited for population statistics. 
In support to our assumption, we note that, some authors (e.g., \citet{gandhi09} and \citet{asmus14} among many others) define the 12-$\mu$m flux as a good indicator of the intrinsic AGN 
continuum emission.\\
The 12MGS sample of AGN benefits from an extensive ancillary dataset, including photometry and spectroscopy from the X-rays to 
radio frequencies, coming from different observational campaigns spread over the past 20 years.
In the next sub-sections we will present the multi-wavelength data available for the 12MGS Seyfert sample, focusing on those used in our analysis.

\begin{figure*}
\includegraphics[width=14cm]{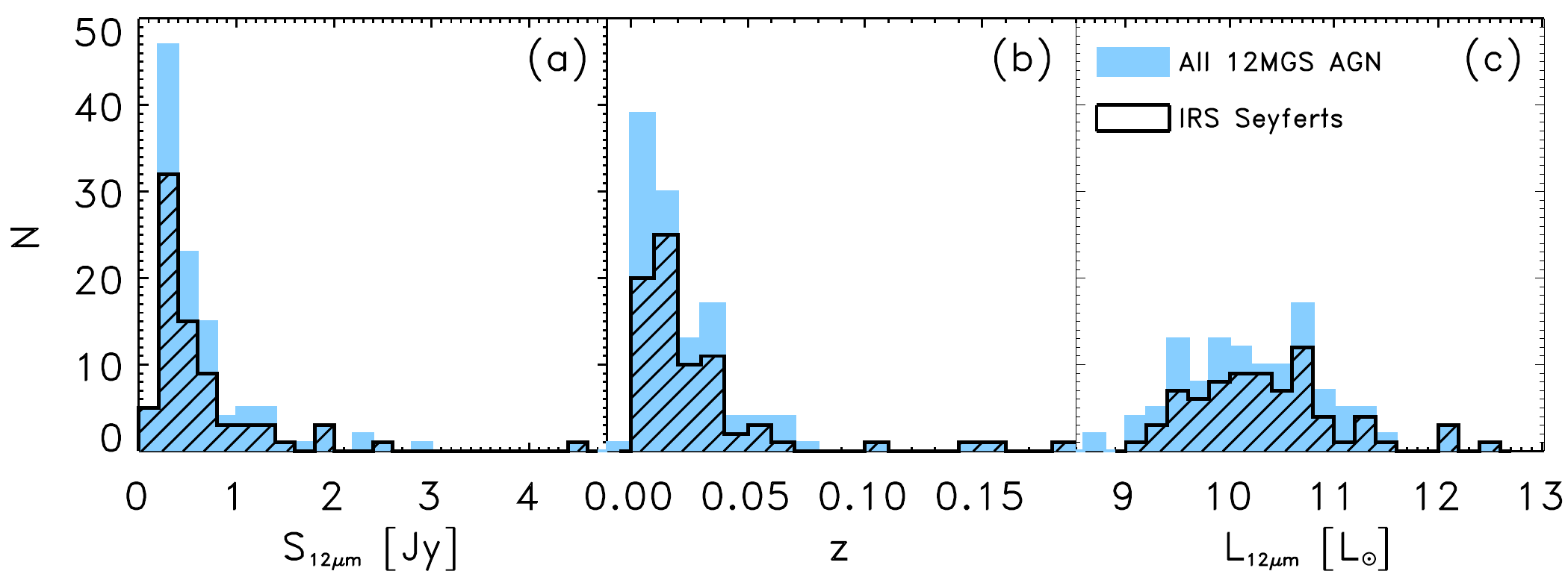}
\caption{Distribution of (a) 12-$\mu$m flux, (b) redshift, and (c) 12-$\mu$m Luminosity of the 76 Seyferts with {\em Spitzer-IRS} low-resolution spectra (black hatched histogram) versus
entire 12MGS sample of optically classified AGN (blue filled histogram).}
\label{fig_fzLhisto}
\end{figure*}

\subsection{Spectroscopic data}
\label{sec_spectroscopy}
As mentioned above, the original AGN classification was based on optical data. Redshifts and optical spectroscopic data from the literature can be found in the IPAC-NED public Database. Using data
from the literature, \citet{brightman11b} re-classified the 116 (excluding the two blazars) 12MGS AGN into Seyfert 1s, intermediate Seyfert 1.2-1.9s, Seyfert 2s and H~II/LINERs, based on optical line ratio diagnostics, 
X-ray, 12 $\mu$m and [O~III] luminosities and X-ray spectral properties.

\citet{tommasin08,tommasin10} presented {\em Spitzer-IRS} high-resolution spectra between 10 and 37 $\mu$m for 91 of the 118 AGN of the 12MGS (some of which changed their classification based on these spectroscopic data; see \citealt{tommasin10} for details), while \citet{wu09} obtained low resolution {\em IRS} spectra in the 5.5--35 $\mu$m range for 103 Seyferts of the 12MGS. 

To obtain a homogeneous {\em IRS} spectroscopy data set we re-reduced the {\em IRS} data. First, we searched for low-resolution {\em IRS} spectroscopy of 12MGS objects in the Spitzer data archive. 
We found 81 and 49 mapping and staring {\em IRS} observations, respectively.
To reduce the data, we used the standard pipeline (version C18.18). In addition to the standard reduction, we subtracted the background emission and removed rogue pixels using IRSCLEAN with the appropriate rogue mask for each observation. For the staring observations, the spectra were extracted assuming point-like emission. For the mapping observations, we created a data cube by projecting the single pointings into a grid in a manner similar to CUBISM (\citealt{smith07b}). The spectra were extracted from the data cubes using 7.7 arcsec$\times$7.7 arcsec and 17.8 arcsec$\times$17.8 arcsec square apertures in the short-low (SL; 5.2-14 $\mu$m) and long-low (LL; 14-36 $\mu$m) cubes, respectively, centered on the nuclei. We applied an aperture correction based on the IRS mapping observations of the stars HR7341 (AOR 16295168) and HR6606 (AOR 16463104). We did this to obtain comparable spectra from both the mapping and the staring data.

In some cases, we found a mismatch in the continuum level between the SL (slit width of 3.6 arcsec) and the LL (slit width of 10.5 arcsec) spectral ranges. This can be produced by two reasons: the LL slit (or aperture) includes noticeable extended emission; or the slits were not well centred on the nucleus. Since we are interested on the {\em IRS} spectra to constrain the AGN torus, we excluded from our analysis all the spectra with a mismatch larger than 20\% in the continuum level at 14 $\mu$m, which may indicate that there is extended emission in the LL spectra. In total, we obtained the {\em IRS} spectra of 76 point-like sources, 56 of them with mapping observations and the rest with staring observations.

For 26 Seyfert galaxies (out of the selected 76) also the far-IR fine-structure lines have been observed with the {\em Herschel--PACS} and {\em --SPIRE} spectrometers, and 
presented by \citet{spinoglio15,pereira13} and Fern\'andez-Ontiveros et al. (in preparation). 

Our final sample used for this study consists of 76 AGN: 42 Seyfert 1 (including 4 quasars), 27 Seyfert 2 and 7 non-Seyfert galaxies (6 H~II and 1 LINER).
In Tables~\ref{tab_mir_lines} and~\ref{tab_fir_lines} we present the fluxes of the mid-IR (from \citealt{tommasin08,tommasin10}) and far-IR lines (from \citealt{spinoglio15,pereira13}, Fern\'andez Ontiveros et al. in preparation) considered in this work.

\begin{table*}
 \caption{12MGS mid-IR lines}
 \rotatebox{90}{
\begin{tabular}{lcrrrrrrrrrrr}
\hline \hline
 Name    &  z     &    [S~IV]10.8  &   [Ne~II]12.8 & [Ne~V]14.3 & [Ne~III]15.5 & [S~III]18.7 &  [Ne~V]24.3 & [O~IV]25.9 & [S~III]33.5 & [Si~II]34.5 & PAH11.2 & PAH6.2 \\ \\
             &          &    \multicolumn{10}{c}{$\times 10^{-14} erg \cdot s^{-1} cm^{-2}$}   \\   \hline    
   3C120          &0.033  &24.1  &7.84   &16.6  &27.6  &7.34    &  29.0 & 123.0& 17.3 &  36.3      &  ...   & $<$13.5 \\
   3C234          &0.1849 &2.86  &$<$1.29  &2.35  &3.4   & ... &  2.91 & 8.97 & ...  &  ...        &  ...   & $<$8.2 \\
   3C273          &0.1583 &3.04  &1.55   &3.38  &6.0   &1.53    &  2.94 & 8.47 & ...  &  ...        &  ...   & $<$21.7\\
   3C445          &0.0562 &2.75  &2.31   &1.95  &6.23  &0.0     &  5.84 & 22.6 & $<$3.16&  $<$3.69      &  ...      &   ...\\
   CGCG381-051    &0.0307 &$<$0.61 &19.1   &$<$0.7  &1.35  & ...   & $<$1.18& $<$1.26& 9.36 &  12.5    &  43.8   & 50.7   \\
   ESO012-G021    &0.03   &2.5   &11.95  &3.19  &6.42  &5.63    &  4.61 & 15.98& 11.17&  26.8    &  88.9   & 109.0  \\
   ESO033-G002    &0.0181 &5.54  &2.13   &6.34  &9.22  &3.99    & 5.27 & 14.39& 2.79 &  4.53    &  10.8  & $<$62.5  \\
   ESO141-G055    &0.0371 &3.45  &2.24   &2.25  &5.62  &1.75     & 1.62 & 7.26 & 5.45 &  8.85    &  19.2   &   ...  \\
   ESO362-G018    &0.0124 &1.7   &12.4   &3.27  &7.49  &3.87       & 2.62 & 8.93 & 7.5  &  11.8    &  71.7   & 66.5   \\
   IC4329A        &0.0161 &29.1  &27.6   &29.3  &57.0  &15.0      & 34.6 & 117.0& 16.0 &  32.5     &  ...   & $<$62.0 \\
   IC5063         &0.0113 &47.4  &26.7   &30.3  &66.3  &21.8      & 23.6 & 114.0& 31.0 &  52.7    &  ...   & 21.4  \\
   IRASF01475-074 &0.0177 &2.14  &13.7   &6.38  &9.95  &$<$4.55     & 1.87 & 6.49 & 3.12 &  $<$6.13   &  30.3  & $<$47.4\\ 
   IRASF03450+005 &0.031  &1.96  &1.09   &$<$1.48 &1.82  &0.0      & $<$1.88& 2.52 & 1.54 &  $<$4.5   &  ...   & $<$65.8 \\
   IRASF04385-082 &0.0151 &2.38  &13.9   &2.28  &7.06  &3.58     & $<$1.44& 8.56 & $<$5.23&  $<$5.36  &  50.5  & $<$80.3 \\ 
   IRASF05189-252 &0.0426 &5.63  &21.12  &17.53 &17.76 &3.18      & 11.73& 23.71& $<$24.0&  11.85   &  ...    & 65.4  \\
   IRASF07599+650 &0.1483 &$<$0.66 &3.92   &$<$0.75 &2.45  &$<$1.9      & $<$3.0 & $<$1.8 & ...  &  ...    &  ...  & 33.6  \\
   IRASF08572+391 &0.0583 &$<$0.5  &7.18   &$<$0.75 &1.99  &1.69    &  $<$5.4 & $<$6.0 & $<$12.0&  $<$25.0  &  ...   & $<$52.5 \\
   IRASF13349+243 &0.1076 &9.3   &4.8    &$<$0.87 &4.21  &0.0      & $<$1.47& 7.35 & ...  &  ...    &  ...   & $<$20.6 \\
   IRASF15480-034 &0.0303 &5.19  &5.57   &6.08  &9.35  &2.72     & 8.9  & 35.0 & 5.2  &  5.13    &  49.2 & $<$52.9 \\ 
   Izw001         &0.0611 &6.95  &2.4    &4.56  &6.26  &0.0     &  $<$1.35& 8.92 & 5.28 &  ...     &  10.8  & $<$29.0  \\
   MCG-02-33-034  &0.0146 &9.87  &7.37   &6.75  &15.93 &7.61     & 20.0 & 81.93& 20.42&  29.46  &  20.7  & $<$51.1\\ 
   MCG-03-34-064  &0.0165 &52.8  &56.2   &62.9  &119.0 &27.0      & 38.1 & 115.0& 17.2 &  24.8   &  71.2 & 42.7  \\
   MCG-03-58-007  &0.0315 &2.75  &8.52   &6.63  &9.29  &5.2       & 3.91 & 8.8  & $<$2.84&  11.7   &  47.1  & 55.8  \\
   MCG-06-30-015  &0.0077 &8.42  &4.98   &5.01  &5.88  &6.48       & 7.37 & 26.0 & 6.51 &  9.26    &  25.5 & $<$63.3 \\ 
   MCG+00-29-023  &0.0249 &$<$0.72 &47.1   &1.01  &4.43  &7.89     & $<$5.42& $<$6.01& 49.3 &  136.0   &  259.0  & 210.0\\  
   MRK0006        &0.0188 &16.69 &28.0   &9.39  &49.34 &14.1    &  10.43& 48.24& 14.09&  36.4   &  20.0 & $<$63.2  \\
   MRK0009        &0.0399 &2.37  &3.23   &2.21  &1.9   &2.38    & 2.24 & 5.55 & 3.94 &  7.32   &  15.2 & $<$81.2  \\
   MRK0079        &0.0222 &10.2  &10.2   &6.55  &19.6  &8.98    &  12.7 & 42.0 & 14.0 &  30.5    &  13.4  & $<$40.5  \\
   MRK0231        &0.0422 &$<$2.1  &19.67  &$<$3.0  &3.05  &$<$4.0     & $<$18.0& $<$9.5 & $<$8.0 &  16.0   &  ...  & 75.0 \\
   MRK0273        &0.0378 &9.58  &41.9   &11.68 &33.57 &13.35   &  15.38& 56.36& 42.56&  14.7   &  ...   & 130.0\\ 
   MRK0335        &0.0258 &0.43  &0.25   &0.38  &0.61  &0.0     & 1.97 & 7.24 & $<$1.21&  $<$1.45  &  ...   & $<$51.7 \\
   MRK0463        &0.0504 &29.86 &9.25   &18.25 &40.78 &15.85   &  19.93& 69.17& 15.5 &  29.79   &  ...  & $<$17.9 \\
   MRK0509        &0.0344 &4.13  &14.0   &4.74  &14.5  &7.19      & 6.82 & 27.5 & 7.41 &  14.5   &  53.9   & 49.2   \\
   MRK0704        &0.0292 &5.09  &$<$3.3   &3.93  &5.63  &$<$4.8     & $<$3.75& 11.8 & 4.3  &  ...     &  ...  & $<$73.8  \\
   MRK0897        &0.0263 &$<$1.5  &24.03  &1.06  &4.38  &14.91   &  $<$0.8 & 0.62 & 22.3 &  21.44  &  117.0& $<$77.7 \\
   MRK1239        &0.0199 &6.08  &9.4    &3.4   &9.38  &1.77    &  3.22 & 15.6 & 9.09 &  10.5   &  34.2  & $<$71.8 \\
   NGC0034        &0.0196 &$<$1.46 &52.1   &$<$2.19 &6.37  &7.56       & $<$0.37& $<$0.66& $<$10.7&  40.5   &  292.0 & 335.0 \\
   NGC0262        &0.015  &7.16  &16.4   &5.82  &20.4  &7.0     &  4.95 & 17.6 & 12.2 &  9.81   &  ...  & $<$57.8 \\
   NGC0424        &0.0118 &8.98  &8.7    &16.1  &18.45 &6.96    &  6.37 & 25.8 & 9.82 &  8.14   &  19.2  & $<$69.5 \\
   NGC0513        &0.0195 &2.77  &12.8   &1.91  &4.43  &6.76     & 1.09 & 6.54 & 14.5 &  27.49   &  74.7   & 95.0  \\
   NGC0526A       &0.0191 &5.32  &5.77   &6.35  &10.4  &$<$2.5      & 5.92 & 19.3 & 5.91 &  8.58    &  ...   & $<$16.0 \\
   NGC0931        &0.0167 &10.7  &5.47   &14.3  &15.41 &4.86    & 13.67& 42.6 & 11.97&  13.72   &  39.5  & $<$74.7 \\
   NGC1056        &0.0052 &$<$1.31 &33.6   &$<$1.8  &10.4  &18.3    & $<$1.23& 1.4  & 36.6 &  49.1    &  214.0  & 269.0 \\
\hline
\end{tabular}
\label{tab_mir_lines}
}
\end{table*}

\begin{table*}
\setcounter{table}{0}
\caption{$-$ Continue}
 \rotatebox{90}{
\begin{tabular}{lcrrrrrrrrrrr}
\hline \hline
 Name    &  z     &    [S~IV]10.8  &   [Ne~II]12.8 & [Ne~V]14.3 & [Ne~III]15.5 & [S~III]18.7 & [Ne~V]24.3 & [O~IV]25.9 & [S~III]33.5 & [Si~II]34.5 & PAH11.2 & PAH6.2 \\ \\
             &          &    \multicolumn{10}{c}{$\times 10^{-14} erg \cdot s^{-1} cm^{-2}$}   \\   \hline    
   NGC1125        &0.0109 &6.07  &16.37  &5.09  &15.55 &10.99   &  9.69 & 40.36& 23.94&  31.32   &  64.2   & 77.5\\  
   NGC1194        &0.0136 &5.05  &3.81   &4.28  &7.37  &$<$2.25      & 3.76 & 15.1 & 3.98 &  3.6    &     ...  &   $<$50.5 \\
   NGC1320        &0.0089 &9.01  &9.58   &10.7  &13.6  &4.15    &  7.46 & 27.4 & 12.0 &  10.4   &     51.5 &    68.6  \\
   NGC1365        &0.0055 &18.6  &143.0  &19.1  &61.3  &51.2    &  97.5 & 365.0& 720.0&  1303.0 &     57.0 &    1730.0\\ 
   NGC1566        &0.0050 &$<$2.04 &17.22  &0.97  &9.42  &6.98      & ...  & 6.12 & 7.37 &  15.0   &     ...  &    276.0  \\
   NGC2992        &0.0077 &26.6  &53.5   &23.3  &49.4  &28.6    &  23.2 & 103.0& 74.7 &  109.0  &     115.0&    153.0  \\
   NGC3079        &0.0037 &$<$1.21 &181.0  &$<$2.57 &24.5  &12.7       & $<$2.66& 12.7 & 42.1 &  166.0  &     540.0&    1110.0 \\
   NGC3516        &0.0088 &13.33 &8.07   &7.88  &17.72 &5.86    &  10.39& 46.92& 9.52 &  22.14  &     25.8 &    $<$65.7  \\
   NGC4051        &0.0023 &4.75  &21.2   &10.7  &17.1  &7.45    &  32.2 & 94.6 & 38.8 &  39.6   &     133.0&   92.0  \\
   NGC4151        &0.0033 &89.6  &122.0  &71.7  &205.0 &71.4    &  81.7 & 261.0& 77.6 &  147.0  &    ...  &   $<$69.8 \\
   NGC4253        &0.0129 &12.2  &23.3   &21.0  &24.1  &12.8    &  18.5 & 46.1 & 21.4 &  15.5   &    58.5 &    30.8  \\
   NGC4388        &0.0084 &45.3  &76.6   &46.1  &106.0 &39.1    &  73.0 & 340.0& 85.1 &  135.0  &   103.0&  150.0\\ 
   NGC4593        &0.0090 &5.5   &7.34   &3.09  &8.13  &3.94    &  $<$4.32& 33.3 & 19.1 &  32.2   &    44.5 &    48.2 \\
   NGC4602        &0.0085 &$<$1.2  &7.57   &0.82  &0.63  &3.2        & $<$1.2 & $<$2.3 & 8.68 &  15.3   &     28.2 &    46.1  \\
   NGC5135        &0.0137 &6.07  &36.7   &4.88  &16.7  &11.4    &  15.2 & 71.3 & 38.3 &  140.0  &     124.0&    416.0 \\
   NGC5256        &0.0279 &2.57  &19.8   &2.31  &10.6  &8.38    &  11.9 & 56.8 & 48.2 &  92.3   &  41.0 &   199.0 \\
   NGC5347        &0.0078 &$<$2.41 &4.17   &2.08  &4.09  &$<$2.86      & $<$1.74& 7.64 & $<$3.38&  $<$4.6   &   33.2 &   $<$43.7  \\
   NGC5506        &0.0062 &25.4  &26.4   &18.5  &45.6  &19.1       & 56.5 & 239.0& 90.1 &  137.0  &    34.8 &   108.0  \\
   NGC5548        &0.0172 &6.37  &8.47   &5.4   &7.27  &5.93      & 3.89 & 17.5 & $<$4.14&  12.46  &     34.9 &  10.8  \\
   NGC5953        &0.0066 &1.52  &67.2   &2.24  &16.7  &22.3       & 6.44 & 21.0 & 80.1 &  145.0  &     239.0&    529.0\\ 
   NGC5995        &0.0252 &5.71  &16.5   &6.13  &8.47  &4.22       & 3.11 & 12.9 & 5.32 &  25.4   &     91.2 &   82.5   \\
   NGC6810        &0.0068 &$<$1.09 &103.0  &$<$1.09 &13.4  &41.0       & $<$2.35& 2.55 & 78.2 &  118.0  &    91.2 &    561.0  \\
   NGC6860        &0.0149 &3.36  &5.6    &2.85  &6.65  &2.8        & 2.41 & 12.1 & 7.93 &  10.4   &  21.7 &   63.2   \\
   NGC6890        &0.0081 &2.92  &11.32  &5.77  &6.57  &4.34       & 3.77 & 10.1 & 16.97&  26.54  &     48.8 &    96.6\\   
   NGC7130        &0.0162 &5.27  &79.3   &9.09  &29.4  &19.6      & 5.22 & 19.7 & 48.2 &  93.9    &  216.0  & 327.0  \\
   NGC7213        &0.0058 &2.32  &25.7   &$<$1.85 &12.0  &5.0        & $<$0.89& $<$13.5& 6.97 &  15.7   &     62.2 &    17.3 \\
   NGC7469        &0.0163 &10.5  &179.0  &17.5  &37.5  &75.1    &  17.1 & 22.9 & 76.8 &  198.0  &     482.0&    608.0\\ 
   NGC7496        &0.0055 &1.3   &48.08  &$<$1.8  &6.67  &23.48      & $<$2.4 & $<$2.4 & 39.47&  44.58  &  153.0&  446.0 \\
   NGC7603        &0.0295 &$<$0.72 &12.5   &$<$0.75 &5.71  &4.94    &  $<$0.93& 3.52 & 9.27 &  18.7   &    65.3 &  70.8  \\
   NGC7674        &0.0289 &16.1  &20.1   &21.2  &35.3  &0.0     &  16.5 & 49.3 & 14.4 &  29.7   &     84.6 &  142.0 \\
   TOLOLO1238-364 &0.0109 &5.7   &45.15  &11.15 &27.0  &16.32   &  5.35 & 21.21& 32.8 &  44.99  &    111.0&    151.0\\ 
   UGC05101       &0.0394 &0.91  &34.13  &2.57  &13.66 &5.57    &  2.82 & 7.35 & 15.46&  32.07  &     ...  &   123.0 \\
   UGC07064       &0.025  &4.03  &8.15   &4.16  &6.65  &4.04    & 5.17 & 14.0 & 9.22 &  10.7   &   42.1  & 79.0  \\ 
   \hline \hline
\end{tabular}
}
\setcounter{table}{1}
\end{table*}

\begin{table*}
\caption{12MGS far-IR lines}
 \begin{tabular}{lrrrrrrr}
\hline \hline
 Name    &    [O~III]52 & [N~III]57 & [O~I]63 & [O~III]88 & [N~II]122 & [O~I]145 & [C~II]158\\ \\
             &          \multicolumn{7}{c}{$\times 10^{-14} erg \cdot s^{-1} cm^{-2}$}   \\   \hline    
           3C120 &   58.38 &     ... &   20.42 &   73.48 &     ... &   59.75 &   44.47\\
           3C234 &     ... &     ... &    1.49 &     ... &     ... &     ... &     ... \\
  IRASF03450+0055 &     ... &     ... &     ... &     ... &     ... &     ... &    2.12\\
  IRASF05189-2524 &     ... &   62.19 &   14.12 &     ... &    3.08 &    2.60 &   17.85\\
  IRASF13349+2438 &     ... &     ... &    8.46 &     ... &     ... &     ... &     ... \\
   MCG-03-34-064 &     ... &     ... &   70.28 &   28.45 &     ... &     ... &   26.59\\
   MCG-06-30-015 &     ... &     ... &     ... &     ... &     ... &     ... &    5.95\\
         MRK0009 &     ... &     ... &     ... &     ... &     ... &     ... &   12.97\\
         MRK0231 &     ... &    2.06 &   29.67 &   14.73 &    3.32 &    3.53 &   44.82\\
         MRK0273 &     ... &   11.46 &   39.62 &   40.33 &   18.54 &    8.06 &   90.75\\
         MRK0509 &     ... &     ... &   25.64 &   16.69 &     ... &    0.92 &   36.15\\
          NGC1365 &     ... &  136.83 &  431.82 &  198.16 &  255.96 &   22.01 & 1102.92\\
         NGC3079 &     ... &     ... &     ... &     ... &     ... &     ... &  726.42\\
         NGC3516 &     ... &     ... &   19.22 &   28.68 &    6.80 &     ... &   21.61\\
         NGC4051 &     ... &     ... &   30.92 &    6.86 &    3.85 &    2.80 &   42.91\\
         NGC4151 &   16.31 &   22.60 &  322.48 &   44.68 &    7.34 &   19.04 &   64.57\\
         NGC4388 &     ... &   19.37 &  221.37 &  137.44 &   10.78 &   15.24 &  238.46\\
         NGC4593 &     ... &     ... &     ... &     ... &    3.03 &    2.94 &   20.53\\
         NGC5135 &     ... &     ... &  140.87 &   65.54 &     ... &   10.17 &  228.33\\
         NGC5347 &     ... &     ... &     ... &     ... &     ... &     ... &   12.78\\
         NGC5506 &   30.16 &   34.92 &  163.30 &  106.16 &   15.00 &   14.29 &  121.69\\
         NGC5548 &     ... &     ... &     ... &     ... &     ... &     ... &   17.63\\
         NGC6860 &     ... &     ... &     ... &     ... &     ... &     ... &   41.92\\
         NGC7213 &     ... &     ... &     ... &     ... &     ... &     ... &   31.98\\
         NGC7469 &     ... &   50.61 &  232.03 &   26.43 &   42.12 &   18.15 &  290.12\\
         NGC7603 &     ... &     ... &     ... &     ... &     ... &     ... &   37.58\\
         NGC7674 &     ... &     ... &   57.82 &   25.15 &     ... &     ... &  111.38\\
   \hline \hline
\end{tabular}
\label{tab_fir_lines}
\end{table*}
         
\subsection{X-ray data}
\label{sec_xray}
\citet{Rush96} studied the soft X-ray spectra of the 12MGS from ROSAT All-Sky Survey observations, while \citet{Barcons95} used hard X-ray data from {\em HEAO 1}. \citet{brightman11a} analysed 
the {\em XMM-Newton} spectra (2--10 keV) of 126 galaxies of this sample, estimating their intrinsic power, continuum shape and obscuration levels. 
In a subsequent paper, \citet{brightman11b} redetermined the activity class of the 12MGS sources by way of optical line ratio diagnostics and characterised the optical class by 
their X-ray, 12-$\mu$m and [O~III] luminosities and X-ray spectral properties.
The X-ray classification has been taken into account, together with the classification based on optical and mid-IR spectra, to constrain the torus model when fitting the SED, as described in Section~\ref{sec_sed}. 

The 12MGS sources of our subsample have been observed in X-rays with different instruments, in different bands, during different campaigns. For the purpose of our work, we have considered the XMM data
in the 2-10 keV range as taken from the IPAC-NED\footnote{\tt http://ned.ipac.caltech.edu} public Database, which provides also the X-ray luminosity.

\subsection{Photometric catalogue}
\label{sec_photometry}
To construct the broad-band SEDs of our subsample of the 12MGS, we have searched in the literature (in the IPAC-NED, Simbad\footnote{\tt http://simbad.u-strasbg.fr/simbad/} and Vizier\footnote{\tt http://vizier.u-strasbg.fr} public Databases) for all the available data, from UV to mm. The public data were mostly patchy and derived from different observational campaigns with different instruments and photometries. When more than one observation was available for the same object in the same band, we have chosen the more recent and/or the one giving the "total flux" (e.g. corrected for aperture).\\
In order to maximise the data coverage in the mid-IR range, crucial for constraining and disentangling the AGN component, we have re-binned the {\em IRS} spectral data in 2-$\mu$m intervals and added those data 
to the photometric SED data. Thanks to the {\em IRS} data, we were able to characterise the dusty torus model and constrain its parameters, e.g., by considering high optical depths for sources showing a strong 9.7-$\mu$m absorption feature in their spectra (see next Section for details).\\
In Tables~\ref{tab_phot_mir}, ~\ref{tab_phot_fir} of Appendix~\ref{appendix} we present the photometric data for the 76 12MGS galaxies considered in this work. We stress that the sample is not a complete sample for statistical purpose. However, since it was randomly selected from a nearly complete sample, it maintains a robust statistical significance. 
Therefore, it can be considered as a representative sample of local the Seyfert galaxy population, for which data are available in virtually all wave-bands, 
from UV to mm, to define their SED, including {\em Spitzer-IRS} spectrum (in the mid-IR range) to constrain the torus component.

Figure~\ref{fig_fzLhisto} shows the distributions of the 12-$\mu$m flux, redshift and 12-$\mu$m luminosity for our 76 Seyfert galaxies, compared to those of the entire 12MGS optically classified AGN sample. We confirm that the objects in our study are representative of the range of properties of the 12MGS AGN: in fact, a two-sided Kolmogorov-Smirnov (K-S) test gives P=0.93 (D=0.079), P=0.67 (D=0.10) and P=0.87 (D=0.082) for the 12-$\mu$m flux, the redshift and the 12-$\mu$m luminosity respectively.

\section{Broad-band SED decomposition}
\label{sec_sed}
\begin{figure*}
\includegraphics[width=5cm]{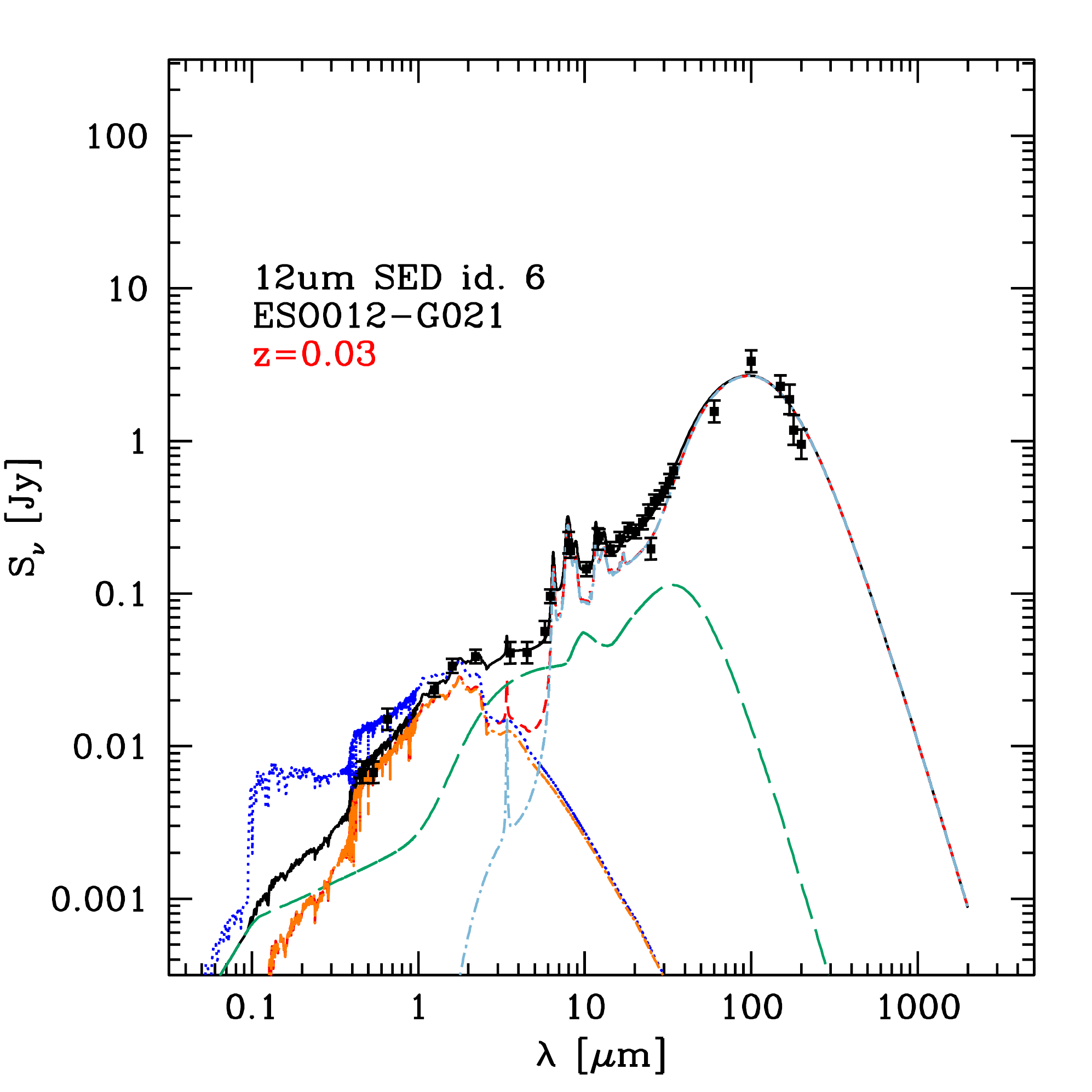}
\includegraphics[width=5cm]{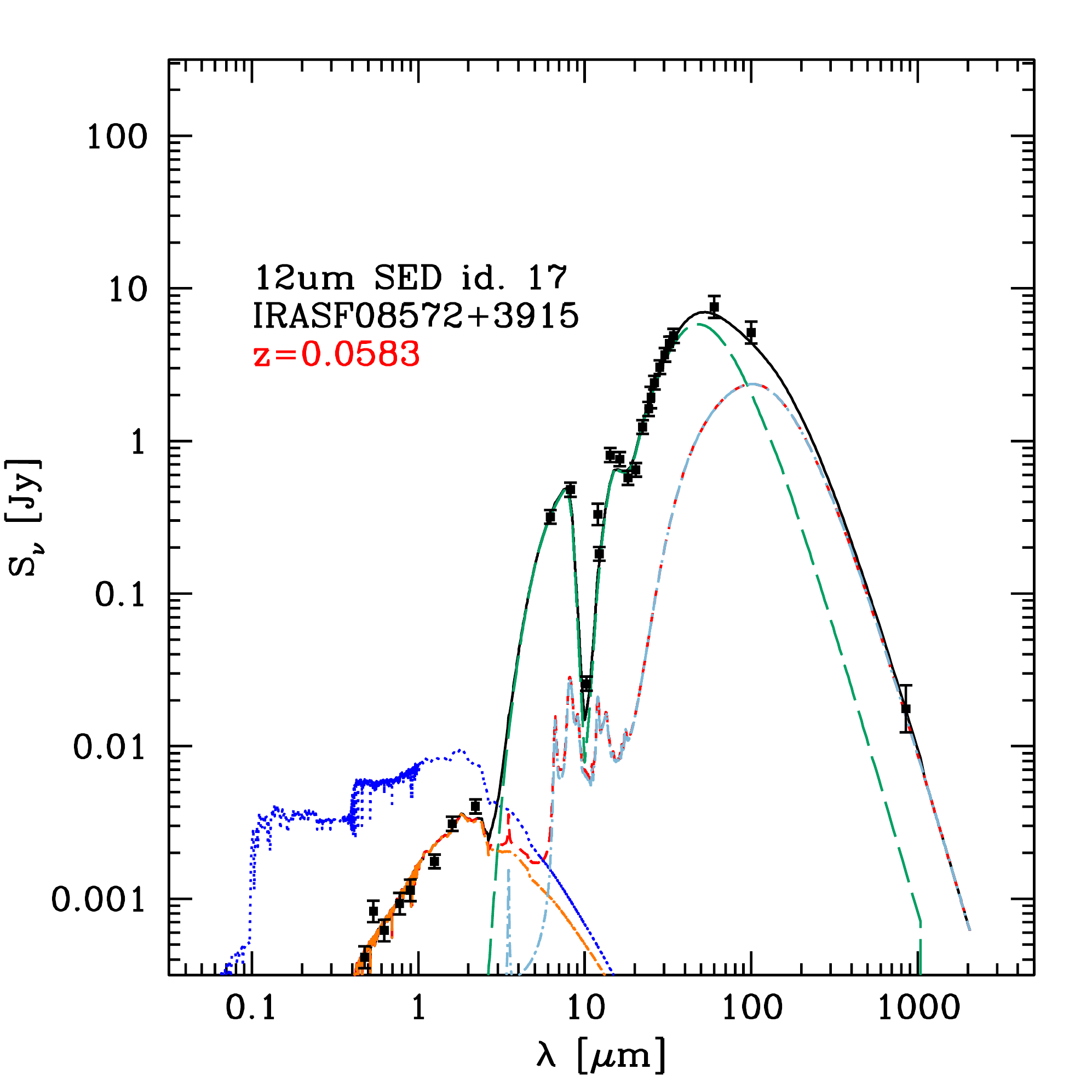}
\includegraphics[width=5cm]{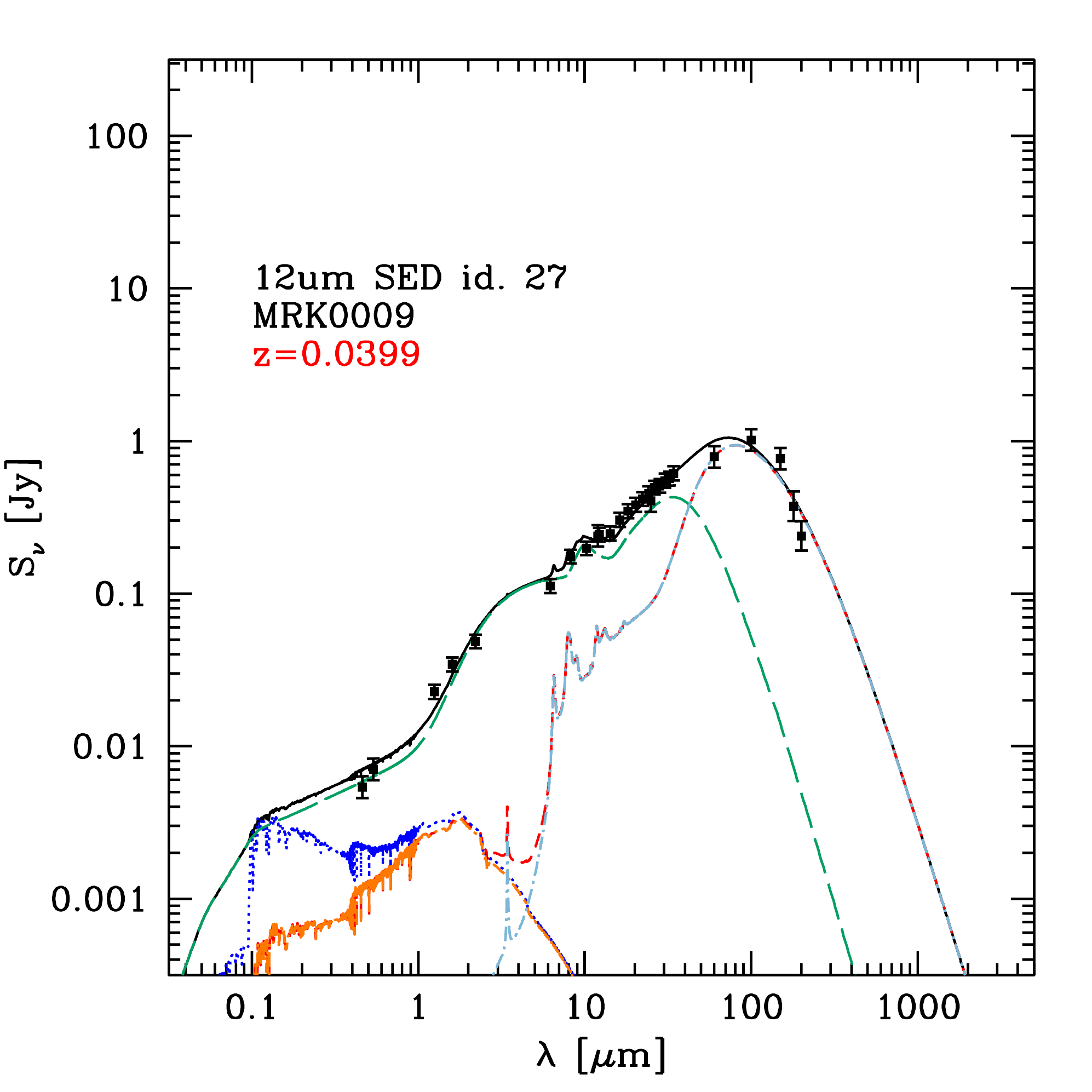}
\includegraphics[width=5cm]{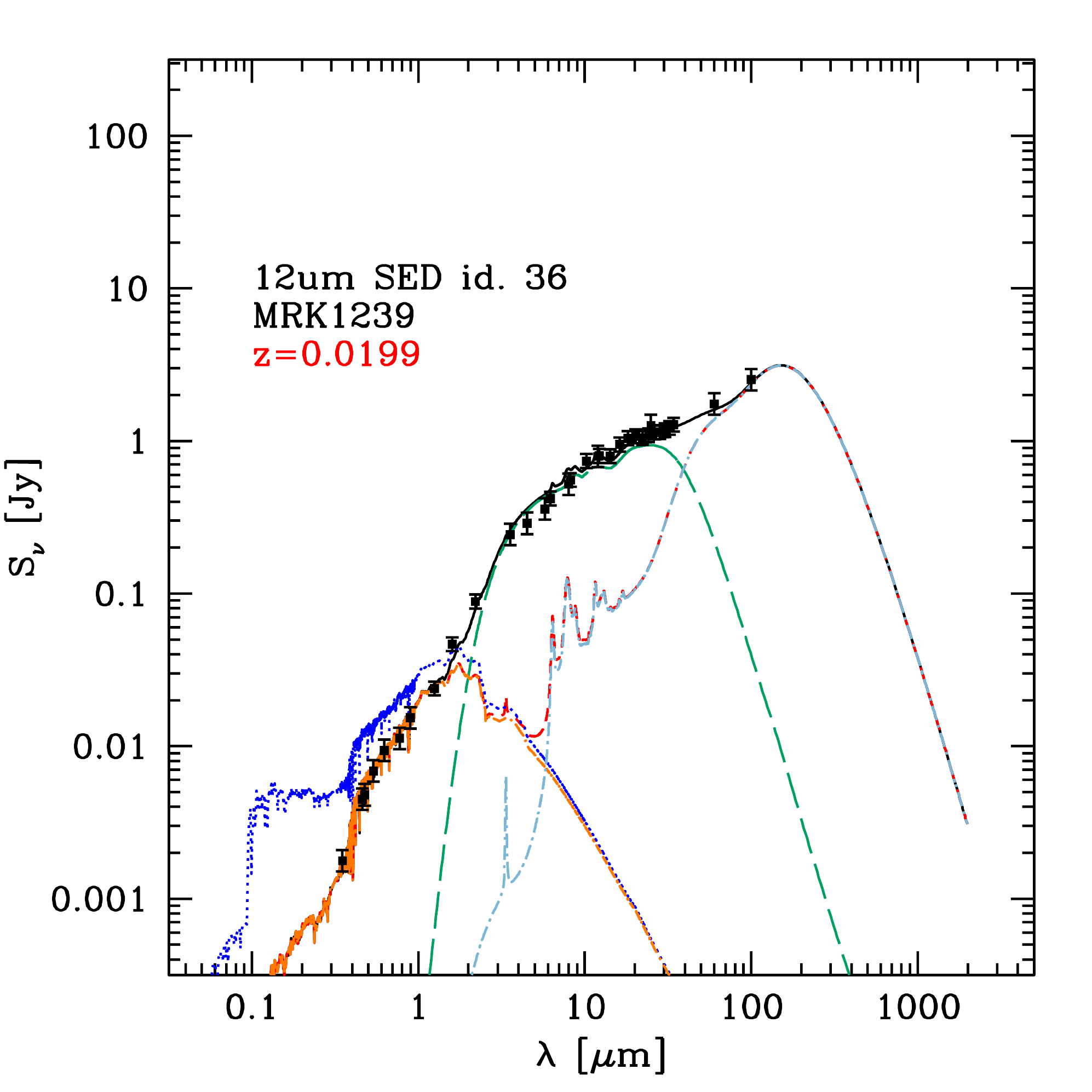}
\includegraphics[width=5cm]{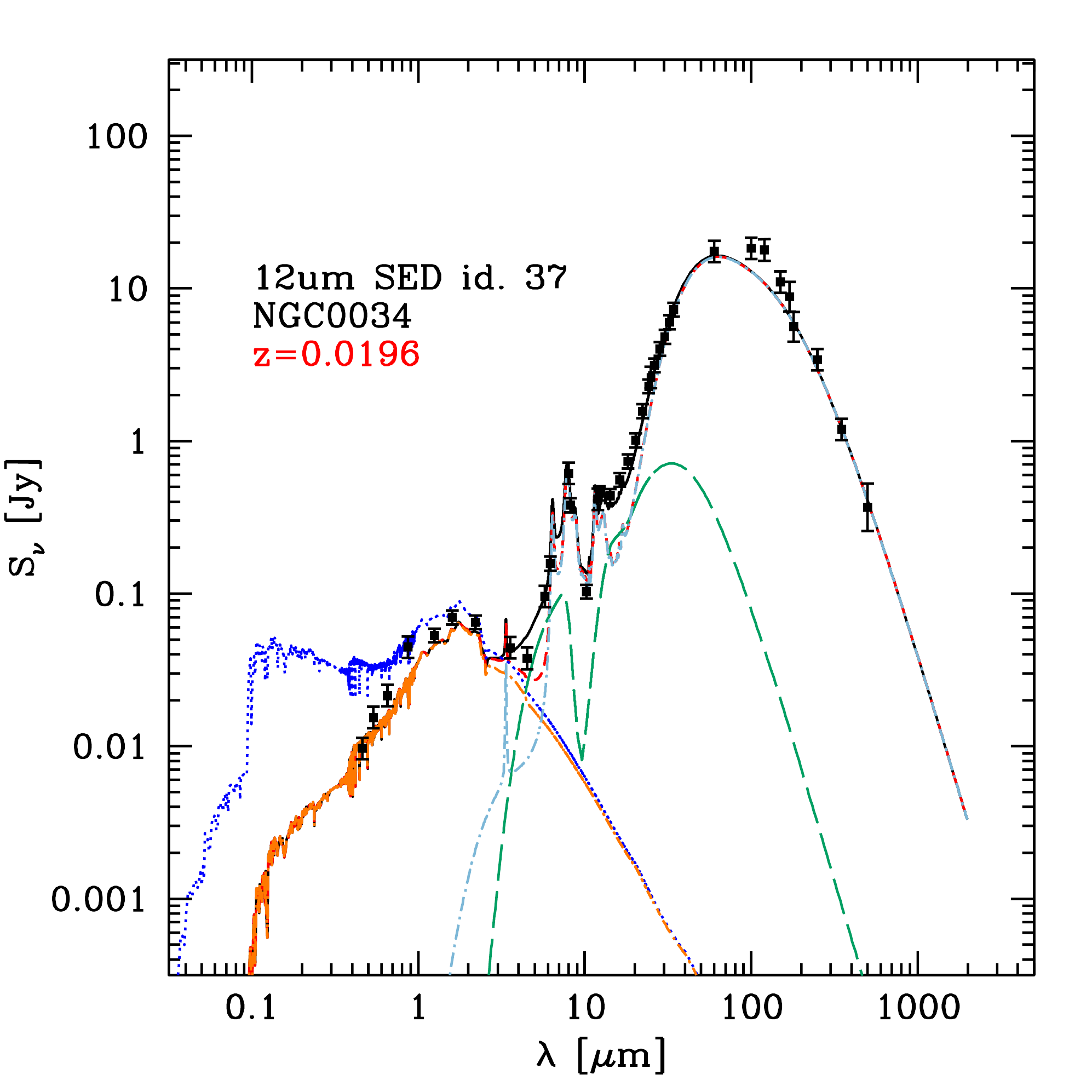}
\includegraphics[width=5cm]{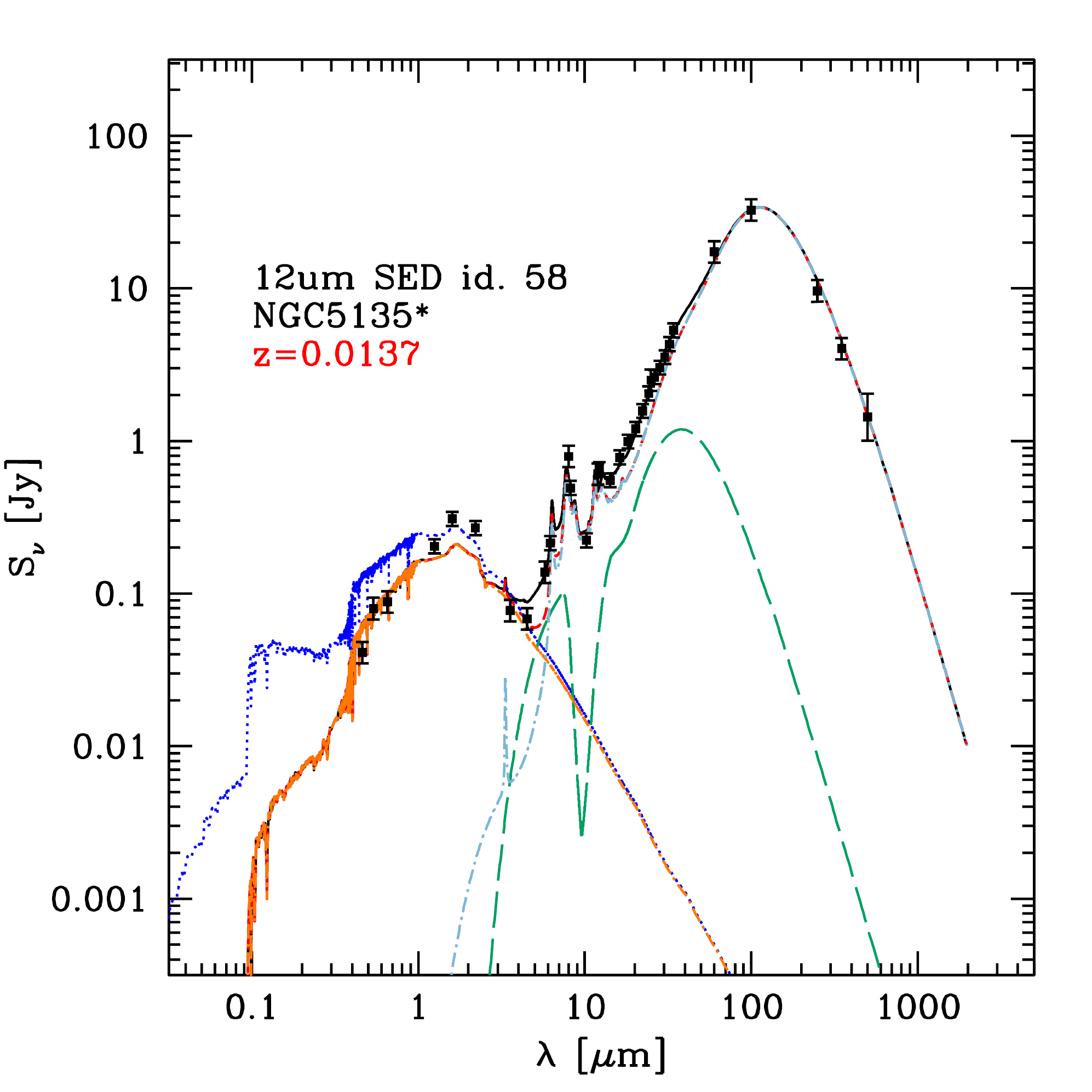}
\includegraphics[width=5cm]{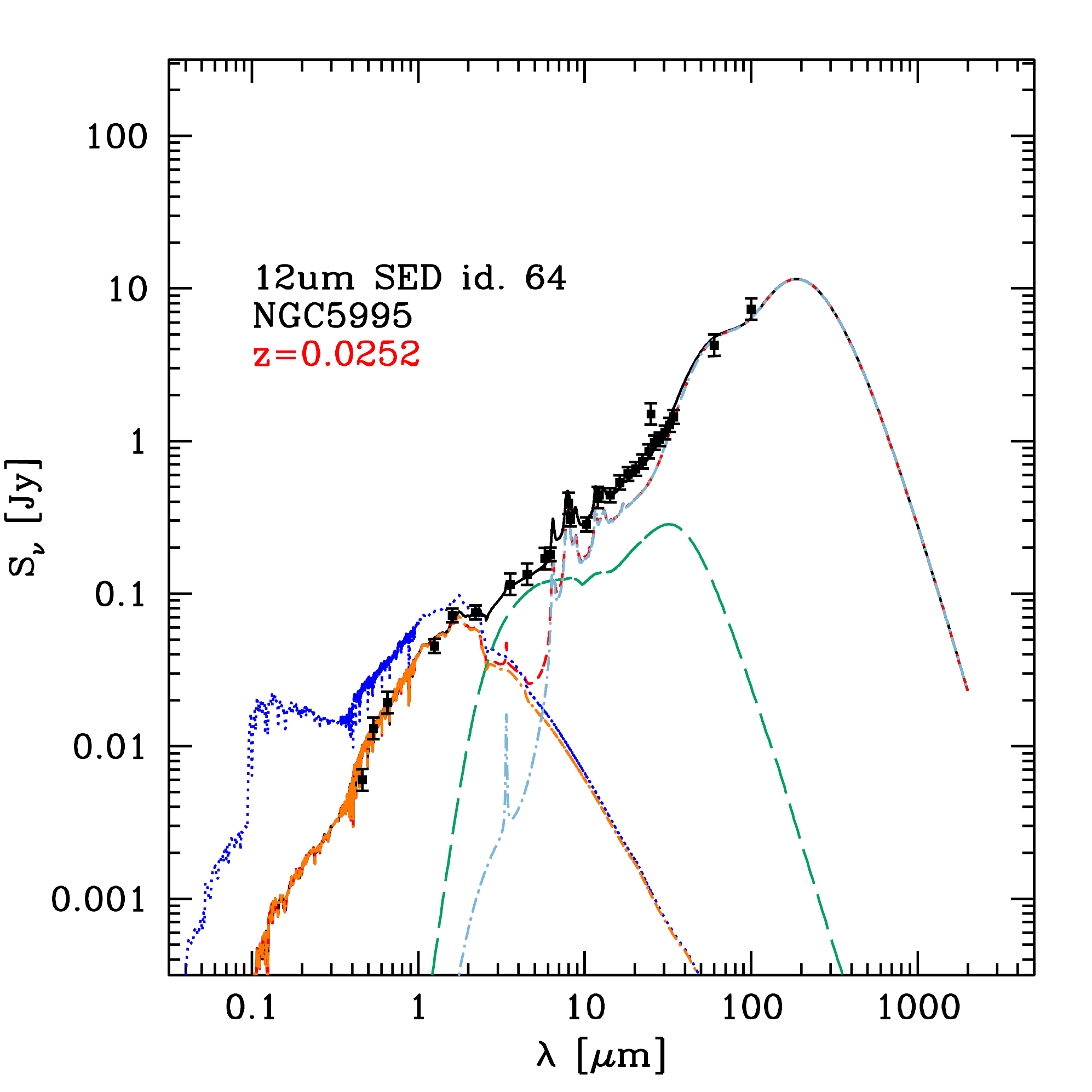}
\includegraphics[width=5cm]{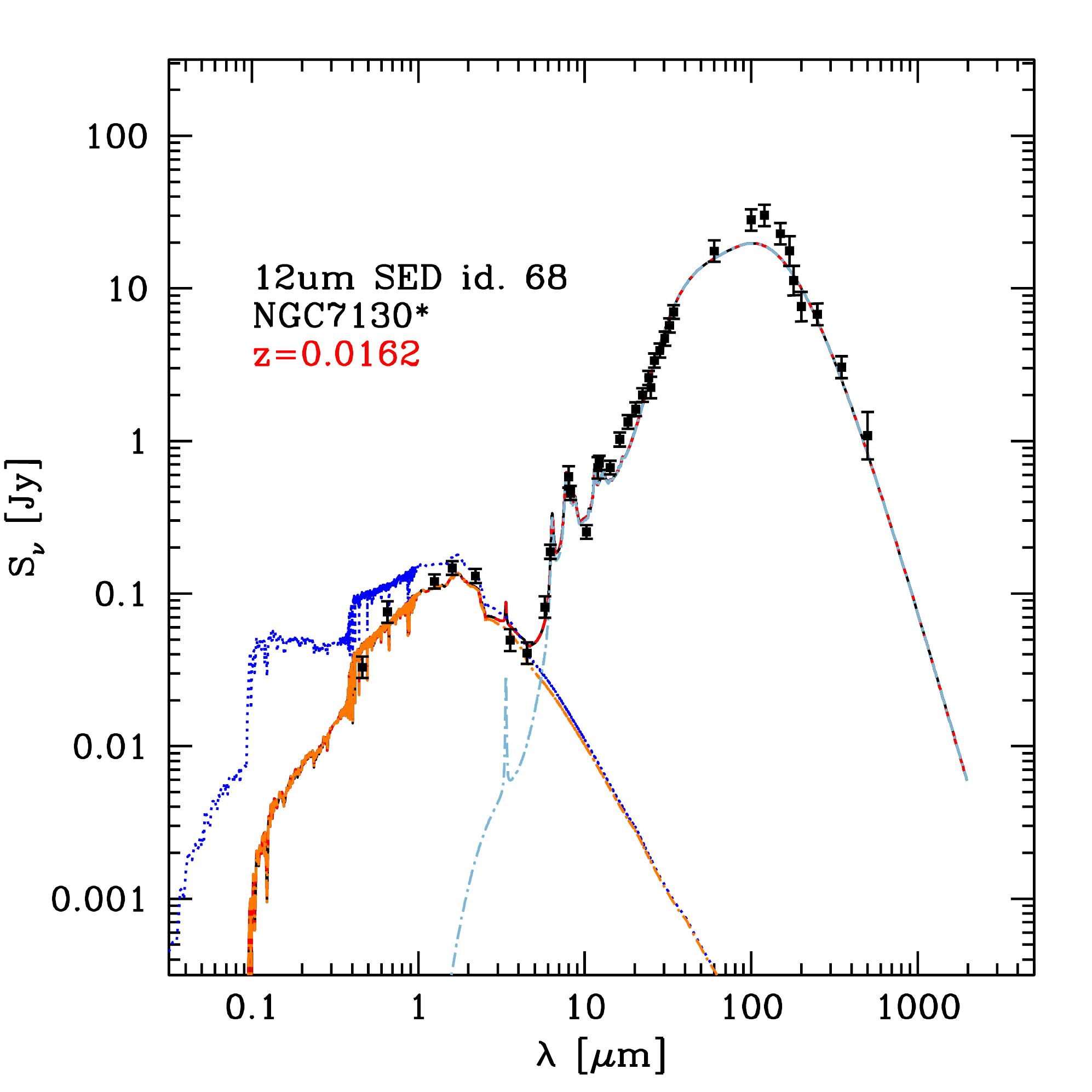}
\includegraphics[width=5cm]{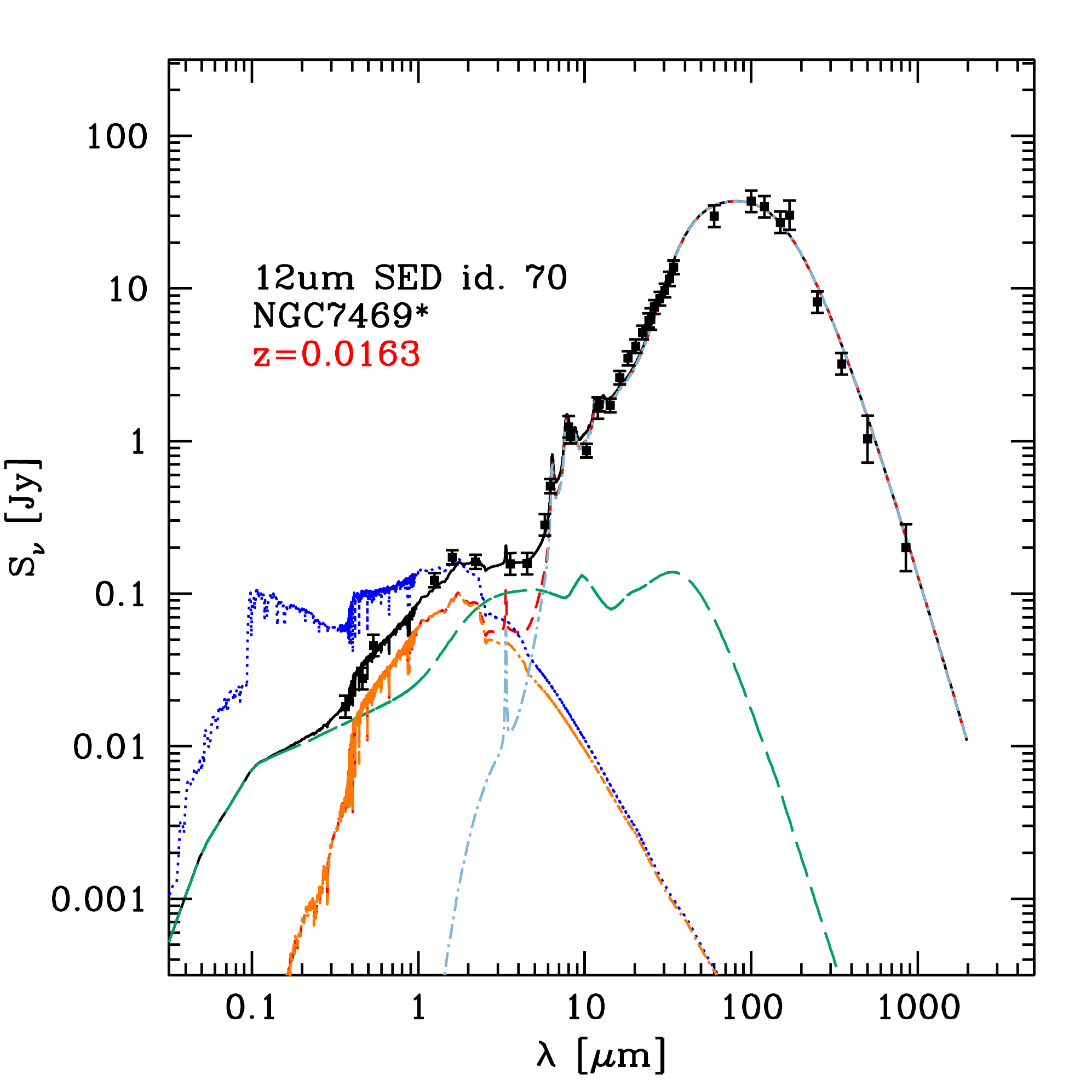}
\caption{Example of observed SEDs decomposed into stellar, AGN and star-formation components using the technique developed by \citet{berta13}. The black filled circles with error bars are our data (with IRS spectra
rebinned, as described in the text). 
Blue dotted lines show the unabsorbed stellar component, red dashed lines show the combination of extinguished stars and dust infrared emission, while long-dashed green lines show the dusty torus emission. Pale-blue dot-dashed lines show the dust re-emission, while black solid lines are the sum of all components (total emission).}
\label{figSED}
\end{figure*}
The rest-frame SEDs of the galaxies in the sample have been 
fitted with a three components model using the approach described by \citet{berta13}. The adopted code was inspired 
by MAGPHYS (\citealt{dacunha08}) for the stellar and dusty star formation part, but it effectively
combines three components simultaneously, i.e. including AGN/torus emission. 
Due to the huge number of possible combinations, the code samples the parameters space randomly, 
allowing for roughly $10^9$ possible combinations at each iteration. 
We defer to \citet{berta13} for further details about the code.

The adopted libraries are the \citet{bruzual03} stellar library \citep[see also][]{dacunha08},  the \citet{dacunha08} 
IR dust-emission library and the library of AGN tori by \citet{fritz06}, updated by \citet{feltre12}. 
The latter includes both the emission of the dusty torus, heated by the central AGN engine, and the emission of
the accretion disk. Note that the considered AGN models are based on a continuous (e.g., ``smooth'') distribution of dust across the
torus; we have not tested the technique with clumpy torus models. For a direct comparison smooth vs. clumpy torus
see \citet{feltre12}.
\begin{figure}
\includegraphics[width=8cm,height=6cm]{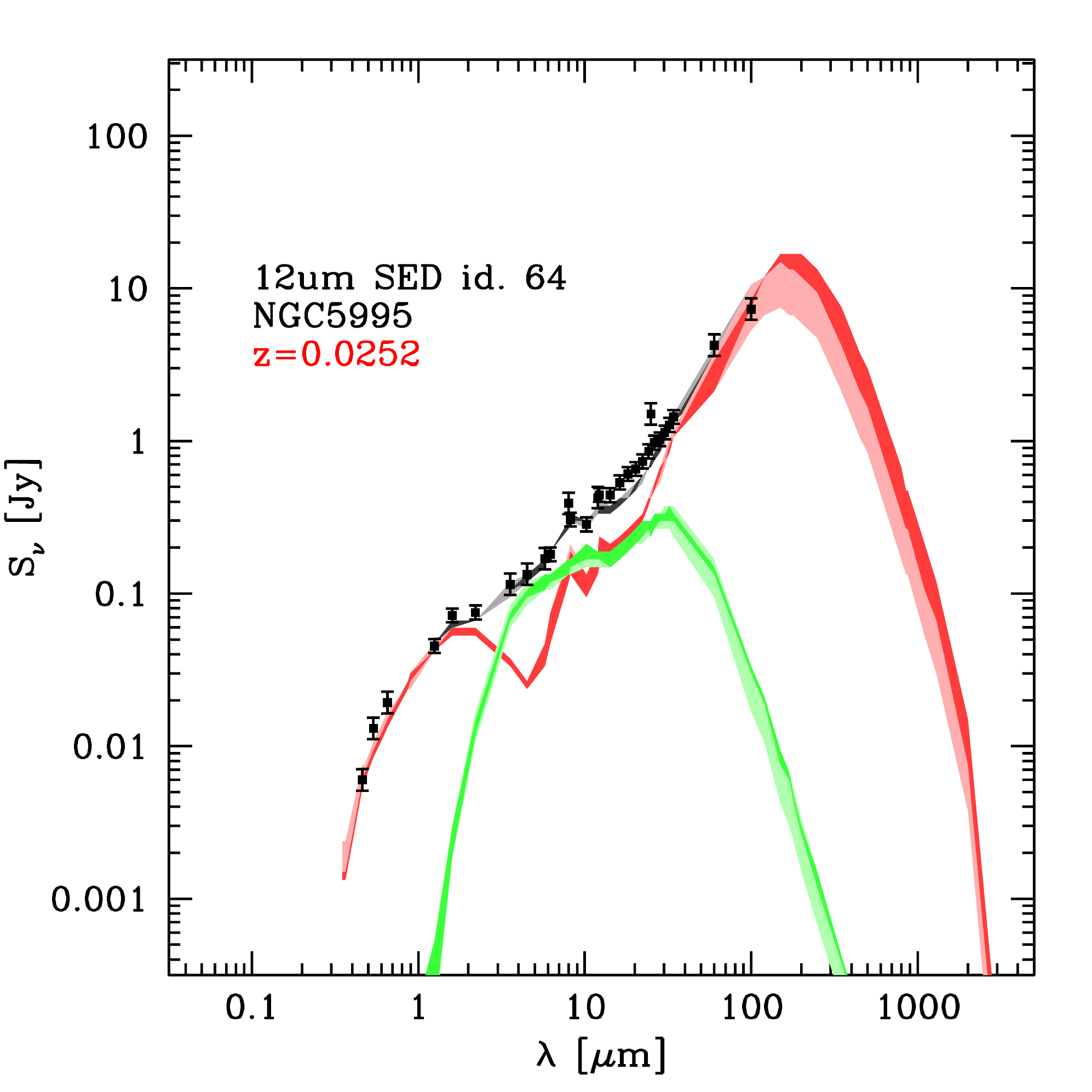}
\includegraphics[width=8cm]{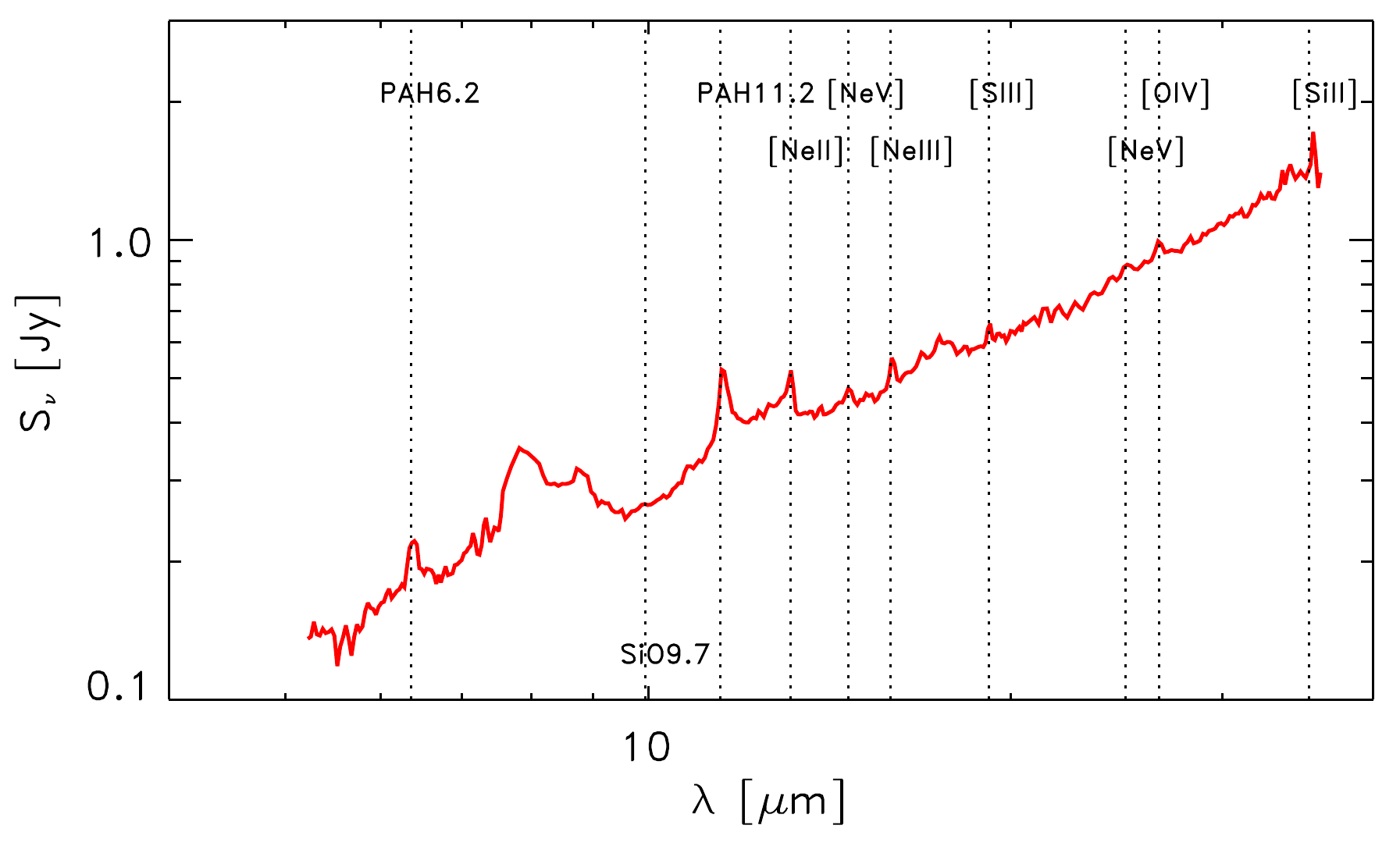}
\caption{Example of decomposed SED, with PDF uncertainty regions associated to the AGN (green) and galaxy (red) 
components highlighted as coloured bands ($top~panel$). $Bottom~panel$: {\em IRS} spectrum associated to the 12MGS source.}
\label{figIRS}
\end{figure}

The AGN library comprises a total of 2376 models, each ``observed'' from 10 different 
viewing angles spanning the 0-90 degrees range with respect to the torus equatorial plane. 
The geometry of the torus varies as parametrised by 
its opening angle of the torus, $\Theta$, and the ratio between the outer and inner radius of the dust 
distribution, $R_m$. The dust density distribution is parametrised as $\rho(r,\theta)=A r^\beta exp(-\gamma cos(\theta))$, 
as a function of the vertical ($\theta$) and radial ($r$) coordinates, with exponents $\gamma$ and $\beta$. 
If $(\gamma,\beta)$$=$(0,0), there is no variation of the dust density in either the vertical or radial direction.
Finally $\tau_{9,7}$ is the reference optical depth of the torus on the equatorial plane, computed 
at 9.7 $\mu$m and integrated over the whole extent of the torus.

Due to degeneracies between the parameters (see \citealt{hatziminaoglou09,pozzi10}), we limit the 
analysis to $\gamma=0.0$ and $\beta=-1.0,\ -0.5,\ 0.0$. 
Moreover, we constrain $R_m\le100$ \citep{netzer07,mullaney11}, in order to avoid too large and cold tori.

The wealth of information available for the 12MGS sample can be used to 
further constrain the torus library and properties, so to 
reduce degeneracies. More specifically, the X-ray spectra allow an estimate 
of the column density $N_H$. In parallel, the presence or absence 
of the 9.7 $\mu$m Silicate feature in absorption in the mid-IR spectra provides 
hints on the most reasonable $\tau_{9.7}$ range to be used. 
Finally optical and X-ray classifications, as well as optical/mid-IR SED shapes 
can put constraints on the most convenient viewing angles to be used.
Hence we combine these three pieces of information to further reduce the number of 
models, and include/avoid those with strong absorption or improper viewing angles in the 
fit, for each specific case. 
In particular, we use the information about $N_H$ from X-rays and the presence/absence of the 9.7 $\mu$m feature to 
reduce the number of possible torus solutions for each source, e.g., by excluding all the torus models with no 9.7 $\mu$m feature in 
absorption for objects with high $N_H$ and/or with that feature observed in
the {\em IRS} spectrum. Viceversa, we have excluded the ``obscured''/``optically thick'' models for objects with low  
column density and/or no 9.7 $\mu$m feature in absorption in the {\em IRS} spectrum.
Analogously, we have used the information from optical spectroscopy (e.g. Seyfert 1 or 2 based on broad or narrow lines and BPT diagrams; 
\citealt{baldwin81}): in case of objects spectroscopically classified as type 1/``broad-line'', we have considered only ``unobscured'' torus models, 
with the optical/UV  
spectrum provided by the AGN/disk and not by the stellar component. Viceversa, for objects classified as type 2, we have considered only  
models with depressed/absorbed optical/UV light, letting it to be reproduced by the host galaxy stellar contribution.\\
In this way the random sampling of the torus library avoids 
to waste computing time on models that will be in any case discarded, and focuses 
on a finer sampling of the appropriate torus parameters range. 

Note that the chosen torus library can in principle be changed and selected among all the 
sets available in the literature, either ``smooth'' (e.g., \citealt{pier92}, \citealt{granato94},
\citealt{fritz06}), ``clumpy'' (e.g., \citealt{nenkova08}, \citealt{honig10}) or ``composite'' (e.g., \citealt{stalevski12}). 

In Figure~\ref{figSED} we show few examples of observed 12MGS SEDs decomposed into stellar, 
AGN and star-formation components using the above described technique. The torus emission,
when needed by the fit, is shown as green dashed line.
The decomposition code provides a probability distribution function (PDF) in each
photometric band and for each fit component, allowing an estimate of the uncertainty related
to each decomposed contribution. In Figure~\ref{figIRS} we show an example of an observed
SED, with the galaxy and AGN components defined by their uncertainty regions obtained
through the PDF analysis (green for the torus and red for the host galaxy). The {\em IRS} spectrum 
associated to the same 12MGS object is also shown (and zoomed) in a separate panel.

\section{Results}
\label{sec_results}
In Table~\ref{tablum} we report the main results obtained from the SED-fitting decomposition described in the previous section. In particular, in Columns $\#$ 2, 3, 4, 5, 6 and 7, for each source we list the 
total IR luminosity (obtained by integrating the SED in the 8--1000 $\mu$m rest-frame, $L_{\rm IR}^{\rm TOT}$), the SFR (obtained through the \citealt{kennicutt98} relation converted to a 
Chabrier IMF), the IR luminosity due to SF ($L_{\rm IR}^{\rm SF}$), the IR luminosity due to AGN accretion($L_{\rm IR}^{\rm AGN}$), the fraction of 5--40$\mu$m luminosity produced by an AGN 
($f_{\rm AGN}(5-40)$) and the AGN bolometric luminosity ($L_{\rm bol}^{\rm AGN}$) respectively. The AGN bolometric luminosity is obtained through a bolometric correction (BC) available for each AGN template. The BC allows us to calculate $L_{\rm bol}^{\rm AGN}$ directly from the IR (1--1000 $\mu$m rest-frame) luminosity of the best-fitting AGN model. In the last three Columns, for comparison, we report the AGN bolometric luminosity derived from the X-ray (by \citealt{brightman11a}), the [Ne~V]14.3 and 24.3$\mu$m and the [O~IV]25.9$\mu$m luminosities (by \citealt{tommasin10}) respectively. 
The uncertainties associated to the parameters derived through the SED decomposition have been obtained from the PDF distribution provided by the fitting routine.

The SED decomposition code provides the fraction of luminosity due to the AGN in several infrared bands (1--5 $\mu$m, 5--10 $\mu$m, 10--20 $\mu$m, 5--40 $\mu$m, 8--1000 $\mu$m): for our
purpose we have chosen to represent the different AGN fractions in the 5--40 $\mu$m band, since it is large enough not to be too sensitive to small scale variations like PAHs, but it is where the
AGN contribution -- if any -- should come out and be maximised with respect to stellar and SF ones. 

\subsection{AGN fraction}
\label{sec_agnfrac}
\begin{figure}
\includegraphics[width=8.5cm]{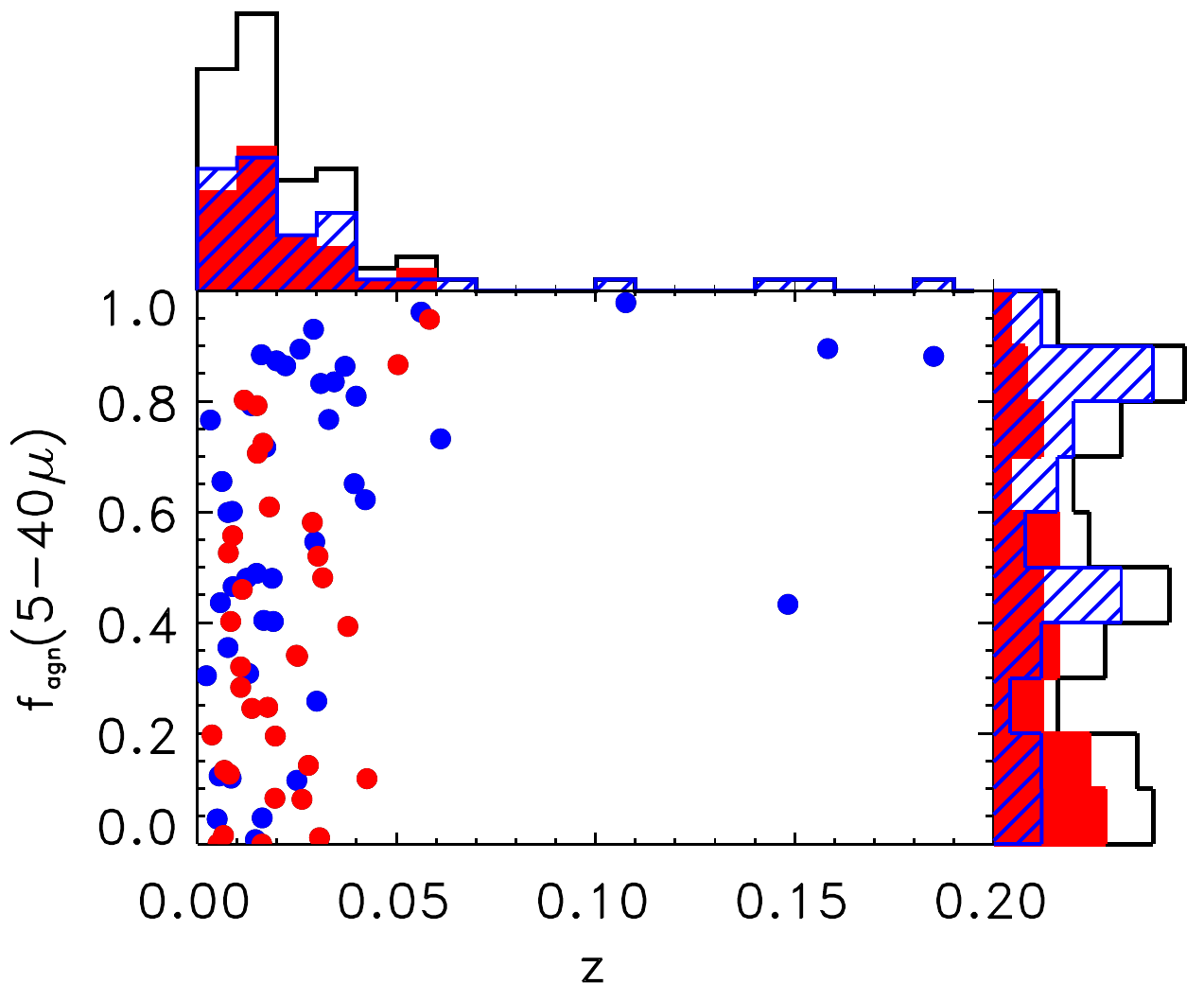}
\caption{{\em Left:} fraction of luminosity due to the AGN in the 5--40 $\mu$m range versus redshift for Seyfert 1 (blue filled circles) and Seyfert 2 (red filled circles) from optical classification. 
The histograms on the top and right of the diagram represent the redshift and AGN fraction distributions respectively: total (black solid), Seyfert 1 (blue hatched) and Seyfert 2 (red filled). }
\label{fig_z_agnfrac}
\end{figure}

\begin{figure}
\includegraphics[width=8.5cm]{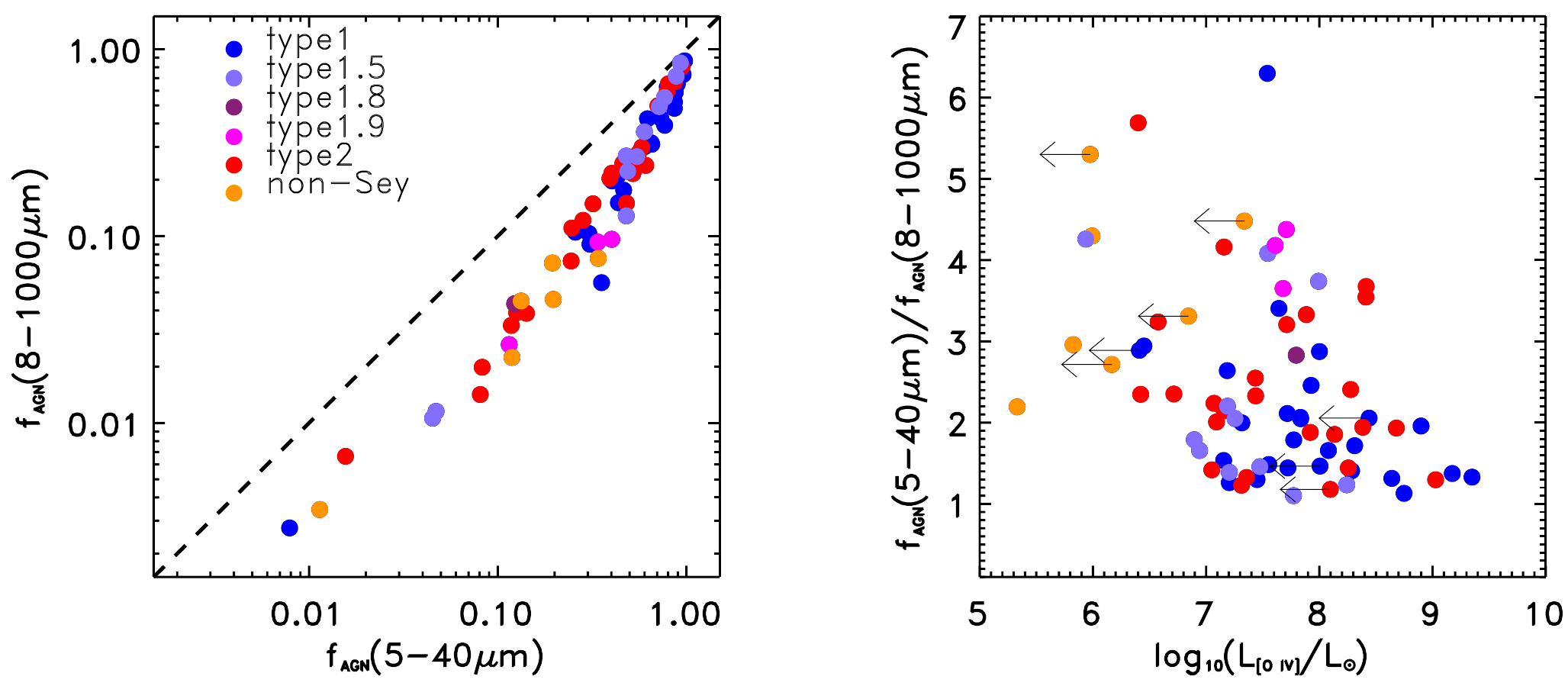}
\caption{{\em Left:} fraction of luminosity due to the AGN in the 5--40 $\mu$m range against that in the 8--1000 $\mu$m range. The dashed line shows the 1-1 relation. {\em Right:} ratio between the fraction of luminosity due to the AGN in the 5--40 $\mu$m range and that in the 8--1000 $\mu$m range, versus the [O~IV] 25.9$\mu$m luminosity. The different types of galaxies -- i.e., Seyfert 1/1.5/1.8/1.9/2 and non-Seyfert (i.e., HII regions and LINERs), as defined through optical spectroscopy (see \citealt{brightman11b}) -- are shown by different colours, as explained in the legend.}
\label{fig_fracMIR_fracFIR}
\end{figure}

 Fig.~\ref{fig_z_agnfrac} we show the fraction of luminosity due to the AGN in the 5--40 $\mu$m range versus redshift for type 1 and type 2 AGN. On the top and along the right axis, we plot
also the redshift distribution and the 5--40 $\mu$m AGN fraction distribution. While there is basically no difference in the $z$-distribution for AGN 1 and 2 (apart from 4 type 1 
sources at 0.1$<$$z$$<$0.2 and no type 2's in that range), type 2 AGN tend to have lower AGN fractions than type 1Õs (more numerous at higher AGN fractions), 
although the total (type1+type2) AGN fraction distribution is almost flat. 
We note that the optical classification is biased against obscuration/torus orientation and often doesnÕt match with X-ray type 1/type 2 definition or with 
IR/SED-fitting classification. However, the general trend of type 1 AGN showing higher intrinsic AGN fractions is not surprising, since it has been shown 
statistically in Table 8 of \citet{tommasin10} for this specific sample, and it is also in agreement with the results of other works 
(i.e., \citealt{veilleux09, brightman11a, netzer15} and references therein). 

In Fig.~\ref{fig_fracMIR_fracFIR} we show the fraction of luminosity due to the AGN in the 5--40 $\mu$m range against that in the 8--1000 $\mu$m range, and the
ratio between the two fractions versus the [O~IV] 25.9$\mu$m luminosity, for the
different types of 12MGS galaxies -- i.e., Seyfert 1/1.5/1.8/1.9/2 and non-Seyfert, as defined through optical spectroscopy and 12 $\mu$m and X-ray luminosity.
From the former diagram it is clear that, even if this sample is biased towards AGN-dominated objects (showing significantly high fractions of the total IR luminosity due to AGN), in the
5--40 $\mu$m band, the AGN contribution is always higher than in the 8--1000 $\mu$m range. The two contributions become similar for objects showing the highest AGN fractions 
(e.g., totally dominated by the AGN). We can also notice that Seyfert 1/1.5 galaxies show, on average, higher $L_{\rm[O~IV]}$ and higher AGN fractions (both at 5--40 and 8--1000 $\mu$m), but 
tend to populate the lower $f_{\rm AGN}(5-40)/f_{\rm AGN}(8-1000)$ part of the diagram (e.g., have higher far-IR to mid-IR luminosity ratios). 
The AGN-dominated objects (those with $f_{\rm AGN}(5-40)$/$f_{\rm AGN}(8-1000)$$\sim$1 and both, $L(8-1000\mu m)$ and $L(5-40\mu m)$ dominated by
the AGN) are more likely to be optically classified as type 1 than type 2.
On the contrary, the non-Seyfert/LINER classified objects are mostly at higher $f_{\rm AGN}(5-40)/f_{\rm AGN}(8-1000)$ (e.g., dominate the mid-IR rather than the far-IR) and 
(as expected) have lower [O~IV] luminosities. Seyfert 2/1.9/1.8 galaxies are spread over the whole range of AGN fraction ratios and luminosities. Low [O~IV] luminosities seem to
correspond to higher mid-to-far-IR fraction ratios. This can be interpreted as the far-IR range to be dominated by SF, as expected for low-luminosity AGN as LINERs, with the AGN
showing up mostly in the mid-IR, where the galaxy SED shows a dip.\\
We have applied a two sample K-S test, to provide an assessment of the likelihood that the Seyfert 1Õs and Seyfert 2Õs samples were drawn from the 
same parent distribution. First we have considered the Seyfert 1's and 
Seyfert 2's distributions plotted in Fig.~\ref{fig_fracMIR_fracFIR}, including the intermediate types (1.5/1.8/1.9) in the Seyfert 2 category. 
We derive that the probability that the two distributions are drawn from the same parent one is $P$$=$0.03 (the test statistic, or the maximum 
difference, is $D$$=$0.299). Then we have performed again the test, by considering all the intermediate types in the Seyfert 1 class. 
As a result, this time we found $P$$=$0.06 ($D$$=$0.359). Therefore, in both cases, the two cumulative frequency 
distributions are more likely to be randomly sampled from populations with different distributions, 
although the significance of this difference is only about 2$\sigma$. 

\begin{table*}
 \caption{Main properties from SED-fitting decomposition of the 76 12MGS Seyfert galaxies}
\begin{tabular}{lcrcccccc}
\hline \hline
 Name    &  log($L_{\rm IR}$)      &    \multicolumn{1}{c}{$SFR$}   &    log($L_{\rm IR}^{\rm SF}$) &   $f_{\rm AGN}$(5$-$40) & \multicolumn{4}{c}{log($L_{\rm bol}^{\rm AGN}$)}  \\ 
  \multicolumn{5}{c}{} &  IR & X-ray  &  [Ne~V]  &   [O~IV] \\ \\
             &   $log(L_{\odot})$ &    \multicolumn{1}{c}{$M_{\odot}/yr$}  &   $log(L_{\odot})$ &    & \multicolumn{4}{c}{$log(L_{\odot})$}  \\   \hline    
                    3C120  & 11.36$\pm$0.01 &  13.83$\pm$0.07 & 11.14$\pm$0.01 &  0.77$\pm$0.05 & 11.50$\pm$0.14 &  12.33$\pm$0.39 &  12.10$\pm$0.24 &  12.44$\pm$0.36 \\
                    3C234  & 12.18$\pm$0.02 &  51.56$\pm$3.45 & 11.71$\pm$0.07 &  0.88$\pm$0.01 & 13.02$\pm$0.01 & ... &  13.11$\pm$0.20 &  13.06$\pm$0.24 \\
                    3C273  & 12.66$\pm$0.02 & 160.80$\pm$0.75 & 12.21$\pm$0.07 &  0.89$\pm$0.03 & 13.06$\pm$0.02 &  14.53$\pm$0.72 &  13.13$\pm$0.20 &  12.83$\pm$0.16 \\
                    3C445  & 11.22$\pm$0.01 &   4.48$\pm$0.20 & 10.65$\pm$0.04 &  0.96$\pm$0.01 & 11.70$\pm$0.01 &  12.67$\pm$0.44 &  11.49$\pm$0.03 &  12.08$\pm$0.22 \\
              CGCG381$-$051  & 11.12$\pm$0.01 &  13.26$\pm$0.07 & 11.12$\pm$0.01 &  0.01$\pm$0.42 &  9.48$\pm$0.27 & ... &$<$10.22 &$<$9.78 \\
              ESO012$-$G021  & 11.02$\pm$0.03 &   9.35$\pm$1.12 & 10.97$\pm$0.12 &  0.26$\pm$0.07 & 10.46$\pm$0.19 &  10.88$\pm$0.20 &  11.06$\pm$0.17 &  11.17$\pm$0.20 \\
              ESO033$-$G002  & 10.58$\pm$0.01 &   2.86$\pm$0.02 & 10.46$\pm$0.01 &  0.61$\pm$0.02 & 10.76$\pm$0.01 &  10.80$\pm$0.12 &  10.89$\pm$0.07 &  10.56$\pm$0.01 \\
              ESO141$-$G055  & 11.14$\pm$0.04 &   7.12$\pm$0.74 & 10.85$\pm$0.10 &  0.86$\pm$0.02 & 11.43$\pm$0.02 &  11.98$\pm$0.30 &  11.12$\pm$0.01 &  10.97$\pm$0.02 \\
              ESO362$-$G018  & 10.23$\pm$0.03 &   1.23$\pm$0.15 & 10.09$\pm$0.12 &  0.48$\pm$0.60 & 10.69$\pm$0.30 &  10.22$\pm$0.01 &  10.13$\pm$0.10 &   9.90$\pm$0.14 \\
                  IC4329A  & 10.85$\pm$0.01 &   2.00$\pm$0.08 & 10.30$\pm$0.04 &  0.88$\pm$0.02 & 11.53$\pm$0.00 &  11.95$\pm$0.27 &  11.66$\pm$0.11 &  11.64$\pm$0.13 \\
                   IC5063  & 10.84$\pm$0.03 &   5.28$\pm$0.63 & 10.72$\pm$0.12 &  0.46$\pm$0.05 & 11.27$\pm$0.30 &  10.51$\pm$0.05 &  11.18$\pm$0.04 &  11.13$\pm$0.04 \\
          IRASF01475$-$0740  & 10.59$\pm$0.03 &   3.49$\pm$0.42 & 10.54$\pm$0.12 &  0.25$\pm$0.60 &  9.97$\pm$0.05 &  10.17$\pm$0.11 &  10.80$\pm$0.20 &  10.04$\pm$0.01 \\
          IRASF03450$+$0055  & 10.94$\pm$0.01 &   4.01$\pm$0.11 & 10.60$\pm$0.03 &  0.83$\pm$0.02 & 11.19$\pm$0.01 & ... &$<$10.67 &  10.19$\pm$0.16 \\
          IRASF04385$-$0828  & 10.83$\pm$0.03 &   3.36$\pm$0.28 & 10.53$\pm$0.08 &  0.71$\pm$0.03 & 11.25$\pm$0.02 & ... &  10.12$\pm$0.20 &  10.07$\pm$0.20 \\
          IRASF05189$-$2524  & 12.21$\pm$0.03 & 155.30$\pm$18.64 & 12.19$\pm$0.12 &  0.12$\pm$0.04 & 11.80$\pm$0.03 &  11.12$\pm$0.01 &  12.45$\pm$0.27 &  11.80$\pm$0.12 \\
          IRASF07599$+$6508  & 12.55$\pm$0.04 & 283.00$\pm$9.00 & 12.45$\pm$0.03 &  0.43$\pm$0.17 & 12.46$\pm$0.05 &   9.29$\pm$0.51 &$<$12.16 &$<$11.84 \\
          IRASF08572$+$3915  & 12.09$\pm$0.02 &  24.31$\pm$2.79 & 11.39$\pm$0.11 &  0.95$\pm$0.60 & 12.90$\pm$0.01 & ... &$<$11.04 &$<$11.41 \\
          IRASF13349$+$2438  & 12.21$\pm$0.01 &  21.89$\pm$1.79 & 11.34$\pm$0.08 &  0.98$\pm$0.01 & 12.76$\pm$0.14 &  11.93$\pm$0.01 &$<$11.85 &  12.26$\pm$0.06 \\
          IRASF15480$-$0344  & 11.11$\pm$0.12 &  10.16$\pm$1.34 & 11.01$\pm$0.13 &  0.52$\pm$0.10 & 11.51$\pm$0.30 &   9.46$\pm$0.32 &  11.46$\pm$0.06 &  11.64$\pm$0.13 \\
                   Izw001  & 11.86$\pm$0.01 &  41.84$\pm$0.64 & 11.62$\pm$0.02 &  0.73$\pm$0.01 & 11.91$\pm$0.06 &  11.68$\pm$0.12 &  12.08$\pm$0.15 &  11.65$\pm$0.06 \\
            MCG$-$02$-$33$-$034  & 10.41$\pm$0.02 &   2.55$\pm$0.04 & 10.41$\pm$0.02 &  0.01$\pm$0.67 &  8.26$\pm$0.88 & ... &  10.76$\pm$0.51 &  11.35$\pm$0.65 \\
            MCG$-$03$-$34$-$064  & 11.06$\pm$0.03 &   5.65$\pm$0.56 & 10.75$\pm$0.10 &  0.72$\pm$0.07 & 11.48$\pm$0.15 &  11.59$\pm$0.18 &  12.14$\pm$0.25 &  11.67$\pm$0.15 \\
            MCG$-$03$-$58$-$007  & 11.37$\pm$0.04 &  20.15$\pm$0.88 & 11.30$\pm$0.04 &  0.48$\pm$0.07 & 11.60$\pm$0.02 & ... &  11.49$\pm$0.05 &  10.85$\pm$0.08 \\
            MCG$-$06$-$30$-$015  &  9.88$\pm$0.03 &   0.49$\pm$0.06 &  9.69$\pm$0.12 &  0.60$\pm$0.02 & 10.26$\pm$0.01 &  10.59$\pm$0.16 &   9.96$\pm$0.06 &  10.07$\pm$0.04 \\
            MCG$+$00$-$29$-$023  & 11.31$\pm$0.03 &  18.71$\pm$0.55 & 11.27$\pm$0.03 &  0.34$\pm$0.13 & 11.23$\pm$0.21 & ... &  10.29$\pm$0.16 &$<$10.48 \\
                  MRK0006  & 10.93$\pm$0.03 &   7.46$\pm$0.02 & 10.87$\pm$0.00 &  0.48$\pm$0.03 & 10.45$\pm$0.03 &  11.30$\pm$0.31 &  11.14$\pm$0.19 &  11.26$\pm$0.23 \\
                  MRK0009  & 11.11$\pm$0.02 &   5.72$\pm$0.35 & 10.76$\pm$0.06 &  0.81$\pm$0.04 & 11.28$\pm$0.04 &  11.15$\pm$0.11 &  11.20$\pm$0.04 &  10.92$\pm$0.01 \\
                  MRK0079  & 10.85$\pm$0.02 &   3.37$\pm$0.10 & 10.53$\pm$0.03 &  0.86$\pm$0.04 & 11.59$\pm$0.31 &  11.35$\pm$0.10 &  11.15$\pm$0.03 &  11.39$\pm$0.05 \\
                  MRK0231  & 12.51$\pm$0.01 & 188.10$\pm$5.50 & 12.27$\pm$0.03 &  0.62$\pm$0.09 & 12.74$\pm$0.04 &  10.21$\pm$0.39 &$<$11.44 &$<$11.29 \\
                  MRK0273  & 11.92$\pm$0.03 &  66.85$\pm$8.02 & 11.83$\pm$0.12 &  0.39$\pm$0.02 & 11.86$\pm$0.30 &  10.75$\pm$0.10 &  12.08$\pm$0.16 &  12.17$\pm$0.21 \\
                  MRK0335  & 10.71$\pm$0.03 &   1.59$\pm$0.16 & 10.20$\pm$0.10 &  0.89$\pm$0.02 & 10.96$\pm$0.02 &  11.15$\pm$0.17 &   9.72$\pm$0.24 &  10.52$\pm$0.05 \\
                  MRK0463  & 11.67$\pm$0.02 &  15.40$\pm$2.03 & 11.19$\pm$0.13 &  0.87$\pm$0.02 & 12.50$\pm$0.30 &  11.20$\pm$0.12 &  12.70$\pm$0.20 &  12.64$\pm$0.22 \\
                  MRK0509  & 11.10$\pm$0.03 &   5.09$\pm$0.45 & 10.71$\pm$0.09 &  0.83$\pm$0.02 & 11.50$\pm$0.02 &  12.17$\pm$0.34 &  11.42$\pm$0.06 &  11.60$\pm$0.12 \\
                  MRK0704  & 10.85$\pm$0.01 &   1.09$\pm$0.07 & 10.04$\pm$0.07 &  0.93$\pm$0.01 & 11.28$\pm$0.00 &  10.88$\pm$0.04 &  11.20$\pm$0.04 &  11.02$\pm$0.02 \\
                  MRK0897  & 11.46$\pm$0.03 &  28.23$\pm$3.39 & 11.45$\pm$0.12 &  0.08$\pm$0.02 & 10.81$\pm$0.01 & ... &  10.24$\pm$0.10 &   9.23$\pm$0.30 \\
                  MRK1239  & 10.93$\pm$0.02 &   3.47$\pm$0.26 & 10.54$\pm$0.08 &  0.87$\pm$0.06 & 11.31$\pm$0.03 &  10.79$\pm$0.01 &  10.71$\pm$0.08 &  10.76$\pm$0.05 \\
                  NGC0034  & 11.42$\pm$0.05 &  24.44$\pm$1.79 & 11.39$\pm$0.07 &  0.19$\pm$0.10 & 11.03$\pm$0.05 &   9.79$\pm$0.17 &$<$10.33 &$<$8.98 \\
                  NGC0262  & 10.53$\pm$0.04 &   1.35$\pm$0.19 & 10.13$\pm$0.14 &  0.79$\pm$0.03 & 11.34$\pm$0.02 &  10.66$\pm$0.03 &  10.55$\pm$0.12 &  10.38$\pm$0.15 \\
                  NGC0424  & 10.56$\pm$0.01 &   1.26$\pm$0.05 & 10.10$\pm$0.04 &  0.80$\pm$0.03 & 11.19$\pm$0.01 &  10.77$\pm$0.03 &  10.84$\pm$0.03 &  10.32$\pm$0.13 \\
                  NGC0513  & 10.83$\pm$0.07 &   6.64$\pm$0.44 & 10.82$\pm$0.07 &  0.08$\pm$0.30 & 10.19$\pm$0.28 & ... &  10.26$\pm$0.02 &  10.16$\pm$0.00 \\
                 NGC0526A  & 10.74$\pm$0.01 &   4.99$\pm$0.03 & 10.70$\pm$0.01 &  0.40$\pm$0.11 & 10.31$\pm$0.17 &  11.22$\pm$0.32 &  10.90$\pm$0.16 &  10.72$\pm$0.11 \\
                  NGC0931  & 10.87$\pm$0.02 &   5.90$\pm$0.71 & 10.77$\pm$0.12 &  0.40$\pm$0.11 & 10.98$\pm$0.09 &   9.90$\pm$0.13 &  10.95$\pm$0.04 &  10.76$\pm$0.06 \\
                  NGC1056  &  9.82$\pm$0.01 &   0.66$\pm$0.01 &  9.82$\pm$0.01 &  0.0 & ... & ... &$<$9.29 &   8.46+-0.05 \\
                  NGC1125  & 10.40$\pm$0.02 &   2.23$\pm$0.06 & 10.35$\pm$0.02 &  0.28$\pm$0.13 & 10.26$\pm$0.08 & ... &  10.11$\pm$0.03 &  10.45$\pm$0.05 \\
                  NGC1194  & 10.21$\pm$0.02 &   0.60$\pm$0.03 &  9.78$\pm$0.05 &  0.79$\pm$0.04 & 11.16$\pm$0.04 &  10.35$\pm$0.07 &  10.30$\pm$0.15 &  10.21$\pm$0.15 \\
                  NGC1320  & 10.10$\pm$0.01 &   0.92$\pm$0.01 &  9.96$\pm$0.01 &  0.56$\pm$0.01 & 10.58$\pm$0.06 & ... &  10.37$\pm$0.02 &  10.12$\pm$0.08 \\
                  NGC1365  & 11.22$\pm$0.03 &  15.74$\pm$0.58 & 11.20$\pm$0.04 &  0.12$\pm$0.60 & 10.47$\pm$0.52 &   9.46$\pm$0.14 &   9.90$\pm$0.11 &  10.71$\pm$0.08 \\
                  NGC1566  & 10.64$\pm$0.02 &   4.32$\pm$0.07 & 10.64$\pm$0.02 &  0.05$\pm$0.33 &  9.23$\pm$0.16 & ... &   7.71$\pm$0.32 &   7.90$\pm$0.30 \\
                  NGC2992  & 10.58$\pm$0.07 &   3.61$\pm$0.26 & 10.56$\pm$0.07 &  0.35$\pm$0.03 & 10.13$\pm$0.06 &  10.40$\pm$0.14 &  10.57$\pm$0.11 &  10.60$\pm$0.11 \\
                  NGC3079  & 10.60$\pm$0.01 &   3.81$\pm$0.04 & 10.58$\pm$0.01 &  0.20$\pm$0.10 & 10.03$\pm$0.25 &   7.65$\pm$0.39 &$<$8.97 &   9.06$\pm$0.21 \\
                  NGC3516  & 10.17$\pm$0.02 &   1.02$\pm$0.05 & 10.01$\pm$0.05 &  0.60$\pm$0.12 & 10.27$\pm$0.01 &  10.69$\pm$0.19 &  10.25$\pm$0.00 &  10.44$\pm$0.05 \\
                  NGC4051  &  9.45$\pm$0.03 &   0.25$\pm$0.03 &  9.40$\pm$0.12 &  0.30$\pm$0.10 &  8.88$\pm$0.30 &   9.02$\pm$0.04 &   9.21$\pm$0.02 &   9.59$\pm$0.08 \\
                  NGC4151  &  9.80$\pm$0.01 &   0.29$\pm$0.01 &  9.46$\pm$0.03 &  0.77$\pm$0.03 & 10.56$\pm$0.02 &  10.38$\pm$0.05 &  10.58$\pm$0.03 &  10.51$\pm$0.02 \\
                  NGC4253  & 10.64$\pm$0.03 &   3.99$\pm$0.17 & 10.60$\pm$0.04 &  0.31$\pm$0.10 & 10.60$\pm$0.04 &  10.70$\pm$0.12 &  11.25$\pm$0.19 &  10.89$\pm$0.10 \\
                  NGC4388  & 10.67$\pm$0.03 &   3.70$\pm$0.11 & 10.57$\pm$0.03 &  0.40$\pm$0.10 & 10.74$\pm$0.20 &   9.70$\pm$0.13 &  10.38$\pm$0.05 &  10.69$\pm$0.03 \\
                  NGC4593  & 10.31$\pm$0.02 &   1.70$\pm$0.03 & 10.23$\pm$0.02 &  0.47$\pm$0.08 & 10.03$\pm$0.01 &  10.67$\pm$0.23 &   9.87$\pm$0.04 &  10.37$\pm$0.07 \\
                  NGC4602  & 10.49$\pm$0.02 &   2.99$\pm$0.07 & 10.48$\pm$0.02 &  0.12$\pm$0.31 & 10.12$\pm$0.11 & ... &   8.97$\pm$0.24 &$<$8.78 \\
                  NGC5135  & 11.23$\pm$0.03 &  15.61$\pm$1.87 & 11.19$\pm$0.12 &  0.25$\pm$0.04 & 10.72$\pm$0.08 &   7.98$\pm$0.44 &  10.51$\pm$0.02 &  11.19$\pm$0.16 \\
                  NGC5256  & 11.52$\pm$0.03 &  31.72$\pm$1.39 & 11.50$\pm$0.04 &  0.14$\pm$0.17 & 11.18$\pm$0.17 &   9.59$\pm$0.23 &  10.83$\pm$0.03 &  11.83$\pm$0.25 \\
 \hline
 \end{tabular}
 \label{tablum}
 \end{table*}
\begin{table*}
 \setcounter{table}{2}
 \caption{$-$ Continue}
 \begin{tabular}{lcrcccccc}
 \hline
 Name    &  log($L_{\rm IR}$)      &    \multicolumn{1}{c}{$SFR$}   &    log($L_{\rm IR}^{\rm SF}$) &   $f_{\rm AGN}$(5$-$40) & \multicolumn{4}{c}{log($L_{\rm bol}^{\rm AGN}$)}  \\ 
  \multicolumn{5}{c}{} &  IR & X-ray  &  [Ne~V]  &   [O~IV] \\ \\
             &   $log(L_{\odot})$ &    \multicolumn{1}{c}{$M_{\odot}/yr$}  &   $log(L_{\odot})$ &    & \multicolumn{4}{c}{$log(L_{\odot})$}  \\   \hline    
                  NGC5347  &  9.96$\pm$0.01 &   0.71$\pm$0.01 &  9.85$\pm$0.01 &  0.53$\pm$0.04 & 10.35$\pm$0.30 &   8.67$\pm$0.27 &   9.49$\pm$0.18 &   9.43$\pm$0.19 \\
                  NGC5506  & 10.45$\pm$0.02 &   1.96$\pm$0.08 & 10.29$\pm$0.04 &  0.65$\pm$0.07 & 10.63$\pm$0.07 &  10.83$\pm$0.15 &  10.38$\pm$0.03 &  10.99$\pm$0.12 \\
                  NGC5548  & 10.62$\pm$0.03 &   2.10$\pm$0.10 & 10.32$\pm$0.05 &  0.72$\pm$0.05 & 11.35$\pm$0.02 &  11.52$\pm$0.19 &  10.78$\pm$0.07 &  10.64$\pm$0.09 \\
                  NGC5953  & 10.41$\pm$0.04 &   2.56$\pm$0.10 & 10.41$\pm$0.04 &  0.02$\pm$0.60 &  8.87$\pm$0.30 & ... &   9.42$\pm$0.06 &   9.84$\pm$0.14 \\
                  NGC5995  & 11.33$\pm$0.03 &  19.19$\pm$2.30 & 11.28$\pm$0.12 &  0.34$\pm$0.05 & 11.32$\pm$0.02 & ... &  11.25$\pm$0.05 &  10.87$\pm$0.03 \\
                  NGC6810  & 10.68$\pm$0.03 &   4.62$\pm$0.55 & 10.66$\pm$0.12 &  0.13$\pm$0.52 & 10.02$\pm$0.25 &   7.07$\pm$0.46 &$<$8.95 &   8.67$\pm$0.28 \\
                  NGC6860  & 10.41$\pm$0.03 &   2.02$\pm$0.12 & 10.30$\pm$0.06 &  0.49$\pm$0.05 & 10.26$\pm$0.06 &  10.51$\pm$0.14 &  10.22$\pm$0.01 &  10.23$\pm$0.01 \\
                  NGC6890  & 10.33$\pm$0.01 &   2.05$\pm$0.25 & 10.31$\pm$0.12 &  0.13$\pm$0.69 &  9.73$\pm$0.57 &   8.12$\pm$0.27 &   9.87$\pm$0.01 &   9.42$\pm$0.09 \\
                  NGC7130$^a$ & 11.32$\pm$0.00 &  20.93$\pm$0.05 & 11.32$\pm$0.01 &  0.0 & ... &   7.89$\pm$0.01 &  10.89$\pm$0.01 &  10.54$\pm$0.01 \\
                  NGC7213  & 10.08$\pm$0.03 &   1.01$\pm$0.12 & 10.01$\pm$0.12 &  0.44$\pm$0.01 &  9.72$\pm$0.02 &   9.85$\pm$0.08 & $<$8.92 & $<$9.20 \\
                  NGC7469  & 11.65$\pm$0.03 &  44.55$\pm$5.35 & 11.65$\pm$0.12 &  0.05$\pm$0.60 & 10.93$\pm$0.41 &  10.94$\pm$0.12 &  11.24$\pm$0.12 &  10.60$\pm$0.03 \\
                  NGC7496  & 10.19$\pm$0.03 &   1.55$\pm$0.19 & 10.19$\pm$0.12 &  0.0 & ... & ... & $<$8.82 & $<$8.27 \\
                  NGC7603  & 11.08$\pm$0.02 &   8.84$\pm$0.26 & 10.95$\pm$0.03 &  0.55$\pm$0.06 & 11.50$\pm$0.02 & ... &$<$10.22 &  10.28$\pm$0.20 \\
                  NGC7674  & 11.53$\pm$0.03 &  23.58$\pm$2.83 & 11.37$\pm$0.12 &  0.58$\pm$0.60 & 11.73$\pm$0.30 & ... &  12.04$\pm$0.17 &  11.71$\pm$0.11 \\
           TOLOLO1238$-$364  & 10.83$\pm$0.01 &   5.76$\pm$0.17 & 10.76$\pm$0.03 &  0.32$\pm$0.60 & 10.60$\pm$0.30 & ... &  10.73$\pm$0.06 &  10.30$\pm$0.04 \\
                 UGC05101  & 11.91$\pm$0.01 &  55.21$\pm$6.63 & 11.74$\pm$0.12 &  0.65$\pm$0.03 & 12.04$\pm$0.13 &   8.73$\pm$0.54 &  11.28$\pm$0.09 &  11.07$\pm$0.11 \\
                 UGC07064  & 11.19$\pm$0.01 &  15.22$\pm$0.04 & 11.18$\pm$0.01 &  0.12$\pm$0.35 & 10.42$\pm$0.34 & ... &  11.02$\pm$0.17 &  10.90$\pm$0.14 \\
    \hline \hline
\multicolumn{9}{l}{$^a$ For this source we do not find evidence of the presence of an AGN from the SED-fitting analysis. However, the detection of mid-IR}\\
\multicolumn{9}{l}{~~~lines as [NeV] and [OIV] in the mid/far-IR spectra (see Table \ref{tab_mir_lines}), and a recent analysis of the X-ray data including new NUSTAR}\\
\multicolumn{9}{l}{~~~data, indicate the possible presence of a heavily buried, Compton-thick AGN (Pozzi et al. in preparation).}
\end{tabular}
\end{table*}

\begin{figure}
\includegraphics[width=8.5cm]{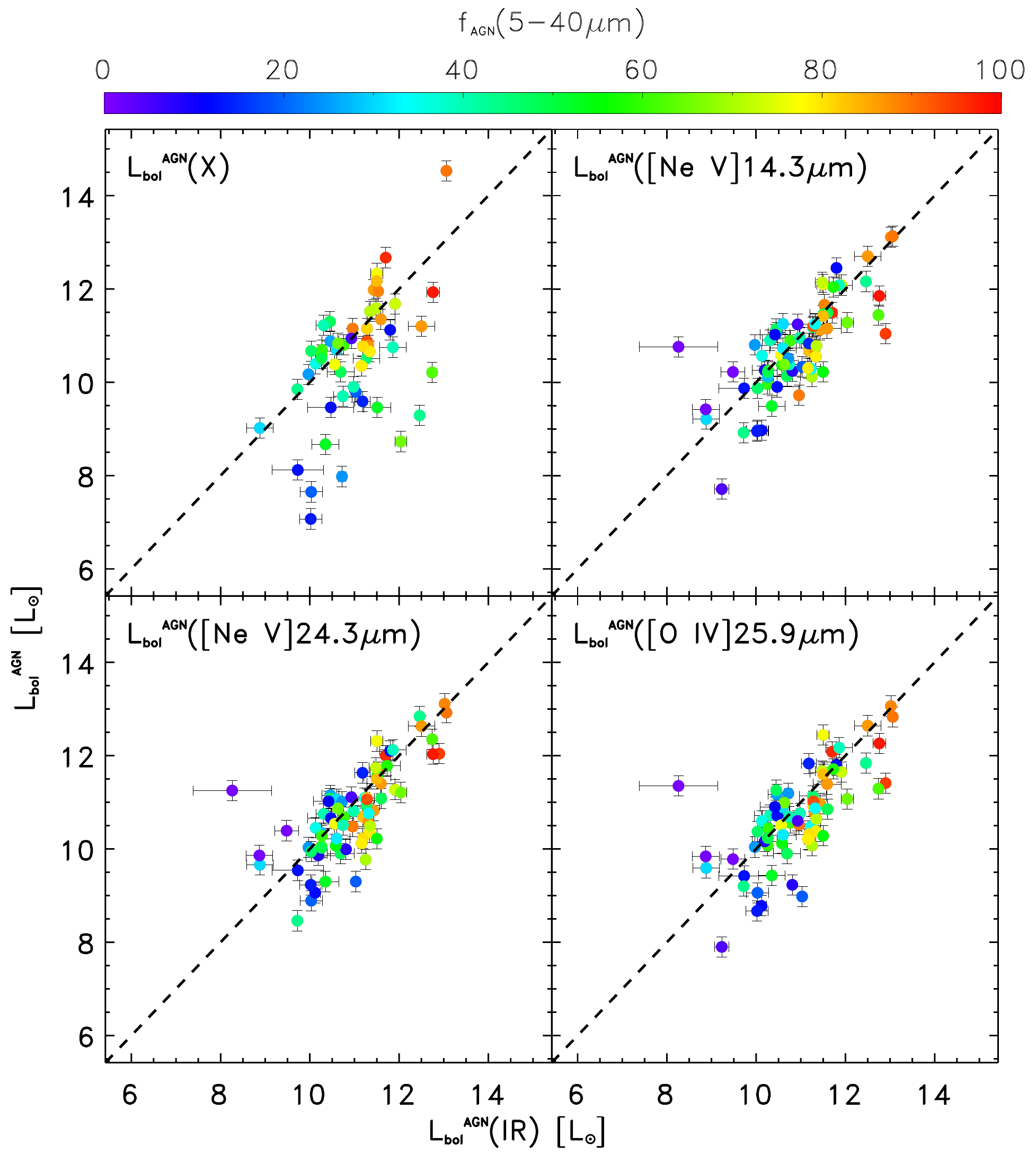}
\caption{AGN bolometric luminosity obtained from the best-fitting torus model versus the AGN bolometric luminosity derived from different indicators:
{\em top left}: X-ray (2-10 keV) Luminosity; {\em top right}: [Ne~V] $14.3\mu m$ emission line; {\em bottom left}: [Ne~V] $24.3\mu m$ emission line; {\em bottom right}: [O~IV] $25.9\mu m$ line.
The symbols shown in colour-gradient represent sources with different fractions of the luminosity produced by the AGN in the 5--40 $\mu$m wavelength range, as shown in the top. 
The dashed line represents the 1-1 relation.}
\label{figLaccLbol}
\end{figure}

\begin{figure*}
\includegraphics[width=15cm]{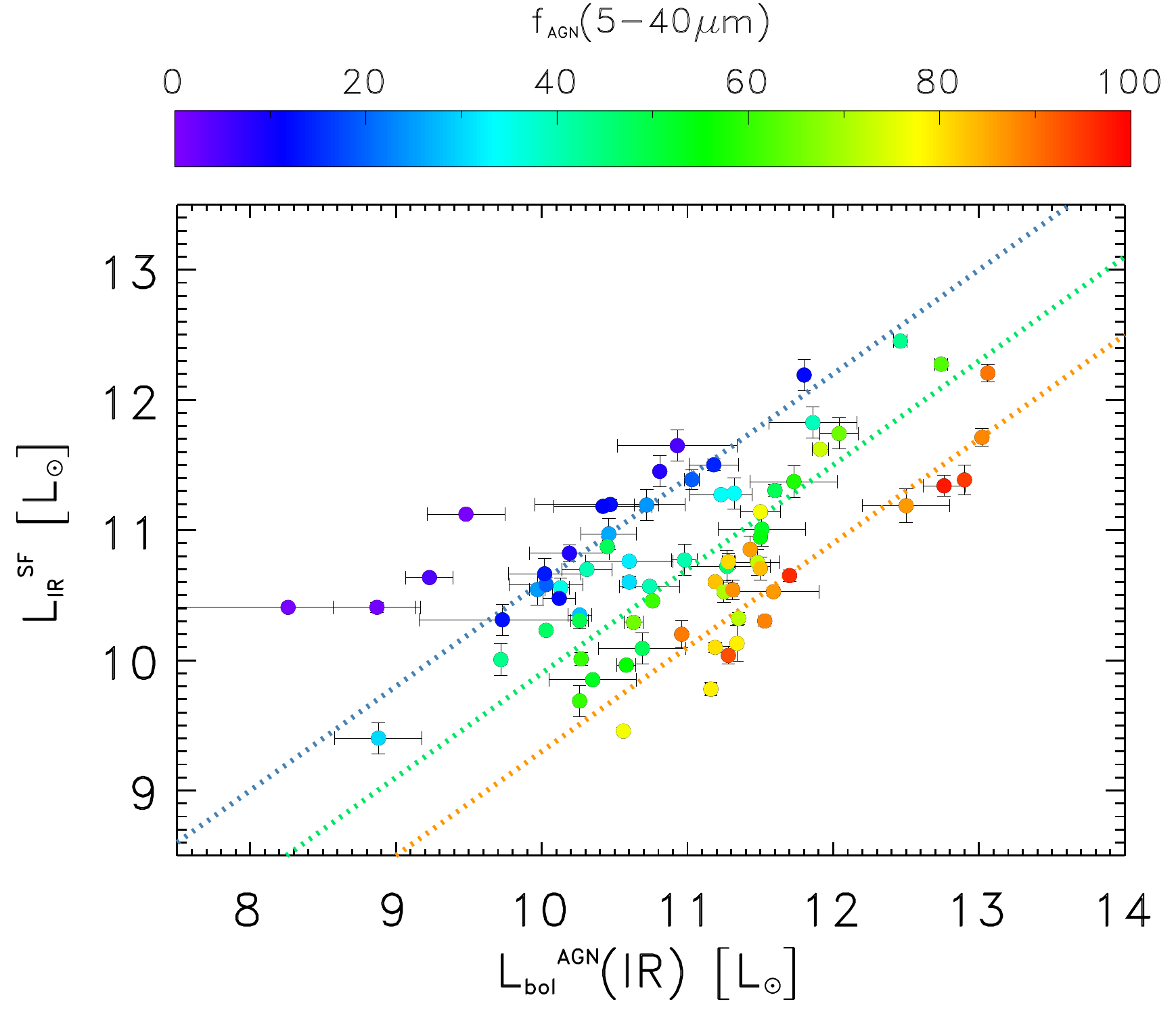}
\caption{IR luminosity due to SF ($L_{\rm IR}^{\rm SF}$) versus AGN bolometric luminosity ($L_{\rm bol}^{\rm AGN}(IR)$) for different fractions of AGN contribution to the 5--40 $\mu$m Luminosity 
(defined by the different colour gradient, as shown in the top). The three dotted lines are the best-fit to data with different f$_{\rm AGN}$(5-40$\mu$m), keeping fixed the slope of the relation at the 
same value found by \citet{netzer07} for local AGN: $L_{\rm SF} \propto L_{\rm bol}^{0.8}$ ($>$70\% orange, 30--70\% green, $<$30\% blue).}
\label{figLir_Lacc}
\end{figure*}

\subsection{AGN bolometric luminosity}
\label{sec_AGNbolL}
In Figure~\ref{figLaccLbol} the AGN bolometric luminosity that we derive from the SED decomposition (e.g., from the best-fitting torus model) is compared to the same quantity obtained by means 
of different methods: i.e. from the 2--10 keV X-ray luminosity (from IPAC-NED public Database), and the AGN mid-IR line luminosity (e.g., [Ne~V] and [O~IV]; \citealt{tommasin08,tommasin10}). 
The agreement is good, especially with the mid-IR line derivations (scatter around the 1-1 relations: $\sigma_X$$=$1.11, $\sigma_{[Ne~V]14.3}$$=$0.61, $\sigma_{[Ne~V]24.3}$$=$0.66, 
$\sigma_{[O~IV]25.9}$$=$0.7).
The significance level associated with these correlations has been derived by calculating the SpearmanÕs partial rank correlation coefficients between two variables independent on a third 
(the distance in our case), by following the method described by \citet{macklin82}. The resulting significance level of these relations is 1.5$\sigma$, 3.4$\sigma$, 3.3$\sigma$ and 3.2$\sigma$ 
for the AGN luminosity derived from X-ray, [Ne~V]14.3, [Ne~V]24.3 and [O~IV]25.9 respectively. 

We note that the relation between SED-fitting and X-ray derived AGN bolometric luminosity is more dispersed and with a lower significance than those with mid-IR AGN lines. This is likely due to a 
significant number of objects showing values of the AGN bolometric luminosity estimated from the X-ray band lower than those derived by decomposing the SED. 
These objects are well known obscured AGN (with weak X-ray emission, i.e., IRASF07599$+$6508, IRASF08572$+$3915, MRK0231, UGC05101), for which the intrinsic X-ray luminosity 
obtained by means of a correction for obscuration may be underestimated.
Moreover, most of the objects deviating from the relations are at relatively low X-ray luminosities ($<$10$^{43.5}$ erg/s), where the X-ray classification as AGN is less certain and the
contamination due to starburst can be significant (e.g., \citealt{brightman11b} consider``unambiguos X-ray AGN'' only sources with an observed 2--10 keV luminosity greater than 10$^{42}$ erg/s).

We are testing these hypothesis by re-analysing the X-ray spectra of the 12MGS, including
also recent data from the {\em Nuclear Spectroscopic Telescope Array} ({\em NuSTAR}; \citealt{harrison13}), when available (Vignali et al. in preparation). Indeed, the objects showing larger deviance 
from the relation are also known to be highly obscured AGN from the literature (e.g., some of them being Compton Thick, CT: $N_{\rm H}$$\gsimeq$$10^{24}$ cm$^{-2}$). 
Note that the objects for which we find a higher fraction of mid-IR luminosity due to the AGN are also those with higher bolometric luminosities.

\subsection{The AGN-SF connection}
\label{sec_AGNSF}
In Figure~\ref{figLir_Lacc} we plot the IR luminosity due to SF ($L_{\rm IR}^{\rm SF}$) versus the AGN bolometric luminosity ($L^{\rm AGN}_{\rm bol}$), using different colours to highlight sources with different AGN fractions. 
A correlation between the two quantities plotted here has been found in the literature for QSOs (at $z$$\sim$0: \citealt{netzer07},
and at $z$$=$2--3: \citealt{lutz08}) and for type II AGN (\citealt{netzer09}), with a slope of about 0.8 ($L^{\rm SF}$$\propto$$(L_{\rm bol}^{\rm AGN})^{0.8}$). 
More recent studies, based on {\em Herschel} data up to $z$$\sim$2--3 (\citealt{Shao10}; \citealt{Rosario12}) demonstrate that the $L^{\rm SF}$--$L^{\rm AGN}$ plane can be divided into 
two regimes with very different behaviours. 
One is the "SF-dominated" regime, where $L^{\rm SF}$$>$$L^{\rm AGN}$, and the two properties are not correlated and $L^{\rm SF}$ exceeds $L^{\rm AGN}$ by a factor that depends on redshift. 
In the second "AGN-dominated" regime, where $L^{\rm AGN}$$>$$L^{\rm SF}$, the two luminosities seem to correlate as $L^{\rm SF}$$\propto$$(L^{\rm AGN})^{0.7}$. 
In all these works the assumption is that the 60 $\mu$m luminosity is totally due to SF ($L_{\rm SF}$$=$$\nu L_{\nu}$(60$\mu$m)); moreover the
way the samples are selected strongly affects  
the distribution in the $L^{\rm SF}$-$L^{\rm AGN}$ plane (e.g., \citealt{Page12}; \citealt{Harrison12}; Lanzuisi et al. in preparation).
Our sample is local, therefore we can compare our result with those found in the local Universe (\citealt{netzer07,netzer09}): indeed a broad correlation is observed in our data between 
$L_{\rm IR}^{\rm SF}$ and $L_{\rm bol}^{\rm AGN}$, with a slope similar to 0.8 and a scatter of $\sim$0.6 around the relation. 
However, as clearly noticeable from the colour gradient of the symbols in figure~\ref{figLir_Lacc}, if we divide the sample in sub-samples with similar AGN fractions, the relation narrows 
significantly (scatter around the relation: $\sim$0.35 for $f_{\rm AGN}(5-40)$$<$30\%, $\sim$0.28 for 30\%$<$$f_{\rm AGN}(5-40)$$<$70\%, $\sim$0.30 for 
$f_{\rm AGN}(5-40)$$>$70\% ), while the normalisation increases with decreasing $f_{\rm AGN}(5-40)$ (e.g., sources with higher AGN fractions 
show lower SF luminosity at fixed AGN bolometric luminosities),
although the slope for different intervals of AGN contribution keeps similar to that found by \citet{netzer07} for local AGN samples.  \\
Indeed, if we calculate the Spearman's partial rank correlation coefficient for this correlation (to derive the significance, independent on redshift), 
we find that $L_{\rm IR}^{\rm SF}$ and $L^{\rm AGN}_{\rm bol}$ for the whole sample are loosely correlated (significance $\sim$1$\sigma$), 
while the correlation becomes significant (at $\gsimeq$3$\sigma$ level) if we divide the sources according to their AGN fraction (i.e., intervals of $f_{\rm AGN}$ discussed above).
These results imply that at fixed SF luminosity, sources with higher AGN luminosity have also higher AGN contribution in the mid-IR. On the other hand, at a fixed AGN fraction, 
the AGN luminosity closely correlates with the SF luminosity, with apparently no dependency on luminosity (i.e., the correlation holds for low and high luminosity sources).\\
Recently, Theios, Malkan \& Ross (in preparation) decomposed the H$\alpha$ luminosity of a representative sample of the 12MGS Seyferts into $L_{\rm H\alpha}^{\rm SF}$ and 
$L_{\rm H\alpha}^{\rm AGN}$ by means of narrow band filter imaging, also finding a correlation between the two quantities, not entirely due to selection effects.
The correlation is different for Seyfert 1 and Seyfert 1.9/2: Seyfert 1s show greater nuclear H$\alpha$ luminosities, due to the presence of broad line regions.\\
Our result confirms previous ones for local AGN, showing that in the local Universe the SF and accretion luminosities are intimately connected. 
They might also reflect a relation between $f_{\rm AGN}(5-40)$ and $L_{\rm IR}^{\rm SF}$/$L_{\rm bol}^{\rm AGN}$, so that galaxies with higher AGN fraction tend to have higher 
AGN luminosities for a given $L_{\rm IR}^{SF}$.
Hints of SF luminosity independent on the accretion one, i.e. the "SF-dominated" regime, are apparently observed for few galaxies with very low AGN contribution 
($f_{\rm AGN}$$\lsimeq$5\%). 

\section{AGN/SB diagnostics}
\label{sec_diagnostics}
\begin{figure}
\includegraphics[width=8.5cm]{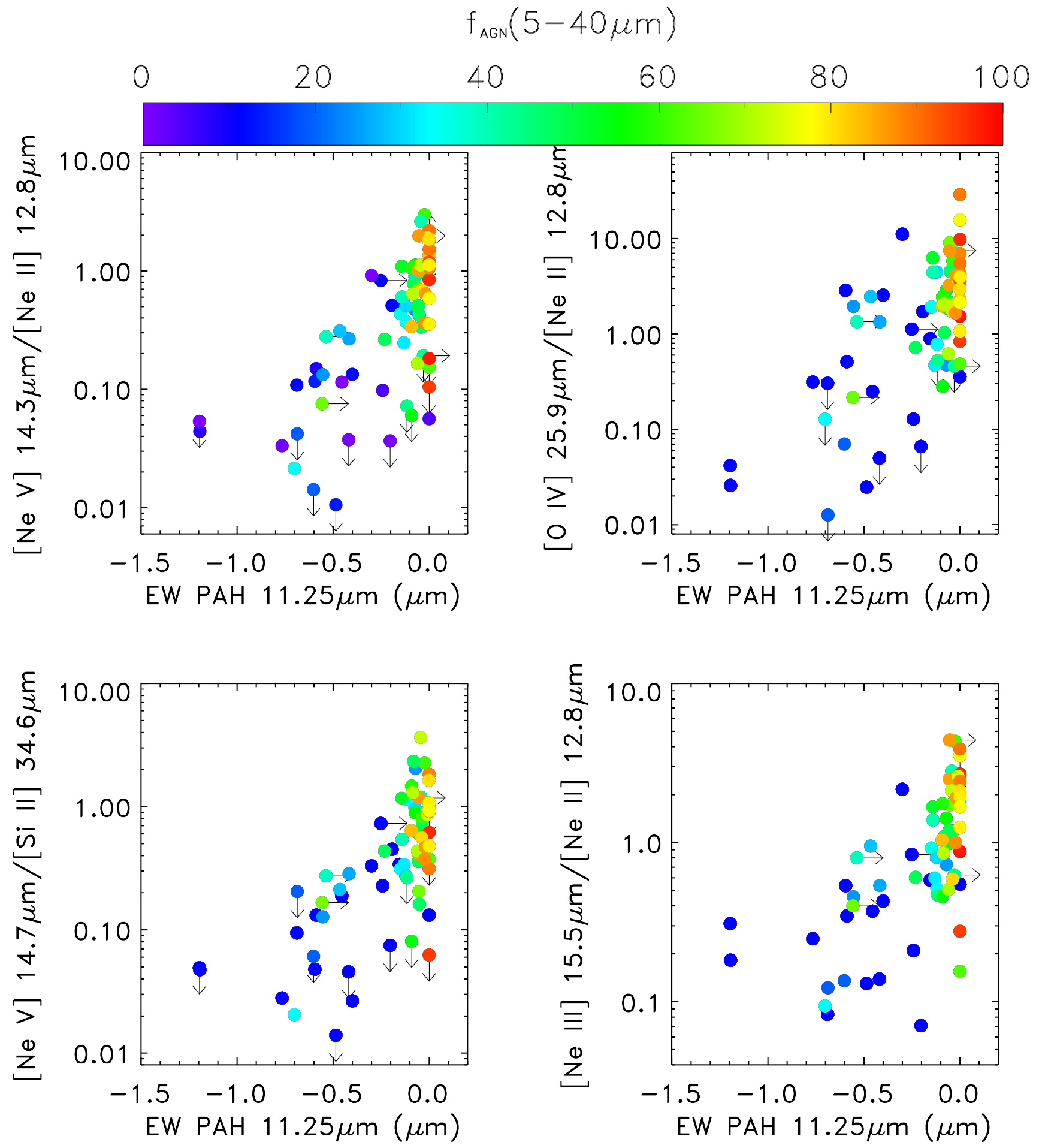}
\caption{Mid-IR line ratios vs. the PAH 11.25-$\mu$m EW (the $top$ and $bottom$ panels show the same diagnostics as reported in Figures~4 and 5 of \citealt{tommasin10}
respectively). As in Fig.~\ref{figLaccLbol} and as shown in the colour-bar on top, the different colours of the symbols represent
different fractions of AGN contribution to the 5--40 $\mu$m Luminosity.}
\label{figLine_pah}
\end{figure}

\begin{figure}
\includegraphics[width=8.5cm]{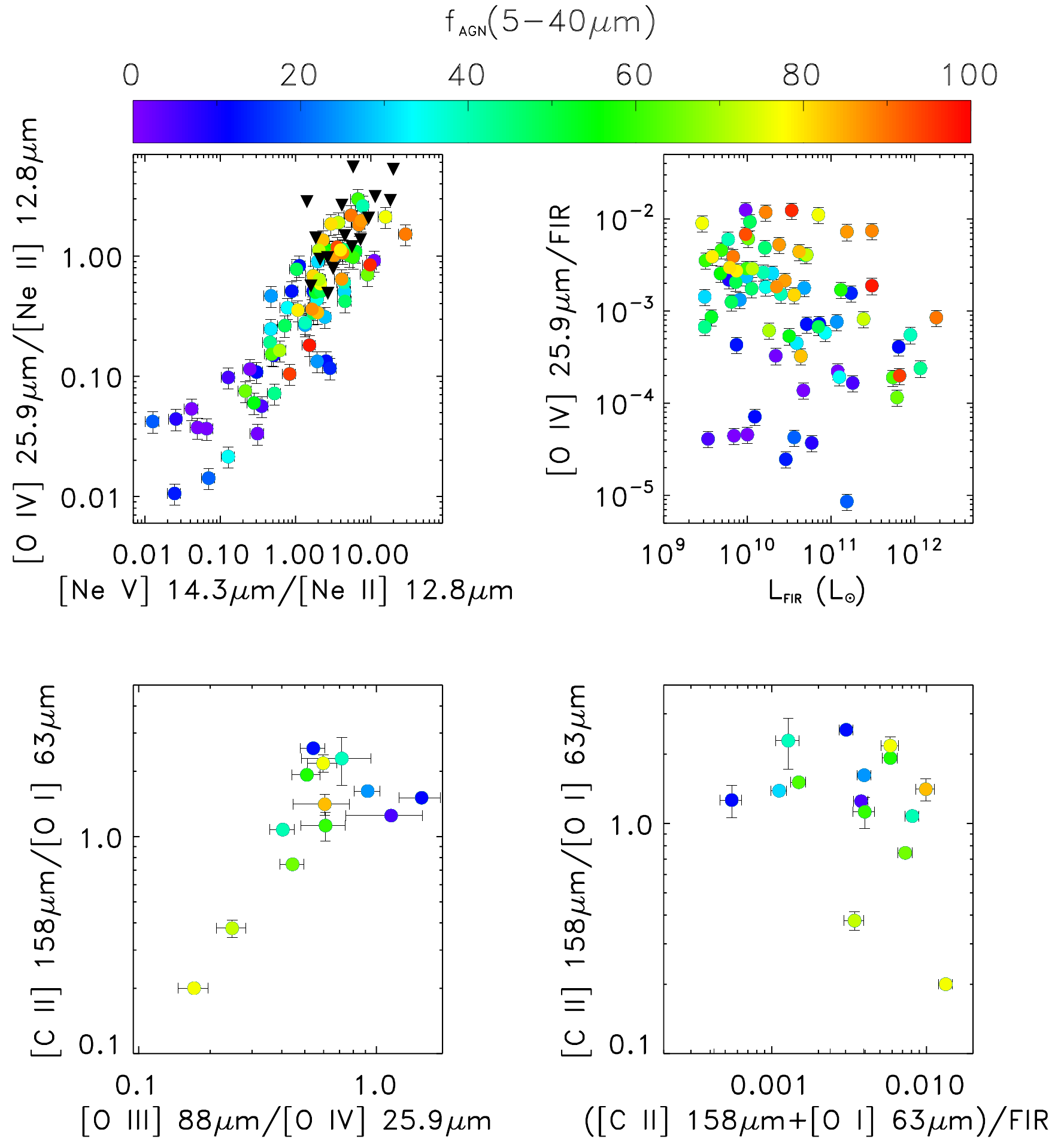}
\caption{Mid-/far-IR line diagnostics commonly used to separate AGN from star-formation powered sources. As in the previous figures and as shown in the top colour-bar,
the different colours of the symbols represent different fractions of the AGN contribution to the 5--40 $\mu$m luminosity, as derived from our SED-fitting decomposition.
The black filled triangles in the {top left} panel show, for comparison, the location of the Palomar Green (PG) QSOs in the [O~IV]/[Ne~II] vs. [Ne~V]/[Ne~II] plane (\citealt{schweitzer06}).}
\label{figLine_diagnostic}
\end{figure}
IR lines can be excited primarily by star formation (SF), by an AGN or by both. The typical AGN lines are three ([Ne~V] 14.3$\mu$m, [Ne~V] 24.3$\mu$m and [O~IV] 25.9$\mu$m), 
while the lines and features that are mainly produced in star forming regions are: PAH 6.2$\mu$m, PAH 11.2$\mu$m, [Ne~II] 12.8$\mu$m, [S~III] 18.7$\mu$m, [S~III] 33.5$\mu$m, [Si~II] 34.5$\mu$m, 
[O~I] 63$\mu$m, [N~II] 121.9$\mu$m, [O~I] 145.5$\mu$m, and [C~II] 157.7$\mu$m. 
The other lines from [S~IV] 10.5$\mu$m, [Ne~III] 15.5 $\mu$m, [O~III] 52$\mu$m, [N~III] 57$\mu$m and [O~III] 88$\mu$m, are excited by both AGN and newly born stars.
All these lines cover a wide parameter space of the critical density versus ionisation potential diagram, tracing different 
astrophysical conditions: from PDRs, to stellar/H~II regions, to the AGN, and coronal line regions (\citealt{spinoglio92}). 
This makes the combination of their ratios very useful for the definition of AGNs versus star formation diagnostic diagrams (e.g., \citealt{spinoglio92}; \citealt{genzel98}; \citealt{smith07a}). 
Higher ionisation lines (i.e., [Ne~III] 15.5 $\mu$m, [O~IV] 25.9 $\mu$m, [O~III]  at 52 and 88 $\mu$m and [N~III] 57 $\mu$m) are excited in H~II regions, but can be excited also in AGN narrow-line regions (NLR) with 
typical conditions (e.g. high densities and ionisation potentials). 
Therefore, in composite objects (like the great majority of our Seyfert galaxies, containing both a SF and an AGN component), the total line emission is the sum of the two components. 
The two [Ne~V] lines at 14.3 and 24.3 $\mu$m are exclusively excited by AGN and can be considered AGN spectral signatures (e.g., \citealt{tommasin10}), while the [O~IV] 25.9$\mu$m line is at 
least a factor of 10 brighter, with respect to the continuum, in AGN than in starburst galaxies (e.g., \citealt{pereira10}). 
\begin{figure*}
\includegraphics[width=8.8cm]{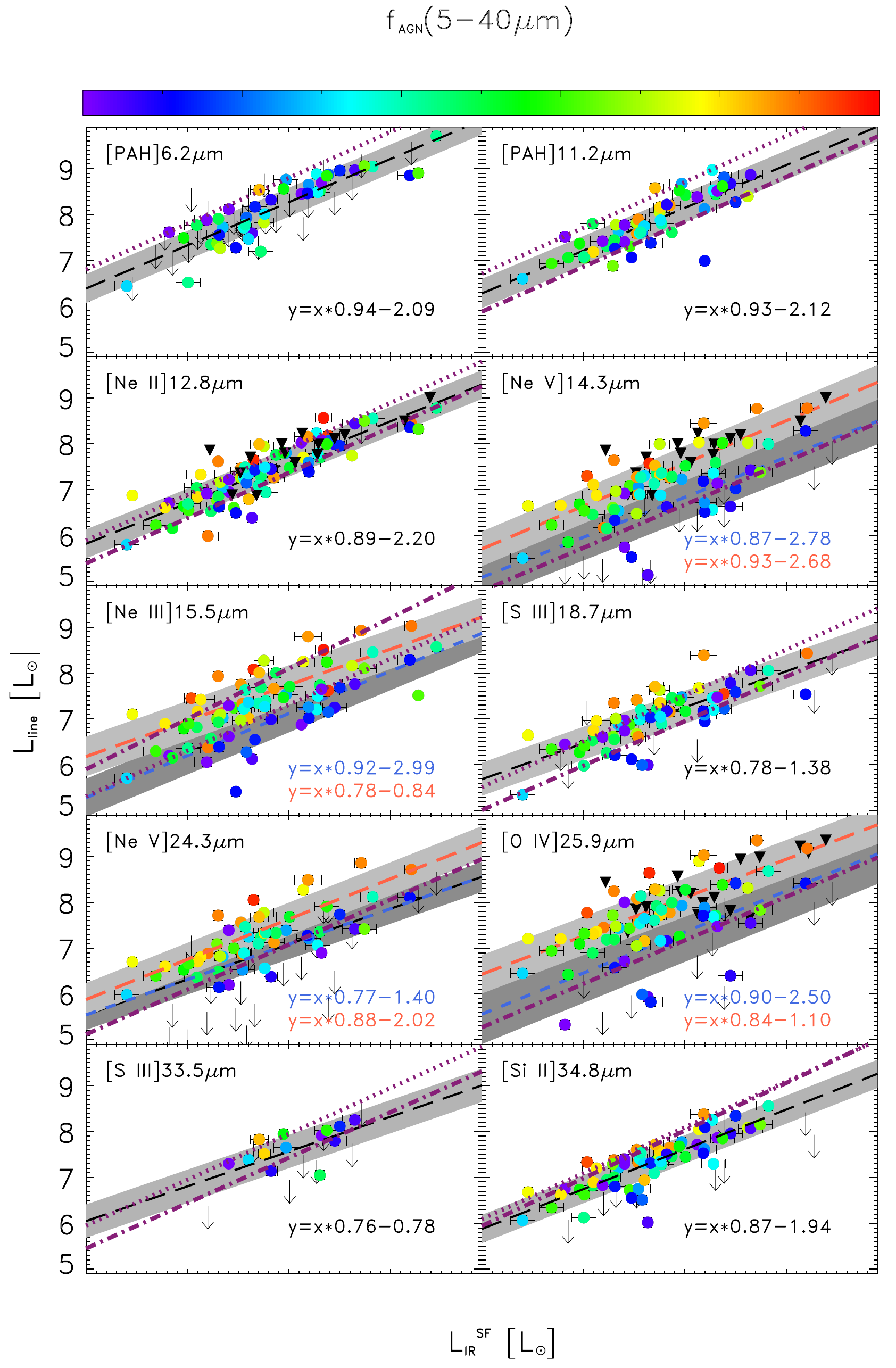}
\includegraphics[width=8.8cm]{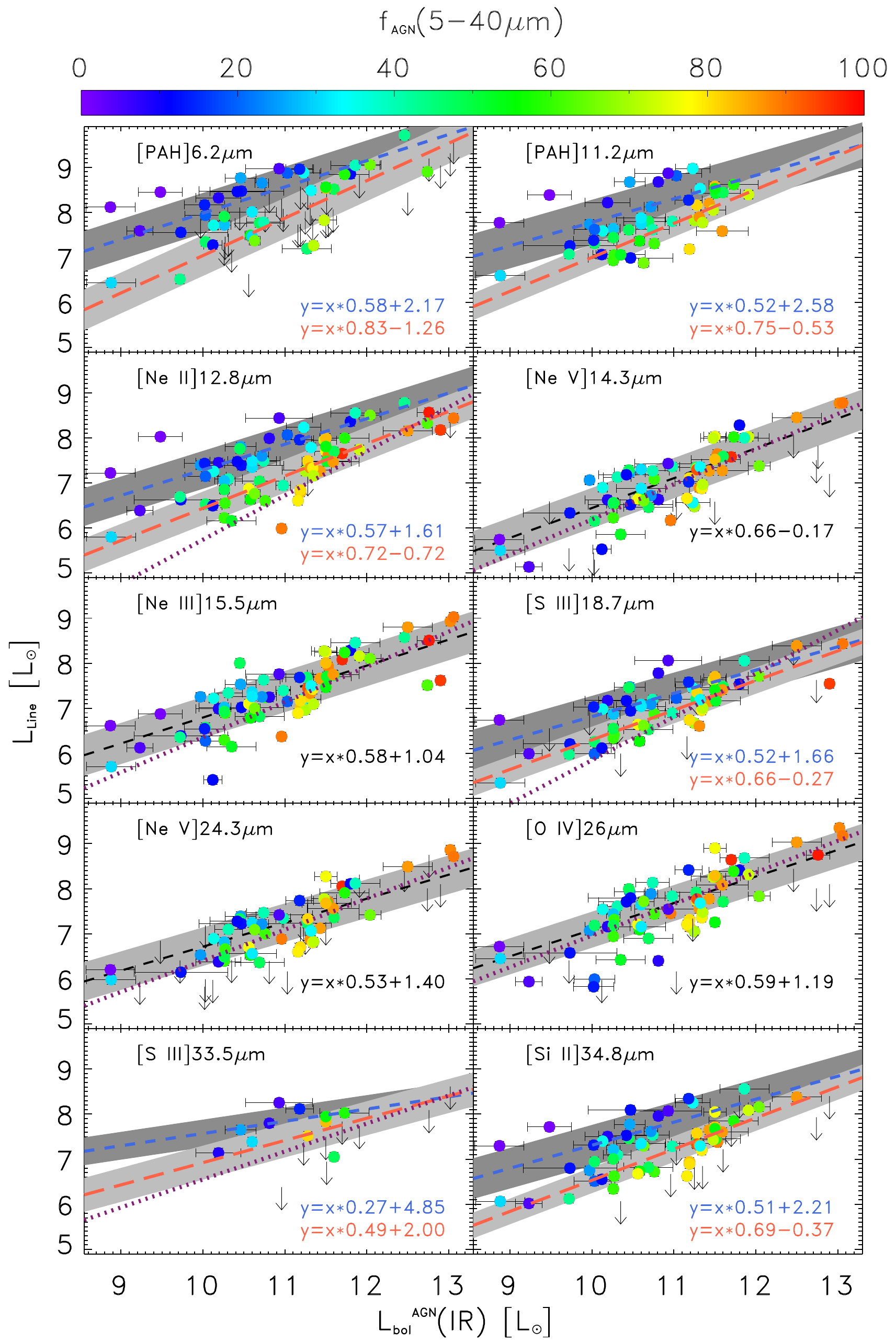}
\caption{Luminosity of the different mid-IR lines as a function of the 8--1000 $\mu$m luminosity due to SF ($L_{\rm IR}^{\rm SF}$, {\it left panels}) and of the AGN bolometric luminosity 
($L_{\rm bol}^{\rm AGN}(IR)$, {\it right panels}) derived from our SED fit.
The different colours of the symbols, as well as in the previous figures, represent the different AGN fractions to the 5--40 $\mu$m luminosity derived through our SED decomposition analysis. 
For comparison, values relative to the PG QSOs, as derived by \citet{schweitzer06}, are shown as black filled triangles in the {\it left} panels (for the available lines and assuming that 
$L_{\rm IR}^{\rm SF}$$\sim$$L_{\rm60 \mu m}$).
The long-dashed orange and the short-dashed blue lines show the best-fitting relations found for sources with a 5--40$\mu$m AGN fraction larger and smaller than 40\% respectively
(the light and dark grey areas are the relative intrinsic dispersion).
For lines showing a unique relation regardless of the AGN fraction, the long-dashed black line (within 
the grey area) shows the ``global'' best-fitting relation (with its intrinsic dispersion).
The purple dotted and dot-dashed lines represent the best-fitting relations found 
by \citep{bonato14a,bonato14b} and \citet{spinoglio12} respectively. Upper limits are shown by down-ward pointing arrows.}
\label{figLlineLirLagn}
\end{figure*}

\begin{figure*}
\includegraphics[width=8.5cm]{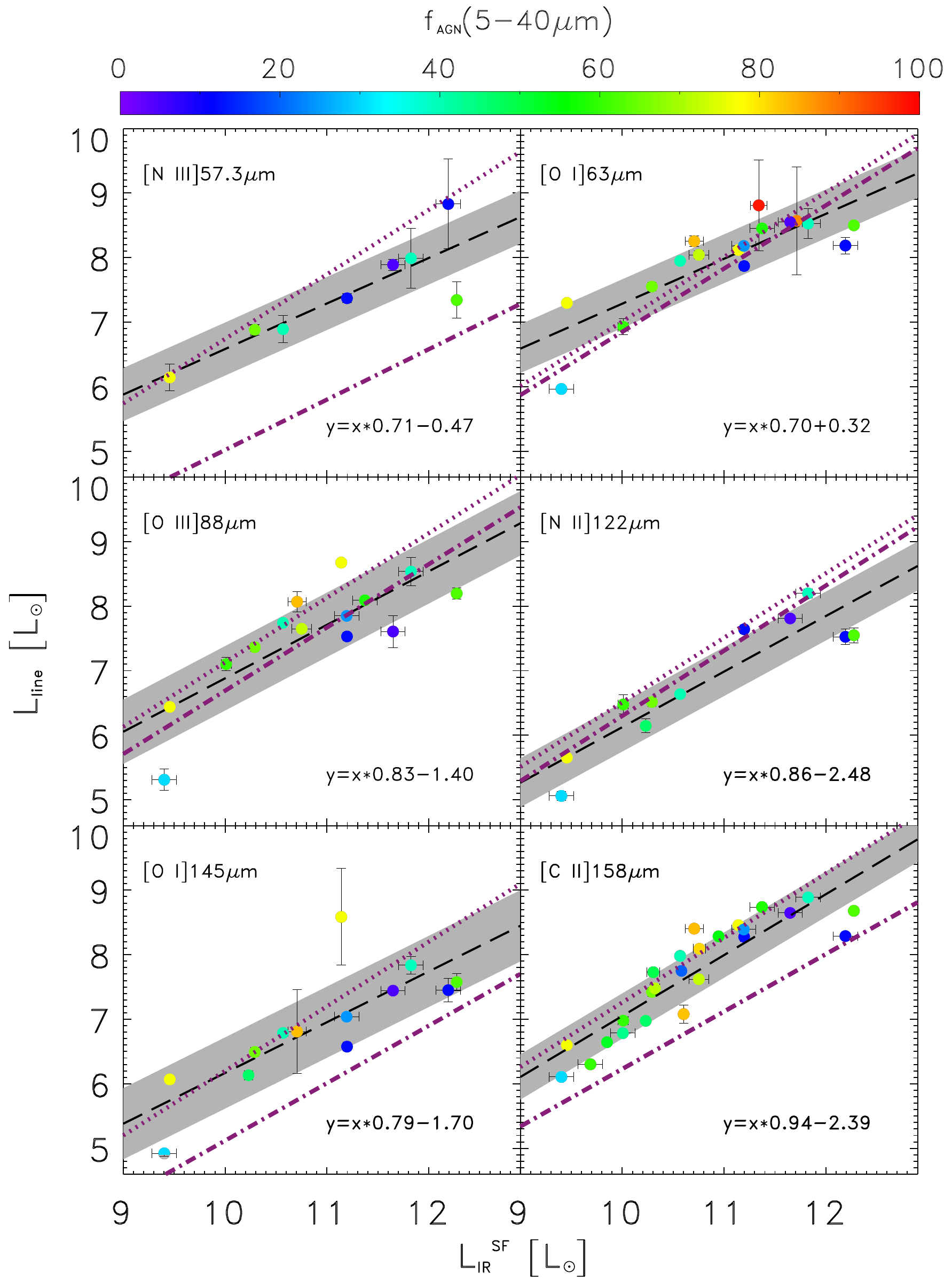}
\includegraphics[width=8.5cm]{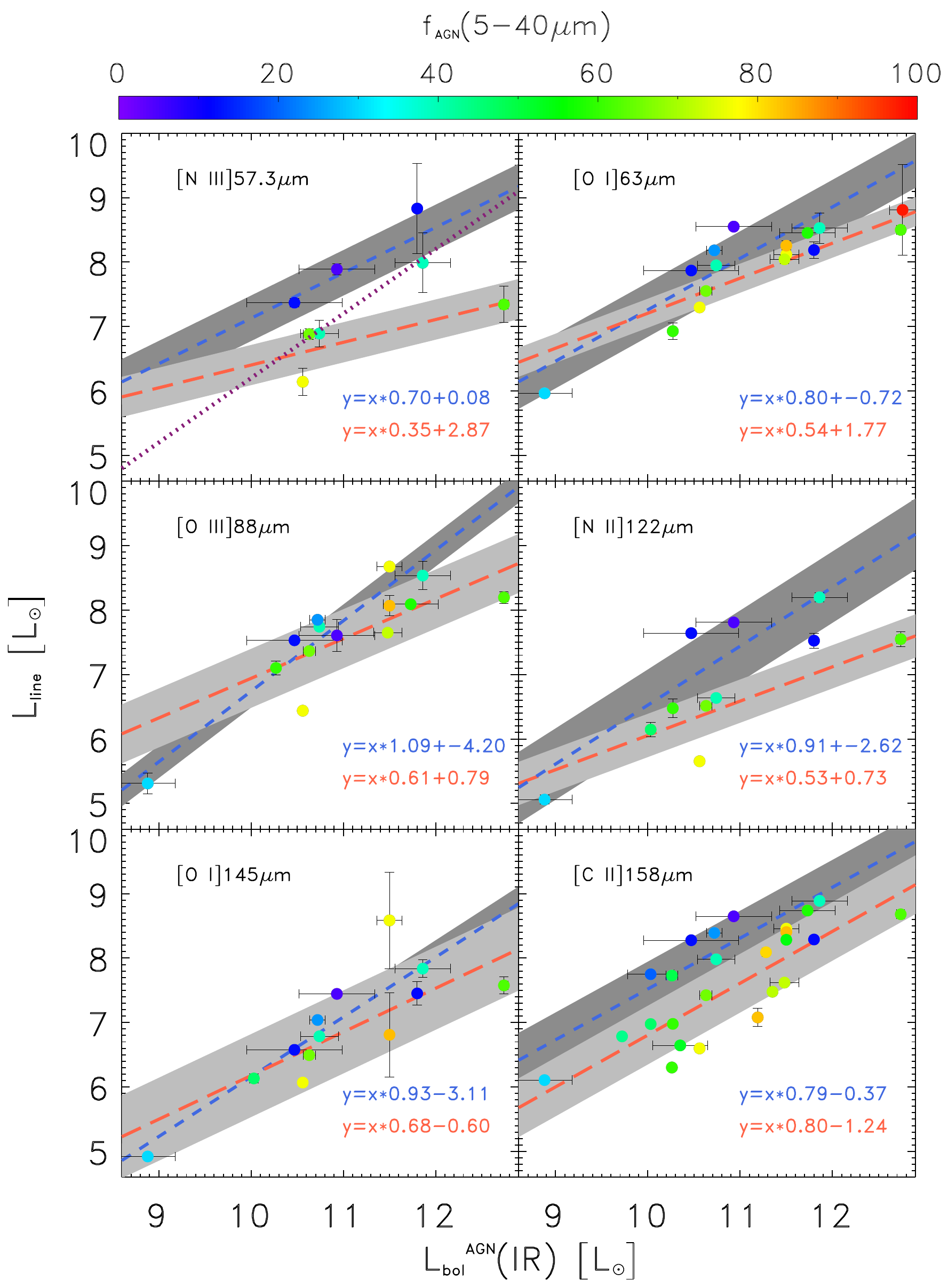}
\caption{Luminosity of the different far-IR lines as a function of the 8--1000 $\mu$m luminosity due to SF ($L_{\rm IR}^{\rm SF}$, {\it left panels}) and of the AGN bolometric luminosity 
($L_{\rm bol}^{\rm AGN}(IR)$, {\it right panels}) derived from our SED fit. 
As in the previous figures, the different colours of the symbols represent the different 5--40 $\mu$m AGN fractions (see top colour-bar). 
The long-dashed black line (within the grey area) shows the best-fitting relation (with its intrinsic dispersion), while the long-dashed orange and the short-dashed blue lines in the $right$
panels show the best-fitting relations found for sources with a 5--40$\mu$m AGN fraction larger and smaller than 40\% respectively
(the light and dark grey areas are the relative intrinsic dispersion). The purple dotted and dot-dashed lines represent the best-fitting relations 
found by \citet{bonato14a,bonato14b} and \citet{spinoglio12} respectively.}
\label{figLine_firLir}
\end{figure*}
  
One of our aims is to test and calibrate our results in terms of accretion luminosity and AGN fractions using mid- and far-IR spectral features. 
First, we test the diagnostic diagrams commonly used
in the literature to estimate the AGN contribution (e.g., through the [Ne~V] to [Ne~II] line ratio) and the star formation contribution (e.g., with the EWs of the 11.25 $\mu$m PAH feature).
In fact, since [Ne~V] 14.3$\mu$m and [O~IV] 25.9$\mu$m are the best AGN tracers in the {\em Spitzer-IRS} spectral range,
their normalisation to the [Ne~II] 12.8$\mu$m (mainly produced in the H~II/SF regions), can provide ionisation-sensitive ratios 
with all the sources (type 1Õs and type 2Õs) lying along the same sequence, as extensively discussed by \citet{tommasin10}. 
This diagram can be used to estimate the ionising power in the NLR, both in type 1 and 2 AGN.
Therefore, the clear positive correlation represents an excitation sequence anchored with the H II/LINERs, moving up in excitation level to the Seyfert 2Õs and Seyfert 1Õs, 
and ending with the PG QSOs (\citealt{veilleux09}).
\citet{pereira10} used the photoionization code MAPPINGSIII (\citealt{groves04}) to interpret some of the AGN line ratios considered in this figure, finding basically no differences between 
type 1 and type 2 AGN.\\
In Figure~\ref{figLine_pah} we show the line ratios [Ne~V] 14.3$\mu$m/[Ne~II] 12.8$\mu$m ({\em top left} panel), 
[O~IV]25.9$\mu$m/[Ne~II]12.8$\mu$m ({\em top right}), [Ne~V] 14.3$\mu$m/[Si~II] 34.8$\mu$m ({\em bottom left}) and 
[NeIII]15.5$\mu$m/[Ne~II] 12.8$\mu$m ({\em bottom right}) as a function of the EW of the PAH feature at 11.25 $\mu$m. The $top$ and $bottom$ diagrams show the same line ratios as reported 
in Fig.~4 and 5 of \citet{tommasin10}, although the AGN in that work were classified based on mid-IR line diagnostics. Our classification based on SED decomposition seems to
agree well with the previous ones, with
all the objects with high AGN fraction (e.g., $>$70\%) showing an absolute value of the EW of the PAH at 11.25 $\mu$m $<$0.1 $\mu$m and a neon ratio [Ne~V]/[Ne~II]$>$0.1. 
On the other hand, most of the objects with low f$_{AGN}$ (e.g., $<$30\%) have $|$EW PAH$|$$>$0.1 $\mu$m and [Ne~V]/[Ne~II]$<$0.5. 
The diagrams PAH EW versus [O~IV]/[Ne~II], [Ne~V]/[Si~II] and [Ne~III]/[Ne~II] show similar characteristics as the previous one: the line ratio (and the AGN fraction) increases when the 
absolute value of the PAH EW decreases. 
All these diagrams show the general inverse relation between the AGN dominance, as measured from the SED fitting and the ionisation sensitive line ratios, and the star formation dominance, estimated 
from the PAH EW. 
We note that the strong PAH emission detected in the spectra of low- and high-$z$ samples of AGN (e.g., \citealt{lutz08}, \citealt{hanami12}, \citealt{castro14}, \citealt{alonso14}), 
providing evidence for intense star formation in the hosts of these sources, are not in contrast with our result, that show basically no feature for sources 
AGN-dominated in the 5--40 $\mu$m range, not depending on the strength of the central AGN. 

Figure~\ref{figLine_diagnostic} shows different diagnostics using mid- and far-IR line ratios and/or far-IR luminosity/flux, commonly used in the literature to segregate AGN-dominated sources from 
SF-dominated ones:
[O~IV] 25.9$\mu$m/[Ne~II] 12.8$\mu$m versus [Ne~V] 14.3$\mu$m/[Ne~II] 12.8$\mu$m ({\em top left}), [O~IV] 25.9$\mu$m/FIR ({\em top right}) versus $L_{\rm IR}$, 
[C~II] 158$\mu$m/[O~I] 63$\mu$m versus [O~III]88$\mu$m/[O~IV] 25.9$\mu$m line ratio ({\em bottom left}) and [C~II]158$\mu$m$+$[O~I] 63$\mu$m versus ([C~II] 158$\mu$m$+$[O~I] 63$\mu$m) / $L_{\rm FIR}$ ({\em bottom right}). As in the previous Figure, the ionisation-sensitive ratios ([O~IV]/[Ne~II] and [Ne~V]/[Ne~II]) show higher values for sources with higher AGN fractions. \\
The diagnostic shown in the {\em upper right} panel is commonly used to estimate the level of contamination due to star-formation in the [O~IV] line (or, rather, the sources for which an AGN is needed to explain the [O~IV] emission): the higher [O~IV]/FIR ratios shown by our AGN dominated sources are consistent with our expectations based on previous results (e.g., \citealt{sturm10}; \citealt{gracia11}).\\
As discussed also by \citet{spinoglio15}, the [C~II]158$\mu$m / [O~I] 63$\mu$m versus the [O~III] 88$\mu$m/[O~IV] 25.9$\mu$m line ratio (see {\em bottom left} panel), although the 12MGS sources 
with measured far-IR lines are just few, is able to separate AGN dominated sources from SF dominated ones (in particular, the [O~III]/[O~IV] ratio). The [C~II]/[O~I] line ratio is only weakly able to segregate sources with 
different fractions of AGN, although we observe on average lower ratio values for AGN-dominated objects. \\
The {\em bottom right} panel shows the [C~II] 158$\mu$m/[O~I] 63$\mu$m versus ([C~II] 158$\mu$m$+$[O~I] 63$\mu$m) / FIR 
diagram: sources with higher AGN fractions are located preferentially at lower [C~II]/[O~I] ratio values and at higher ([C~II]$+$[O~I])/FIR values. [O~I] and [C~II] are the brightest cooling lines of 
the cool ISM in galaxies. However, {\em ISO} showed the [C~II] line in local ULIRGs to be about an order of magnitude lower relative to the FIR continuum than in normal and starburst galaxies (e.g., 
\citealt{Malhotra01}; \citealt{Luhman03}). The [C~II]/[O~I] ratio is known to be sensitive to gas density in star-forming galaxies (e.g., \citealt{genzel00}), with the ratio decreasing with increasing density 
(PDRs account for more than half of the [C~II] emission in starburst galaxies, while strong shocks could boost the [O~I] emission relative to [C~II]), while the larger values of the ([C~II]$+$[O~I])/FIR ratio
shown by our AGN dominated sources imply normal values and not a deficit in the major PDR cooling lines relative to the FIR.
A detailed discussion on the line ratios characteristics and sensitivity to ionisation and density is given by \citet{spinoglio15}, but it is far from the scope of this work: here we plot these diagnostics 
uniquely with the aim of checking our decomposition results and their reliability (in particular to verify that objects with different AGN fractions are located where expected). 
Indeed, we see a trend in our data, in all panels and figures, with AGN-to-SF dominated line ratios increasing with increasing AGN fraction.

\section{Line-luminosity relations}
\label{sec_relations} 
\begin{table*}
\caption{Best-fitting parameters of the Line - Luminosity relations: 
$log_{10}(L_{\rm line})=a\cdot log_{10}(L_{\rm IR})+b$ [eq.~\ref{eq_lline_lirtot}]; $log_{10}(L_{\rm line})=c\cdot log_{10}(L_{\rm bol}^{\rm AGN})+d$ 
[eq.~\ref{eq_lline_lagn}]; $log_{10}(L_{line})=e\cdot log_{10}(L_{\rm IR}^{\rm SF})+f$ [eq.~\ref{eq_lline_lsf}]}
 \begin{tabular}{lccccccc}
\hline \hline
  Line & $a$ & $b$ & $c$ & $d$ & $e$ & $f$ & f$_{\rm AGN}$(5--40$\mu$m) \\ \hline
  PAH 6.2   & 0.92$\pm$0.02 & $-$1.91$\pm$0.27 & 0.58$\pm$0.03 & 2.17$\pm$0.29 & 0.94$\pm$0.02 & $-$2.09$\pm$0.27 & $<$40\% \\
                 &                               &                                 & 0.83$\pm$0.03 & $-$1.26$\pm$0.37 &                      &                                  &$>$40\% \\ \hline                 
  PAH 11.2 & 0.95$\pm$0.03  & $-$2.41$\pm$0.31 & 0.52$\pm$0.04 & 2.58$\pm$0.30  & 0.93$\pm$0.03 &  $-$2.12$\pm$0.30 & $<$40\% \\
                  &                             &                                 & 0.75$\pm$0.04 & $-$0.53$\pm$0.40 &                        &                                 & $>$40\% \\ \hline
  [Ne~II] 12.8 & 0.87$\pm$0.19 & $-$2.17$\pm$0.21 & 0.57$\pm$0.02 & 1.61$\pm$0.25 & 0.89$\pm$0.019  & $-$2.20$\pm$0.21 & $<$40\% \\
                   &                             &                                &  0.72$\pm$0.02 & $-$0.72$\pm$0.24 &                       &                                  & $>$40\% \\  \hline     
  [Ne~V] 14.3 &  0.93$\pm$0.04 & $-$3.55$\pm$0.40 &  0.58$\pm$0.02 & 0.66$\pm$0.25  & 0.91$\pm$0.04 & $-$3.34$\pm$0.40 & $<$40\% \\  
                    &  0.78$\pm$0.02 & $-$1.36$\pm$0.25 &                              &                                & 0.77$\pm$0.02 & $-$1.08$\pm$0.25 & $>$40\% \\ \hline
  [Ne~III] 15.5 &  0.92$\pm$0.04 & $-$2.99$\pm$0.40 & 0.52$\pm$0.02 & 1.68$\pm$0.25  & 0.92$\pm$0.04 & $-$2.99$\pm$0.40 & $<$40\% \\
                      &  0.78$\pm$0.02 & $-$1.10$\pm$0.25 &                              &                             & 0.78$\pm$0.02 & $-$0.84$\pm$0.25 & $>$40\% \\ \hline
  [S~III] 18.7 &  0.79$\pm$0.02  & $-$1.59$\pm$0.22  & 0.52$\pm$0.03  & 1.68$\pm$0.25   & 0.78$\pm$0.02 & $-$1.38$\pm$0.25 & $<$40\% \\
                    &                             &                                  & 0.66$\pm$0.03 & $-$0.27$\pm$0.28 &                       &                                   & $>$40\% \\ \hline
  [Ne~V] 24.3 & 0.80$\pm$0.04 & $-$1.95$\pm$0.41 & 0.44$\pm$0.03 & 2.48$\pm$0.28 & 0.63$\pm$0.03 & $-$0.02$\pm$0.36 & $<$40\% \\
                     & 0.90$\pm$0.02 & $-$2.69$\pm$0.25 &                            &                            & 0.88$\pm$0.02 & $-$2.3$\pm$0.26 & $>$40\% \\ \hline
  [O~IV] 25.9 & 0.74$\pm$0.04 & $-$0.87$\pm$0.40 & 0.59$\pm$0.01 & 1.19$\pm$0.16    & 0.72$\pm$0.04 & $-$0.66$\pm$0.40 & $<$40\% \\
                     & 0.73$\pm$0.02 & $-$0.21$\pm$0.25 &                             &                             & 0.70$\pm$0.02 & 0.31$\pm$0.25 & $>$40\% \\ \hline
  [S~III] 33.5 & 0.72$\pm$0.05 & $-$0.53$\pm$0.61 &  0.27$\pm$0.05  & 4.85$\pm$0.47 & 0.76$\pm$0.06 & $-$0.78$\pm$0.66 & $<$40\% \\
                     &                            &                                &  0.49$\pm$0.33  & 2.00$\pm$3.79 &                              &                                  & $>$40\% \\ \hline
  [Si~II] 34.8 &  0.86$\pm$0.02 & $-$2.03$\pm$0.23 & 0.51$\pm$0.03 & 2.21$\pm$0.27 & 0.87$\pm$0.03  & $-$1.94$\pm$0.27 & $<$40\% \\
                   &                              &                                 & 0.69$\pm$0.03 & $-$0.37$\pm$0.34 &                         &                                  & $>$40\% \\  \hline
  [N~III] 57.3 & 0.71$\pm$0.09 & $-$0.67$\pm$0.50 &  0.70$\pm$0.09  & 0.08$\pm$1.04 &  0.71$\pm$0.04 & $-$0.47$\pm$0.46 & $<$40\% \\
                     &                            &                               &  0.35$\pm$0.06   & 2.87$\pm$0.67 &                             &                                   & $>$40\%  \\ \hline
  [O~I] 63 & 0.72$\pm$0.04 & $-$0.08$\pm$0.35 &   0.80$\pm$0.04  & $-$0.72$\pm$0.48  & 0.70$\pm$0.03 & 0.32$\pm$0.35 & $<$40\% \\
                      &                           &                               &   0.54$\pm$0.04   & 1.77$\pm$0.41 &                                   &                              & $>$40\%  \\ \hline
  [O~III] 88 & 0.88$\pm$0.04 & $-$2.11$\pm$0.40 &   1.09$\pm$0.05   & $-$4.20$\pm$0.53 & 0.82$\pm$ 0.035 & $-$1.40$\pm$0.39 & $<$40\% \\
                     &                          &                                 &   0.61$\pm$0.05   & 0.79$\pm$0.56    &                                   &                                 & $>$40\%  \\ \hline
  [N~II] 122 & 0.88$\pm$0.04 & $-$2.76$\pm$0.37  &   0.91$\pm$0.04   & $-$2.62$\pm$0.48  &  0.86$\pm$0.03 & $-$2.48$\pm$0.36 & $<$40\% \\
                     &                           &                                  &   0.53$\pm$0.05   & 0.73$\pm$0.54       &                                 &                               & $>$40\%   \\ \hline
  [O~I] 145 & 0.84$\pm$0.03 & $-$2.38$\pm$0.38  &   0.93$\pm$0.04  & $-$3.11$\pm$0.48 &  0.79$\pm$0.03 & $-$1.70$\pm$0.37 & $<$40\% \\
                     &                           &                                  &   0.68$\pm$0.05   & $-$0.60$\pm$0.55 &                         &                                 &  $>$40\%  \\ \hline
  [C~II] 158 & 0.96$\pm$0.03 & $-$2.73$\pm$0.31 &  0.79$\pm$0.04   & $-$0.37$\pm$0.46  & 0.94$\pm$0.03 &  $-$2.39$\pm$0.30 & $<$40\% \\
                     &                          &                                 &   0.80$\pm$0.03 & $-$1.24$\pm$0.38  &                              &                                   & $>$40\% \\                        
                   \hline \hline
 \end{tabular}
 \label{tab_relation}
 \end{table*}                   
We find that the luminosities of the mid-/far-IR lines of our 12MGS subsample correlate with both the total IR luminosity due to SF ($L_{\rm IR}^{\rm SF}$) and the bolometric luminosity of the AGN ($L_{\rm bol}^{\rm AGN}$).  The best-fitting parameters (the intercept and slope constants) of the linear relations (on a log-log scale) and their associated probable uncertainties have been derived by minimising the chi-square error statistic, considering the data errors on both axes.
In Figure~\ref{figLlineLirLagn} we plot the mid-IR line luminosity versus the total IR luminosity due to SF ({\em left} panels) and versus the AGN bolometric luminosity ({\em right} panels), as derived from the
SED decomposition (Section~\ref{sec_results}). 
The linear relations best-fitting the data are reported, together with their intrinsic scatter (grey regions around the relations). 
Note that the intrinsic scatter considered here represents the dispersion, $\sigma$, given by the rms deviation of the data from the mean relationship, obtained by considering a Gaussian distribution
around the mean (see \citealt{bonato14a,bonato14b}). 
Therefore, the shaded bands shown in the plots represent the $\pm$1$\sigma$ uncertainty regions that we need to take into account when we derive
the line LFs by converting the total IR/AGN LFs using these relations (see next Section).\\
The relations between the (fewer) far-IR line and $L_{\rm IR}^{\rm SF}$ ($left$) and $L_{\rm bol}^{\rm AGN}$ ($right$) luminosities 
are shown in Figure~\ref{figLine_firLir}, with their intrinsic scatter as in the previous figure.
We stress that, although for some lines and sources there are upper limits, they are available not for all the considered lines (e.g., the collection of data is not always uniform), having just no data or no observation in some cases (mainly for the far-IR lines, but also for some mid-IR ones). Therefore, for homogeneity, we decided not to consider any upper limits in our analysis, but only detections, although when available, we plot the limits as pointing-down arrows in the figures.\\ 
From the {\em left} panels of Fig.~\ref{figLlineLirLagn}, we note that for the SF-dominated lines (i.e., PAHs, [Ne~II], etc.) a unique relation is derived 
between the line and the SF luminosity, regardless of the AGN fraction; while for the AGN-dominated lines (i.e., [Ne~V], [O~IV], etc.) the relation changes as the AGN fraction increases.
In particular, by dividing the sample in two (according to f$_{\rm AGN}$(5--40$\mu$m)$>$ and $<$40\%), we find that galaxies in the higher f$_{\rm AGN}$ sub-sample
show higher (AGN-dominated) line luminosity at a given SF luminosity. 
These results are confirmed by the location of the PG QSOs (AGN-dominated) in some of these diagrams (shown for comparison as black filled triangles), as derived by \citet{schweitzer06}:
they always occupy the same region of our sources with higher AGN fraction (i.e., are at higher line luminosities in the [Ne~V] and [O~IV] vs. SF luminosity diagrams and mixed up with the
other sources regardless the AGN fraction in the [Ne~II] vs. SF luminosity one).
The opposite happens to the relation between the line luminosity and the AGN bolometric luminosity (see the {\em right} panels of Fig.~\ref{figLlineLirLagn}):  
at a given $L_{\rm bol}^{\rm AGN}$, sources with higher AGN fraction show lower (SF-dominated) line luminosity, but the same line luminosity if the lines are AGN-dominated
(just one correlation, regardless their $f_{\rm AGN}$).
This reflects the fact that lines mostly excited by SF have a major contribution not related to the AGN, with their luminosity enhanced for higher SF fraction (lower AGN fraction) with
respect to sources with similar AGN luminosity, but lower SF fraction (higher AGN one). 
The significance level associated to these luminosity-luminosity correlations, 
independent on redshift, is derived through the Spearman partial rank correlation coefficients (\citealt{macklin82}).
We find a high significance for the correlations between all the mid-IR lines and the IR luminosity due to SF (between $\sim$3 and 7$\sigma$) and 
between the AGN mid-IR lines (i.e., [Ne~V], [Ne~III] and [O~IV]) and the AGN bolometric luminosity (at $\gsimeq$3.5$\sigma$). 
Instead, the significance is lower between SF-dominated lines and AGN bolometric luminosity (between $\sim$1 and $\sim$2$\sigma$), but increases to 
2.5$\sigma$ -- 5$\sigma$ for the sub-samples defined according to f$_{\rm AGN}$(5--40$\mu$m)$<$ or $>$40\%.

Note that, since a measure of the far-IR lines are available only for a limited number of sources, the results (in terms of linear coefficients and significance) obtained by 
dividing the sample in two (according to the $f_{AGN}$(5--40$\mu$m) $<$ or $>$40\% criterion) are not wholly meaningful, given the very large uncertainties. 
However, the trend of sources with higher AGN fraction showing brighter IR lines (due to SF) found for the mid-IR lines (more numerous and statistically significant sample) 
is confirmed also for the far-IR lines.

The best-fitting parameters and their probable uncertainties found for the different line-luminosity relations, expressed in the form 
\begin{equation}
log_{10}(L_{\rm line})=a\cdot log_{10}(L_{\rm IR})+b
\label{eq_lline_lirtot}
\end{equation}
\begin{equation}
log_{10}(L_{\rm line})=c\cdot log_{10}(L_{\rm bol}^{\rm AGN})+d
\label{eq_lline_lagn}
\end{equation}
\begin{equation}
log_{10}(L_{line})=e\cdot log_{10}(L_{\rm IR}^{\rm SF})+f
\label{eq_lline_lsf}
\end{equation}

\noindent are listed in Table~\ref{tab_relation}.

\section{The line luminosity functions}
\label{sec_llf}
Our understanding of the cosmological evolution of IR sources has dramatically improved in recent years, but studying the co-evolution of star formation and black hole accretion 
is still one of main scientific goals of the future astrophysical missions, both in the X-rays and in the IR.
Observational determinations of the evolution of the IR population up to $z$$\simeq$4 are now available thanks to {\em Herschel} (e.g., \citealt{gruppioni13}), although the 
major step forward will be the characterisation and the evolution with redshift of their physical properties. 
This will be possible only through IR spectroscopy, provided by planned or forthcoming facilities, like, e.g., {\em JWST} and {\em SPICA}, that will explore the distant Universe in the near-/mid-IR
and mid-/far-IR respectively. 
In order to understand what will be spectroscopically observable for galaxies and AGN, we need to estimate the redshift and galaxy luminosity ranges that can be measured, 
based on the line luminosity functions. 
A key ingredient to make such estimates is the relationship between the line and the total IR (or SF/AGN) luminosity, that will allow us to rely on the evolution of the IR (SF/accretion) luminosity function:
to this purpose, we use the ratios obtained in the previous section, between the line luminosity and the total IR (SF/AGN) luminosity, to derive the mid-/far-IR line LFs.

Note that, converting the accretion LFs (e.g., \citealt{hopkins07}; \citealt{delvecchio14}) using the relations found between the line luminosity and the 
AGN accretion luminosity, and the SFR LFs (e.g., \citealt{gruppioni15}) is probably the most appropriated approach for the AGN dominated lines (or for 
objects containing an AGN, at any level of dominance) and the SF dominated ones respectively. 
However, the AGN bolometric luminosity and the SFR depend on several assumptions: the former i.e., on the bolometric correction, needed whatever the AGN selection would be, 
which depends on the observing band and - strongly - on the considered AGN model; the latter on the stellar population synthesis models, their SF histories, etc..
For this reason, they are not quantities directly measurable by the observations as, instead, is the total IR luminosity (very robust if far-IR data are available). 
Therefore, here we decide to use the total IR LF (and the line - total IR luminosity relations: e.g., coefficient $a$ and $b$ in Table~\ref{tab_relation}) 
instead of the SFR or accretion LFs, since the adopted approach is more direct and applicable to virtually any galaxy samples observed in the IR, for which 
a measure of the total IR LF is available.
As a quality test, we have also followed the other approach, converting the AGN bolometric LFs derived 
by \citet{delvecchio14} into line LFs through the line -- AGN luminosity relations (coefficient $c$ and $d$ in Table~\ref{tab_relation}) for the 
AGN-dominated lines ([Ne~V], [Ne~III], [O~IV]), and the SFR LFs by \citet{gruppioni15} and the line -- SF luminosity relations (coefficient $e$ and $f$ in Table~\ref{tab_relation}) 
for the other lines (SF-dominated). 
The line LFs derived with this second approach are well consistent (e.g., within 1 $\sigma$) with those obtained by converting the total IR LF. \\

The line LFs can be easily computed from the total IR ones by convolving the latter with the distribution of the IR-to-line luminosity ratios, assumed to be a Gaussian with 
mean value given by the best-fitting relations and standard deviation equal to the intrinsic dispersion (which must be added to the data errors to obtain the total uncertainty in the relations).  
The detailed LF analysis performed by \citet{gruppioni13} and the evolutions found for the different populations, allow us to convert the total IR LF not ``globally'', with an average relation, 
but considering the {\em Herschel} populations separately by using the line-to-total IR luminosity relations corresponding to their median AGN fraction (in the 5--40 $\mu$m range, as derived by \citealt{delvecchio14} through an accurate SED decomposition). 
We therefore convert the total IR LFs applying the appropriate relation to each population of IR objects observed by {\em Herschel}, given the average AGN fraction 
of that population. 
In particular, \citet{gruppioni13} divided the IR population in five main classes, based on their broad-band SEDs: {\tt spiral}, {\tt starburst}, {\tt SF-AGN} (sub-divided into 
{\tt SF-AGN(Spiral)} and {\tt SF-AGN(SB)} as they represent low-luminosity and obscured AGN respectively), {\tt AGN1} and {\tt AGN2}. 
Successively, \citet{delvecchio14} performed a SED decomposition analysis (as the one performed in this work), deriving the AGN bolometric luminosity and the fraction of AGN luminosity
in different wavelength ranges for each source considered by \citet{gruppioni13}. Since the average $f_{\rm AGN}(5-40)$ found for the {\tt AGN1} and {\tt AGN2} populations (e.g., those with AGN-dominated SEDs) 
is larger than 40\%, we have considered the relation found for sources with an AGN contribution 
to the 5--40 $\mu$m luminosity larger than 40\%. For all the other populations (i.e., {\tt spiral}, {\tt starburst}, {\tt SF-AGN(Spiral)} and {\tt SF-AGN(SB)}), we have
instead considered the relation for sources with less than 40\% of AGN (since their average $f_{\rm AGN}(5-40)$ resulted lower).
 
Note that, since there is virtually no contribution of star forming galaxies to the [Ne~V] lines, we have not converted to [Ne~V] LFs the total IR LFs of the two populations powered by star-formation
(i.e., {\tt spiral} and {\tt starburst} galaxies), while for the [Ne~III] 15.5 $\mu$m and [O~IV] 25.9$\mu$m lines we have derived (and applied to those two population LFs) a relation for the SF 
component only by using the pure starburst sample of \citet{bernard09}. 
For all the other lines, we have considered the emission to be dominated by star-formation and we have applied the derived relations to all the lines 
(according to the different AGN fractions).

\begin{figure*}
\includegraphics[width=16cm,height=14cm]{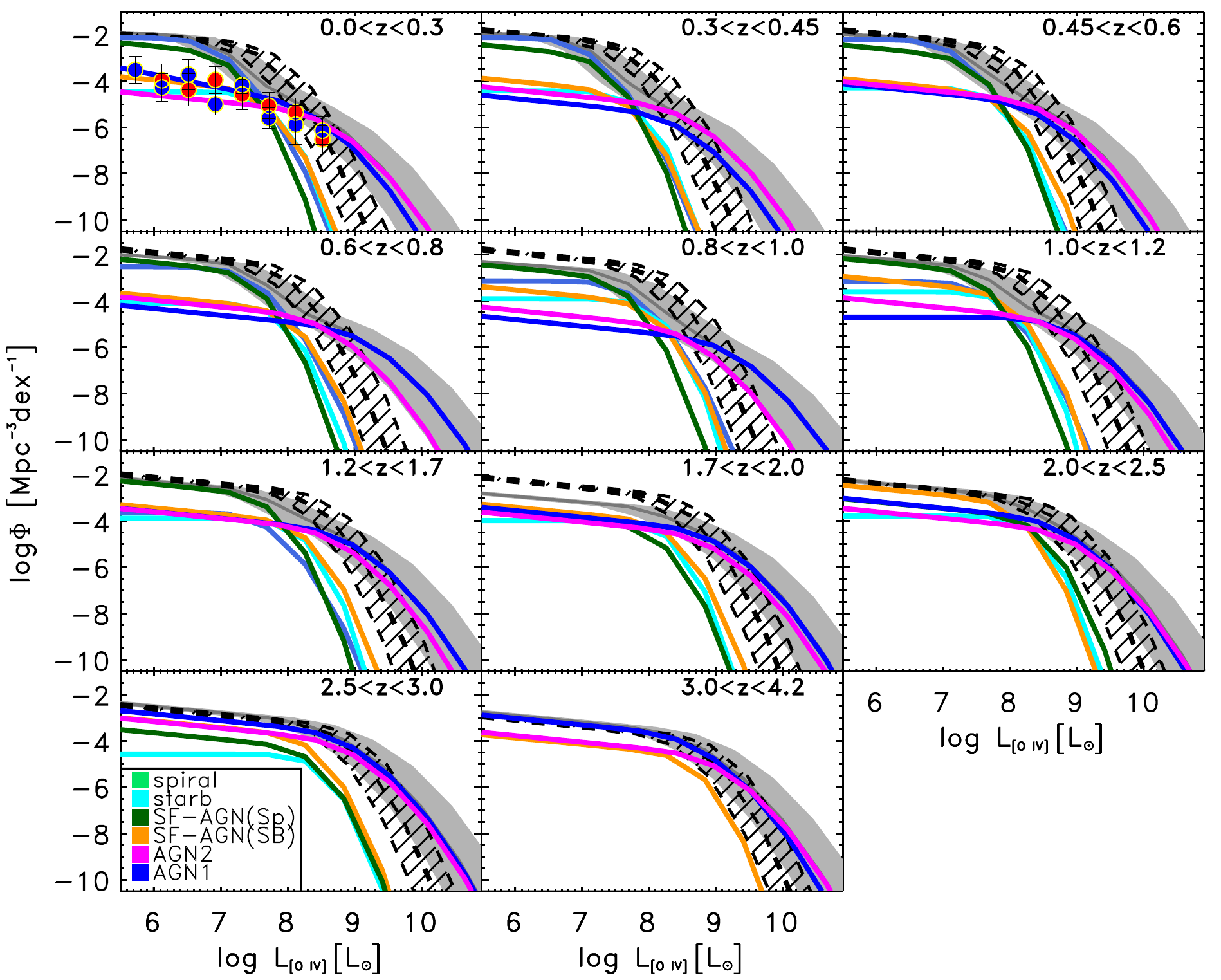}
\caption{Luminosity function of the [O~IV] ${25.9\mu m}$ line in different redshift intervals, obtained by converting the total IR LF of \citet{gruppioni13} through the $L_{\rm line}-L_{\rm IR}$ relations found in Section~\ref{sec_relations}. The different coloured lines represent the different IR populations (evolved and converted separately, according to their average AGN fraction), whose sum is shown by the grey band. The black dashed line shows the LF obtained by converting the total IR LF using the average $L_{\rm [O~IV]}-L_{\rm IR}$ relation. The filled symbols in the 0.0$<$$z$$<$0.3 panel show the [O~IV] LF derived
by \citet{tommasin10} for Seyfert 1 (red circles) and Seyfert 2 (blue circles) galaxies.}
\label{figOIV_LF}
\end{figure*}

In Figure~\ref{figOIV_LF} we show an example of line LF ([O~IV] 25.9 $\mu$m), where the contributions of the different populations are hilighted by different colours (described in the legend)
and the total contribution (sum of all the populations), with its uncertainty region (obtained by considering the intrinsic and measurement errors in the relations)
being shown as grey filled area and compared to the result we would have obtained by applying to the total IR LF a ``global'' average relation for all the populations 
(described by the $a$ and $b$ parameters in Table \ref{tab_relation}: black dashed line with uncertainty region defined by the hatched area).\\
We note that the LFs obtained by using the average relation for all the sources or different relations for different AGN fractions, are almost consistent within 2$\sigma$, although at $z$$<$2 and 
$L_{\rm[O~IV]}$$>$$10^9$ L$_{\sun}$, the two LFs differ by more than 2$\sigma$ (especially at 0.6$<$$z$$<$0.8). 
The main difference in this particular case can be noticed at the bright end of the line LF, where the AGN populations, if treated separately with their specific line-to-total IR luminosity relation, give
a major contribution, mainly at $z$$<$2. A slight difference in the LF knee is also observed in the same range of redshift, with the ``globally'' converted line LF showing a higher $\Phi^*$ and a lower $L^*$
with respect to the LF obtained by summing the different population contributions. At $z$$>$2, where AGN-dominated populations dominate the IR LFs (especially the bright-end), the difference
is negligible, since the average relation reflects the relation found for sources with high AGN fraction. For comparison, at the lower redshift (0.0$<$$z$$<$0.3) we have plotted also the values
of the [O~IV] LF found by \citet{tommasin10} for the Seyfert 1 and 2 of the 12MGS. The agreement with our estimates for {\tt AGN1} and {\tt AGN2}$+${\tt SF-AGN(SB)} is very good, although we 
stress that the faint-end of our reference total IR LF for AGN-dominated objects (i.e., {\tt AGN1} and {\tt AGN2}) is not constrained at low redshift (e.g., the faint-end slope is taken from a higher 
$z$ interval, more populated by AGN at low luminosities). The faint-end of our derived [O~IV] LF is dominated by low-luminosity AGN ({\tt SF-AGN(GAL)}) and normal galaxies ({\tt spiral}), which
in principle should not be included in the 12MGS.
Note that for lines mainly produced by SF, for which an average relation for all sources is found regardless of the 
AGN fraction (i.e., PAHs, [Ne~II], [S~III], [Si~II] and all the far-IR lines; see Figs.~\ref{figLlineLirLagn} and \ref{figLine_firLir}), the two approaches here described are equivalent. 

For the sake of clarity, we note that the local relations found in this work - then applied to the {\em Herschel} populations up to high $z$ to derive the evolution of the line LFs - are valid mainly at 
$L_{\rm line}$$<$$10^9$ L$_{\sun}$, while the observed discrepancy occurs mainly at brighter luminosities. 
Unfortunately, at present, no data are available to constrain the local bright-end of the line LFs. 
The purpose of this work (and Section) is to present a method based on the best 
currently available data and models -Ð improved with respect to the previous ones (see Section \ref{sec_sed}) -Ð to provide expected numbers of detections in the different IR lines. 
All this will be crucial for planning spectroscopic surveys with the future facilities (see next Section).
\begin{figure*}
\includegraphics[width=8cm]{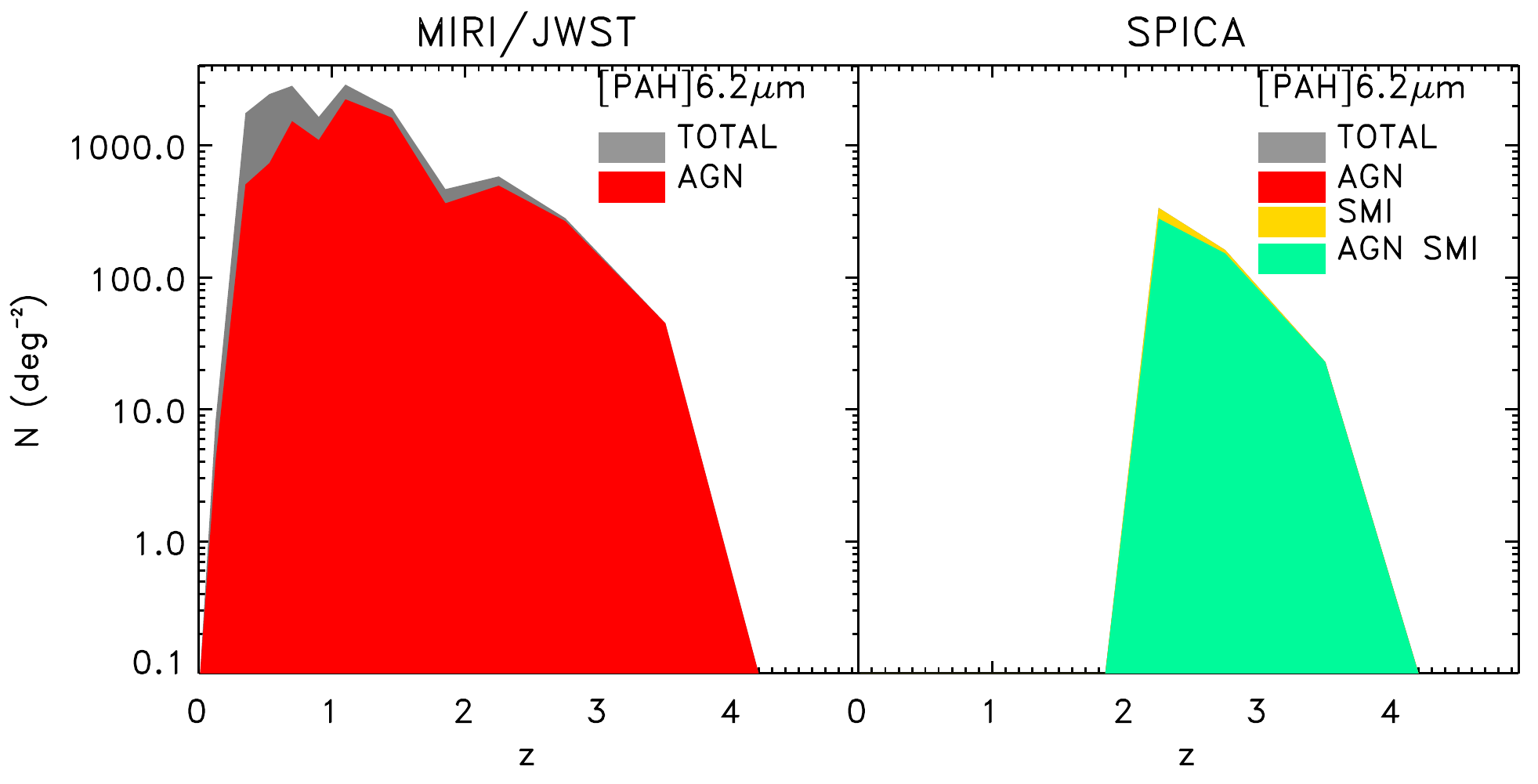}
\includegraphics[width=8cm]{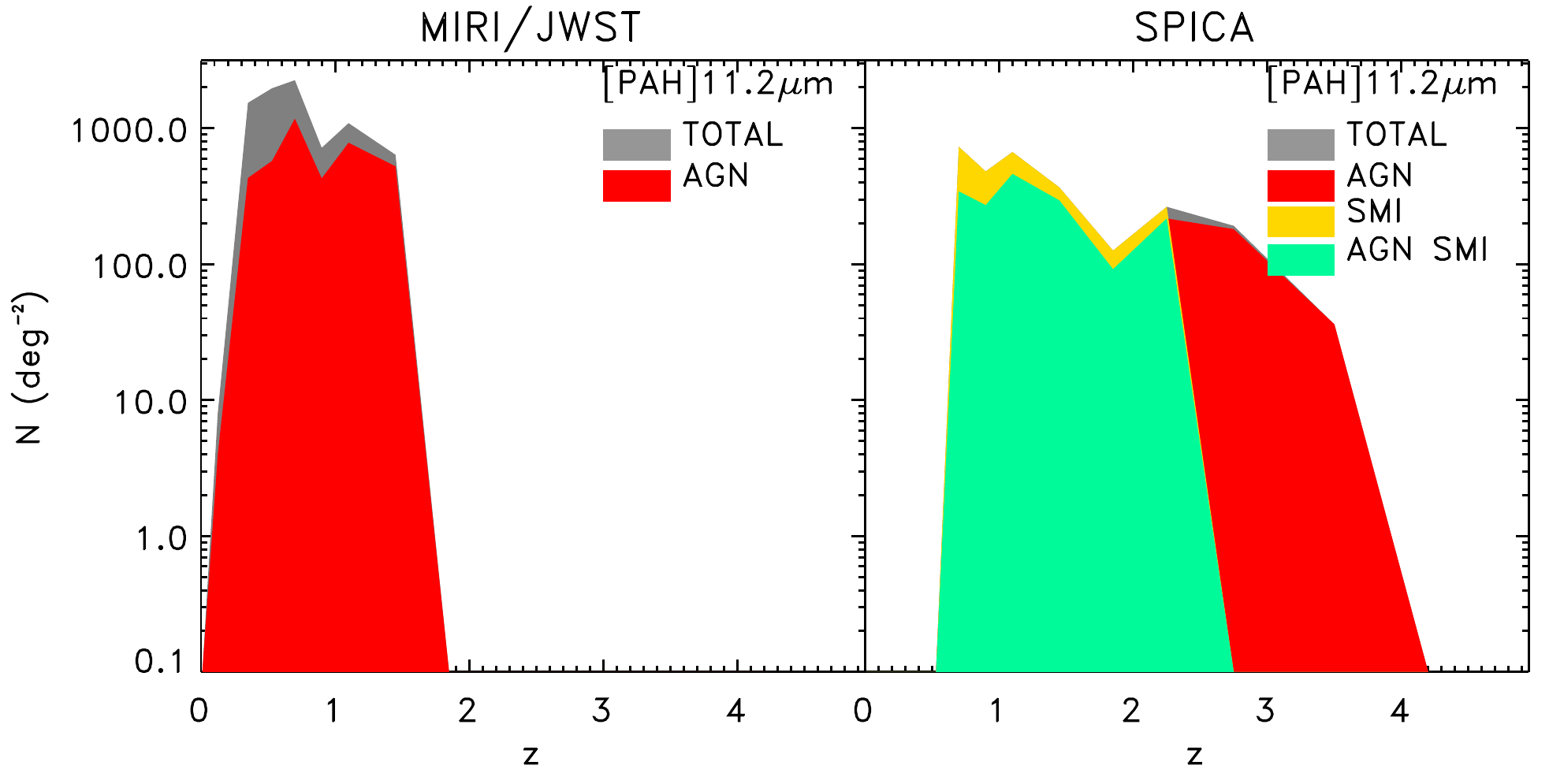}
\includegraphics[width=8cm]{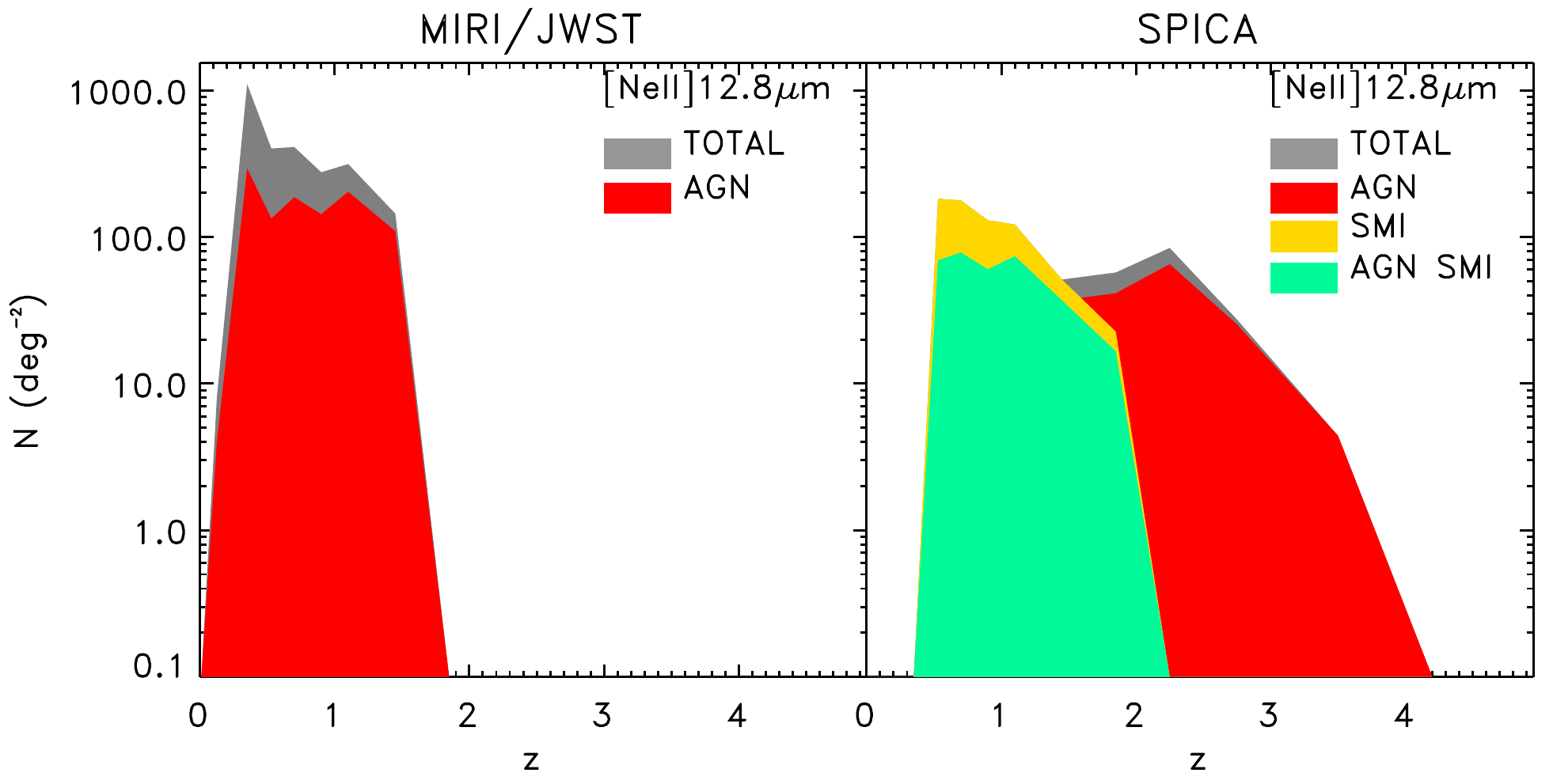}
\includegraphics[width=8cm]{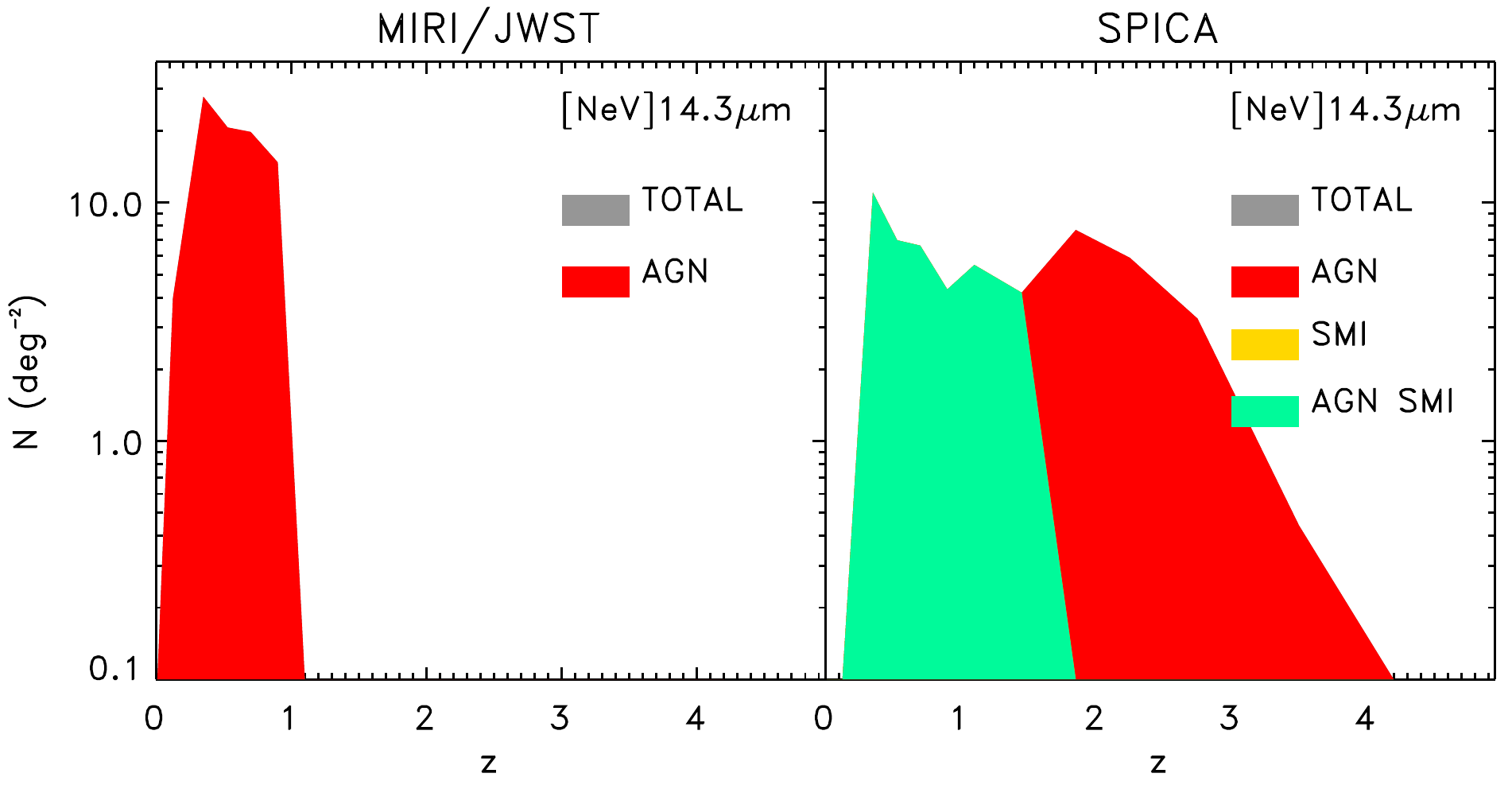}
\includegraphics[width=8cm]{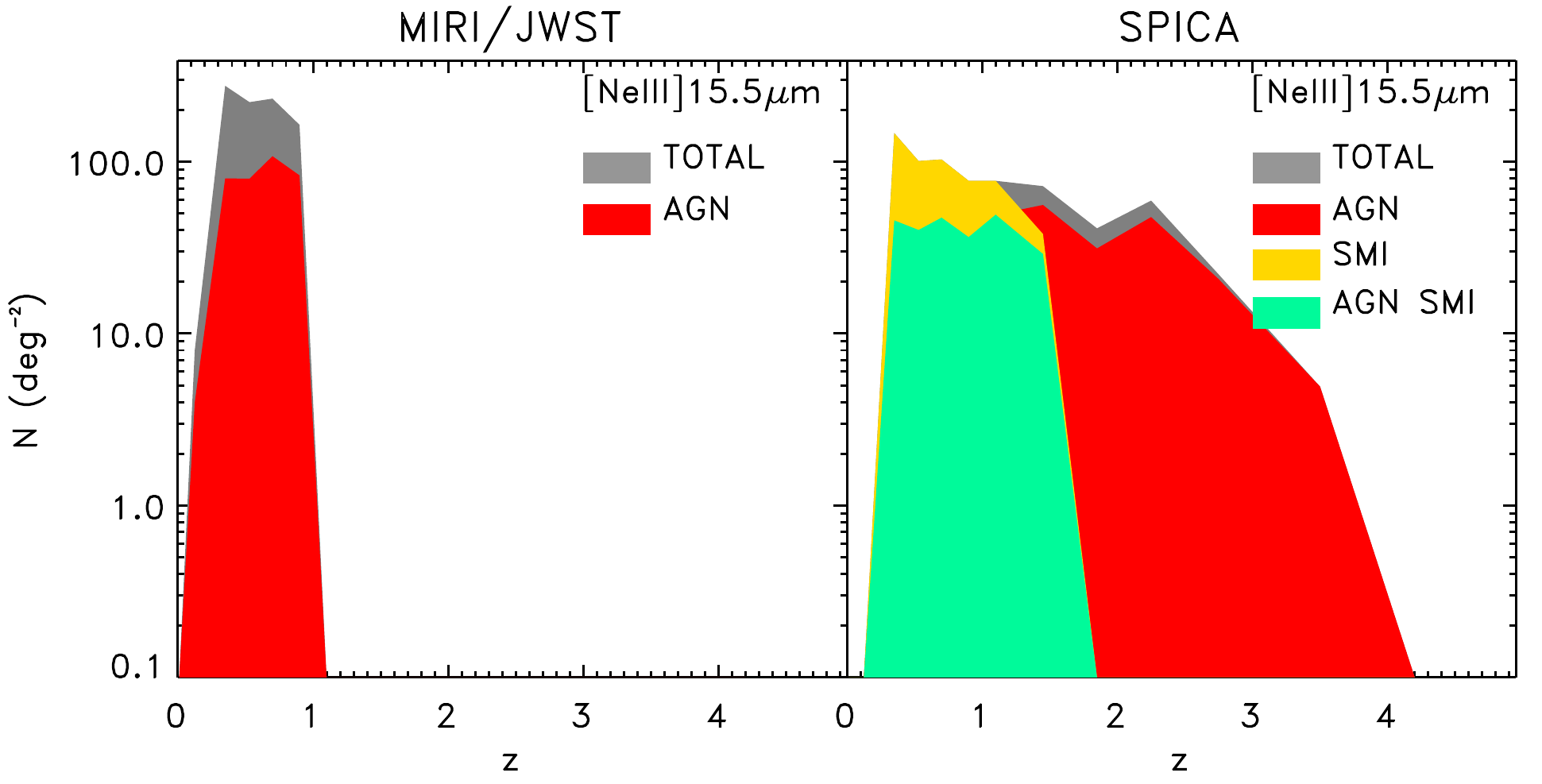}
\includegraphics[width=8cm]{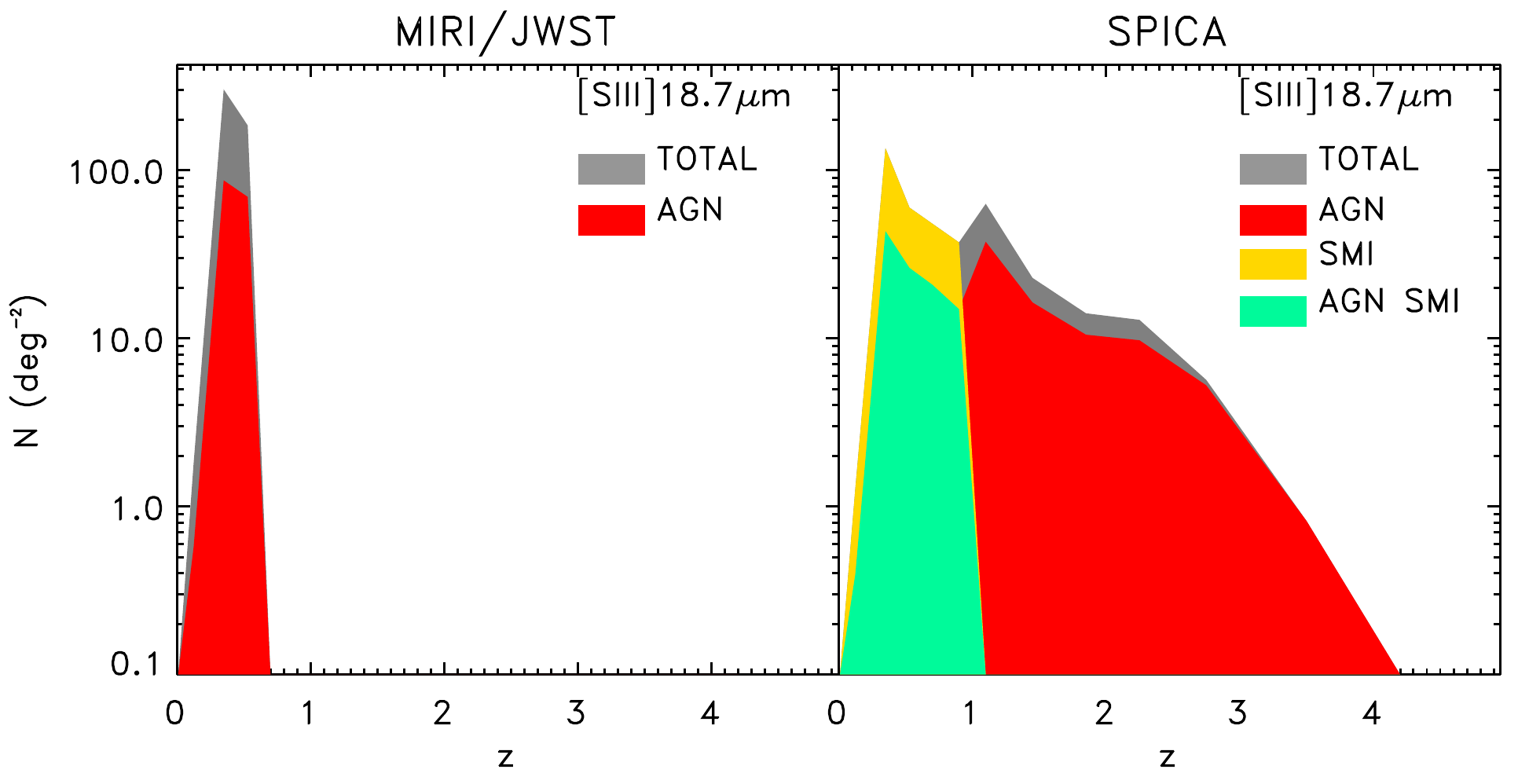}
\includegraphics[width=8cm]{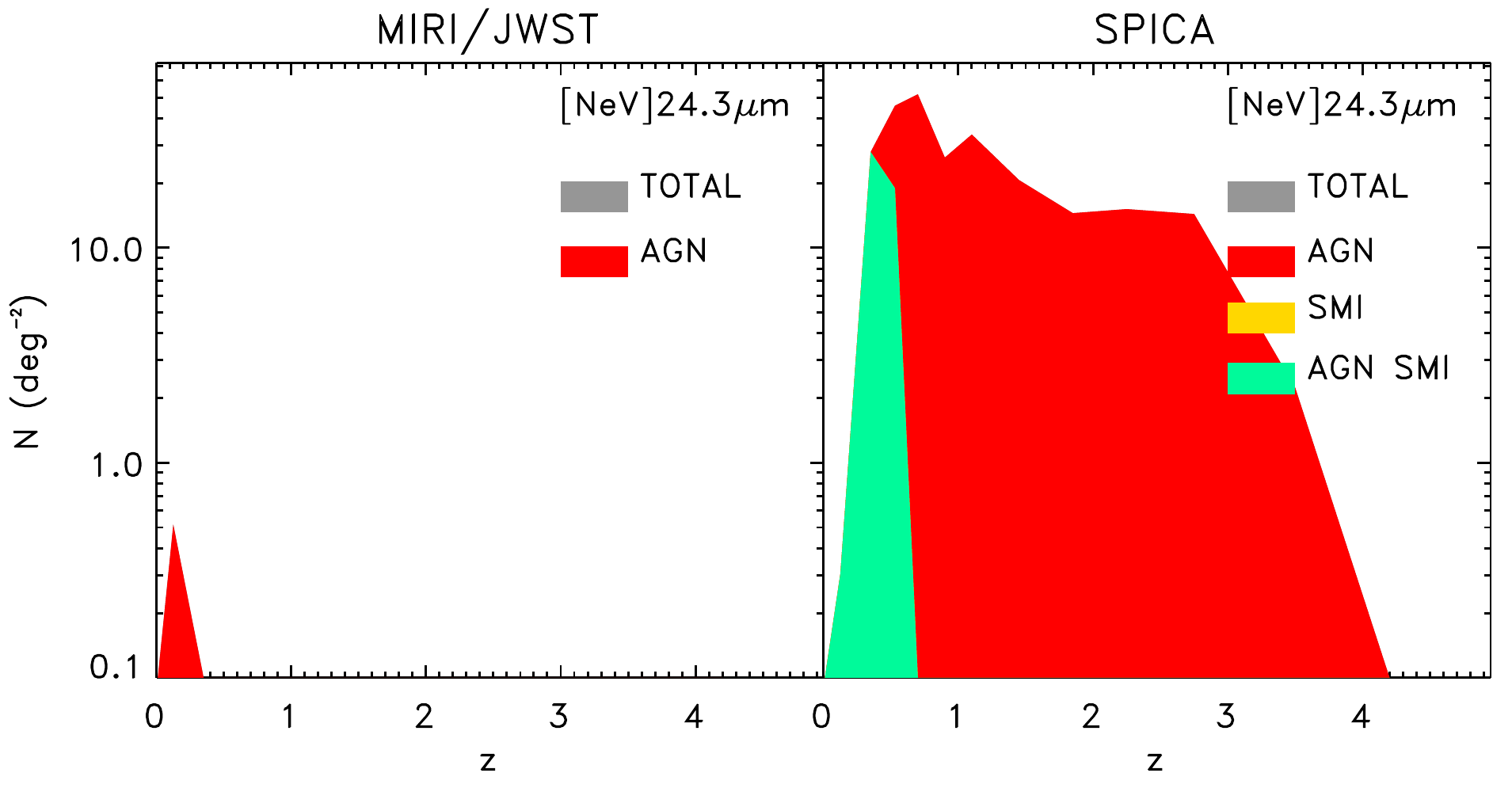}
\includegraphics[width=8cm]{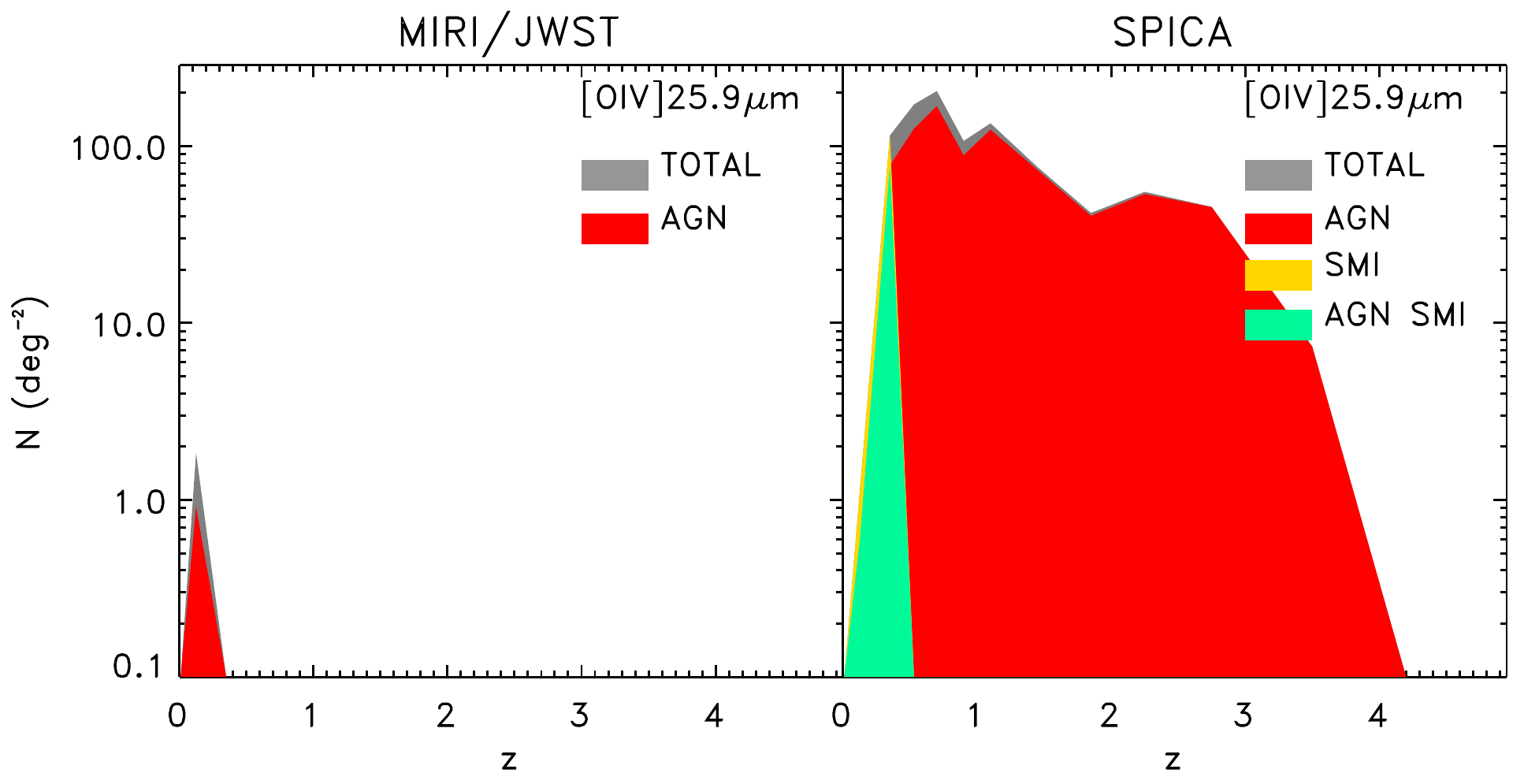}
\includegraphics[width=8cm]{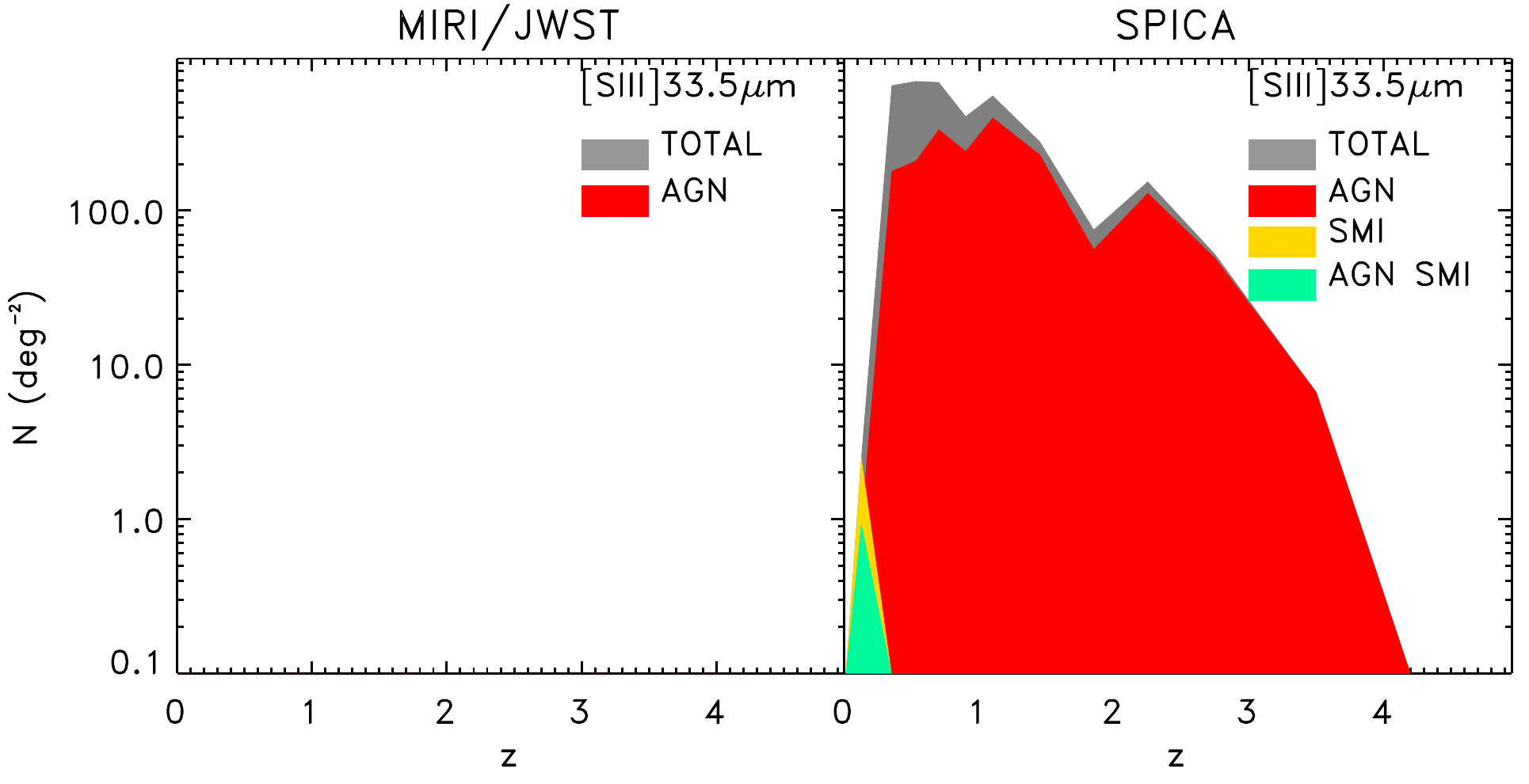}
\includegraphics[width=8cm]{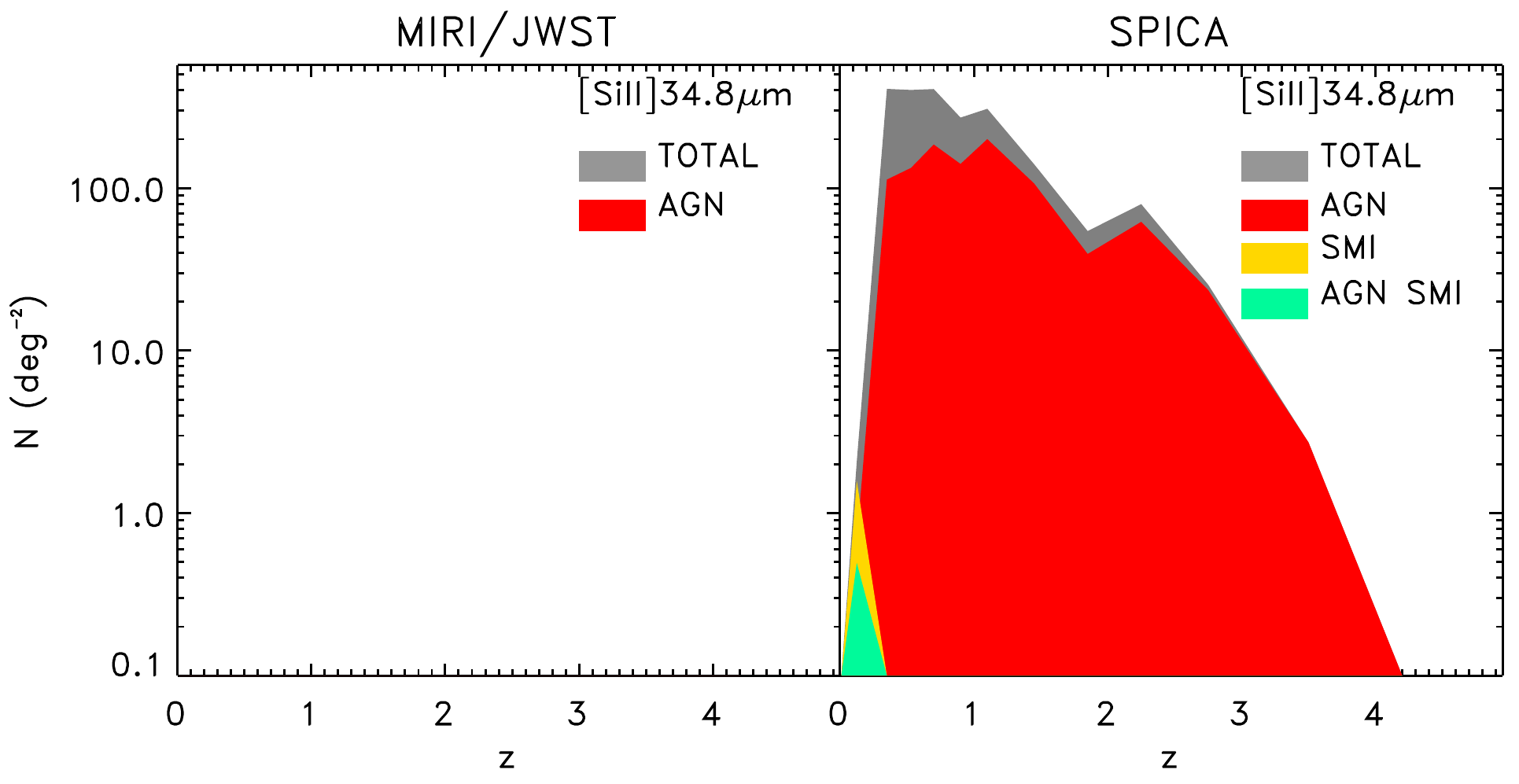}
\caption{Redshift distribution (per unit area) of sources with fluxes of the different mid-IR lines larger than the {\em MIRI-JWST} sensitivities (5$\sigma$, 10000s; {\em left}) and the {\em SPICA} ones (5$\sigma$, 10,000s; {\em right}), as expected based on the method described in this work.
The total (e.g., AGN$+$galaxies) number of sources per unit area (for {\em SPICA} it is the number detectable by both {\em SMI} and {\em SAFARI} instruments) is shown as grey-filled histogram, while the AGN contribution is shown in red. The yellow and green histograms highlight the {\em SMI} total and AGN contribution respectively.}
\label{figzdistr}
\end{figure*}

\subsection{Predictions for future IR facilities}
\label{sec_pred}
The line LFs obtained in the previous Section provide an important tool to estimate how many sources will be observable in each of the considered mid-/far-IR lines
with any forthcoming/future/planned/eventual facilities in the IR/sub-mm. With our LFs, we can not only derive numbers, but redshift and luminosity distributions 
of the detectable sources to any flux limits, making it possible to plan future spectroscopic surveys in a crucial, but totally unexplored wavelength range for high-$z$ sources.

In the next decades a few space IR facilities will be put in orbit.
Due to be launched in 2018, the {\em JWST} (\citealt{gardner06}) will be the next NASA observatory that will explore the Universe at near-/mid-IR wavelengths.
The {\em MIRI} instrument (\citealt{wright10}) is the only mid-IR instrument for {\em JWST}, performing photometric imaging in between 5 and 27 $\mu$m  over a 2.3 arcmin$^2$ field-of-view (FOV), and spectroscopy  (including medium resolution, R $\sim$1500 to 3500, integral field spectroscopy over a 13 arcsec$^2$ FOV at 5--28.5 $\mu$m). 

Complementary to {\em JWST}, {\em SPICA} (\citealt{nakagawa12}), to be proposed for the ESA 5th Medium missions call (due launch in
2028-2030), will observe the Universe at mid-/far-IR, from 10 to 37 $\mu$m with the {\em SMI} camera and spectrograph (\citealt{kataza15}) and continuously from 34 to 230 $\mu$m with the {\em SAFARI} grating spectrograph (\citealt{roelfsema14}). The {\em SPICA} 2.5-m diameter mirror,
cooled down to $\lsimeq$8 K, will allow us to reach unprecedented sensitivities over an unexplored wavelength range, bridging the gap between {\em JWST} and {\em ALMA}. 
\begin{table*}
\caption{Estimated redshift distributions (per unit area) of IR galaxies detectable by {\em JWST} and {\em SPICA} to 5$\sigma$ with 10,000 sec integration per FoV in each of the mid-IR emission lines. In parenthesis we give the number of detections of galaxies containing an AGN. }
 \rotatebox{90}{
 \begin{tabular}{llcccccccccccc}
\hline \hline
Spectral line         & Instrument             & 0.01--0.3 & 0.30--0.45 & 0.45--0.6 & 0.6--0.8         & 0.8--1.0    & 1.0--1.2          &  1.2--1.7         & 1.7--2.0    & 2.0--2.5    & 2.5--3.0    & 3.0--4.2 &All $z$ \\ \hline
                              &                               &           \multicolumn{12}{c}{(deg$^{-2}$)}     \\
{PAH 6.2} & {\em MIRI} & 8(4)         & 1763(507)  &  2457(737) & 2844(1534) & 1651(1103)&  2900(2244) & 1888(1626) & 468(365) & 584(498) & 282(268) & 45(45)  & 14890(8931)  \\
                             & {\em SMI} & 0(0)          & 0(0)           &  0(0)            &  0(0)            &  0(0)         &    0(0)            &    0(0)           &  0(0)        & 338(280) & 161(152)    & 23(23) &  552(455) \\        
                             & {\em SAFARI} &  0(0)   & 0(0)           &  0(0)            &  0(0)            &  0(0)         &    0(0)            &    0(0)           &  0(0)        &   0(0)        &   0(0)        &   0(0)   &  0(0) \\
{PAH 11.2} & {\em MIRI} & 8(4)          & 539(431)   & 1971(576)    & 2258(1177)  & 718(428)  & 1091(783)      &  638(525)    &  0(0)        &   0(0)        &   0(0)        &   0(0)   &  8223(3924) \\
                             & {\em SMI} & 0(0)          & 0(0)           &  0(0)            & 731(344)      & 481(272) & 668(462)       & 367(294)     &  126(92)   & 264(217)  &   0(0)        &   0(0)   &  2637(1682) \\
                             & {\em SAFARI} &  0(0)   & 0(0)           &  0(0)            &  0(0)            &  0(0)         &    0(0)            &    0(0)           &  0(0)        &   0(0)        & 192(181)    &  36(36) & 228(217) \\ 
{[Ne~II] 12.8} & {\em MIRI} & 8(4)      & 1107(294)   & 402(134)    & 412(187)      & 277(143)   & 315(204)       & 145(110)        &   0(0)        &   0(0)        &   0(0)        &   0(0)   & 2664(1076)  \\  
                             & {\em SMI} &  0(0)         & 0(0)           &   183(69)     & 178(78)        & 130(60)    & 122(74)          & 51(37)           &  23(17)      &   0(0)        &   0(0)        &   0(0)   & 687(336) \\ 
                             & {\em SAFARI} &  0(0)   & 0(0)           &  0(0)            &  0(0)            &  0(0)         &    0(0)            &  0(0)             &  57(41)      & 84(65)     & 27(25)      &  4(4)    & 151(120) \\
{[Ne~V] 14.3} & {\em MIRI} &  4(4)      &  28(28)      & 21(21)          & 20(20)        &  15(15)        &     0(0)           &    0(0)           &  0(0)        &   0(0)        &   0(0)        &   0(0)   &  87(87) \\
                             & {\em SMI} &   0(0)         &  11(11)         &  7(7)             &  7(7)             &  4(4)         &   5(5)              & 4(4)             &  0(0)        &   0(0)        &   0(0)        &   0(0)   &  39(39) \\               
                             & {\em SAFARI} &  0(0)   & 0(0)           &  0(0)            &  0(0)            &  0(0)         &    0(0)            &  0(0)             &  8(8)        &  6(6)          &  3(3)         &  1(1)    & 17(17) \\
{[Ne~III] 15.5} & {\em MIRI} &  8(4)     &  276(80)   & 222(80)      & 233(108)      &  164(83)   &     0(0)           &    0(0)           &  0(0)        &   0(0)        &   0(0)        &   0(0)   &  903(355) \\  
                             & {\em SMI} &     0(0)     & 147(46)       & 101(40)       &  103(47)        &  77(36)     &  77(49)          &  38(29)        &   0(0)        &   0(0)        &   0(0)        &   0(0)   &  544(248) \\           
                             & {\em SAFARI} &   0(0)   & 0(0)           &  0(0)            &  0(0)            &  0(0)        &    0(0)            &  34(27)         & 41(31)       & 59(48)     &  22(21)       &  5(5)  & 161(132) \\
{[S~III] 18.7} & {\em MIRI} &  2(1)         &  302(87)    & 184(69)       &  0(0)            &  0(0)         &    0(0)            &    0(0)           &  0(0)        &   0(0)        &   0(0)        &   0(0)   &  489(157) \\ 
                             & {\em SMI} &  2(1)          &  135(43)      & 60(26)       & 48(21)           &  37(15)     &    0(0)            &    0(0)           &  0(0)        &   0(0)        &   0(0)        &   0(0)   &  281(106) \\ 
                             & {\em SAFARI} &   0(0)   & 0(0)           &  0(0)            &  0(0)            &  0(0)        &  63(37)          &  23(16)         &  14(11)       &   13(10)      &  6(5)        &   1(1)     &  119(80) \\
{[Ne~V] 24.3} & {\em MIRI} &  1(1)   & 0(0)           &  0(0)            &  0(0)            &  0(0)         &    0(0)            &    0(0)           &  0(0)        &   0(0)        &   0(0)        &   0(0)   & 1(1) \\
                             & {\em SMI} & 0.5(0.5)  &  28(28)      & 19(19)            &    0(0)            &  0(0)         &    0(0)           &    0(0)           &  0(0)        &   0(0)        &   0(0)        &   0(0)   & 47(47) \\
                             & {\em SAFARI} & 0(0)   & 0(0)           &  27(27)         &   52(52)        &   26(26)    &   34(34)        &  21(21)         &   14(14)     &  15(15)      &  14(14)       &   2(2)    & 205(205) \\
{[O~IV] 25.9} & {\em MIRI} & 2(1)     &  0(0)           &  0(0)            &  0(0)            &  0(0)         &    0(0)            &    0(0)           &  0(0)        &   0(0)        &   0(0)        &   0(0)   &  2(1) \\ 
                             & {\em SMI} &   2(1)       & 116(78)       &   0(0)            &  0(0)            &  0(0)         &    0(0)            &    0(0)           &  0(0)        &   0(0)        &   0(0)        &   0(0)   &  116(78) \\
                             & {\em SAFARI} &  0(0)  &  0(0)          & 172(125)      & 204(168)     &  107(89)    &  134(124)      &  76(73)         & 42(40)    &  55(53)     & 45(45)      & 7(7)   & 842(724) \\
{[S~III] 33.5} & {\em MIRI} &   0(0)   & 0(0)           &  0(0)            &  0(0)            &  0(0)         &    0(0)            &    0(0)           &  0(0)        &   0(0)        &   0(0)        &   0(0)   &  0(0) \\
                             & {\em SMI} &   3(1)     &   0(0)           &  0(0)            &  0(0)            &  0(0)         &    0(0)            &    0(0)           &  0(0)        &   0(0)        &   0(0)        &   0(0)   &  3(1) \\
                             & {\em SAFARI} & 0(0)  &  647(180)   & 690(210)    & 680(336)    &  408(243)  & 556(401)     & 280(230)    &  75(56)    & 154(130)    &  52(49)    & 7(7)   &  3550(1841) \\
{[Si~II] 34.8} & {\em MIRI} &   0(0)   & 0(0)           &  0(0)            &  0(0)            &  0(0)         &    0(0)            &    0(0)           &  0(0)        &   0(0)        &   0(0)        &   0(0)   &  0(0) \\
                             & {\em SMI} &  2(1)     &   0(0)           &  0(0)            &  0(0)            &  0(0)         &    0(0)            &    0(0)           &  0(0)        &   0(0)        &   0(0)        &   0(0)   &  2(1) \\ 
                             & {\em SAFARI} & 1(0)  &  408(113)  &  402(133)    & 407(185)      & 272(141)   & 308(200)    &  140(107)     &  54(39)      & 80(62)     & 25(24)     &  3(3)   & 2100(1007) \\
                           \hline \hline
 \end{tabular}
 }
 \label{tab_zdistr}
 \end{table*}                    
The resulting number of sources (per unit area) that will be detectable in each line as a function of $z$ with the {\em MIRI} IFU spectrograph (by considering the nominal sensitivities) and {\em SPICA} 
(by considering the nominal sensitivities for the resolutions R$\sim$1300--2300 and R$\sim$300 for $SMI$ and $SAFARI$ respectively), to 5$\sigma$ in 10,000sec (time on-source only) 
are presented in Table~\ref{tab_zdistr} and are plotted in the Figure~\ref{figzdistr}.

Figure clearly shows the complementarity of {\em JWST} and {\em SPICA}, the former being more efficient in detecting PAH features, but only the 6.2 $\mu$m one to high redshifts; the other mid-IR lines will 
run out of the {\em MIRI} range at much lower redshift, with [Ne~V] 24.3 $\mu$m and [O~IV] 25.9 $\mu$m being barely detectable up to $z$$<$0.1--0.2. Similarly, the {\em SMI} contribution becomes less and less
relevant for longer wavelength lines (such as [Ne~V] 24.3 $\mu$m and above) and is more important at low redshifts, while {\em SAFARI} will allow us to detect mid-IR lines up to $z$$\sim$3--4. Note that, even 
for lines produced mainly by SF, objects containing an AGN provide a significant contribution (green and red coloured area in Fig.~\ref{figzdistr}).

We must point out that the comparison between {\em JWST} and {\em SPICA} based on source numbers per unit area (depending only on sensitivity) is not totally realistic, it can only tell us the redshift 
ranges covered by the two missions. 
In order to discuss about the real efficiency of detecting sources in the different lines, we have to take into account the 
different FOV of the different instruments. 
Given its small FOV, {\em MIRI} will not be a survey instrument, however, it will be possible to perform very deep pencil-beam surveys over small areas of the sky.
The slit length of {\em SMI} (medium resolution, R$\sim$1300--2300) on {\em SPICA} is planned to be $60^{\prime \prime} \times 3.7^{\prime\prime}$, while {\em MIRI} (medium resolution spectrograph, R$\sim$2400) on {\em JWST} will have a FOV ranging from 3$^{\prime\prime}$.6$\times$3$^{\prime\prime}$.6 to 7$^{\prime\prime}$.6$\times$7$^{\prime\prime}$.6 from 6.4 $\mu$m to 22.5 $\mu$m. Therefore, within the same amount of time, {\em SMI} will scan an area of the sky $\sim$3.8 times larger than {\em MIRI}, even considering the longest wavelength band of the latter (with the larger FOV). As an example, in terms of observing time, to cover an area like the HST eXtreme Deep Field (XDF, 10.8 arcmin$^2$) to the sensitivities considered in our simulations (5$\sigma$, 10,000s on source), {\em MIRI} will have to observe $\sim$ 1,800 hours, against the $\sim$470 hours of {\em SMI}. 
The {\em SAFARI} grating spectrograph will only make pointed observations of single sources at a time, therefore a direct comparison with the areal coverage of the mid-IR instruments is not possible.
However, we have estimated that in $\sim$1,000 hours with {\em SAFARI} we will be able to observe (simultaneously over the whole 32-230 $\mu$m range) in at least three lines 
$\sim$20 $L_{IR}$$\leq$10$^{11}$ L$_{\odot}$ sources at $z$$\sim$1, $\sim$20 $L_{IR}$$\leq$10$^{11.5}$ L$_{\odot}$ sources at $z$$\sim$2 and 
$\sim$20 $L_{IR}$$\leq$10$^{12}$ L$_{\odot}$ sources at $z$$\sim$3. These luminosities are below the knee of the IR LFs at the corresponding redshifts.

Note also that {\em MIRI/JWST} has spectroscopic coverage only up to 28 $\mu$m, and therefore most of the atomic and
molecular diagnostic features will be inaccessible beyond $z$$\sim$2. ALMA, which is designed for the sub-mm wavelengths,
has coverage only beyond $\sim$330 $\mu$m and can be used to trace important far-IR fine structure lines (e.g. [C~II] 157$\mu$m
and [O~III] 88 $\mu$m) at high-redshift, but is blind to many of the mid-infrared atomic, H$_2$ and dust features. {\em SPICA} will
provide a unique window on the physical processes fuelling star formation and BH growth in the dusty Universe.

\section{Conclusions}
We have exploited all the information -- both photometric and spectroscopic, from X-rays to mm -- available for 76 sources from the extensively studied 12MGS local sample of Seyfert 
galaxies, to perform a broad-band SED-fitting and decomposition and obtain new relations between mid- and far-IR line luminosity and AGN, SF and total IR luminosities,
also investigating the variations of these relations for different fractions of AGN contribution. Thanks to the use of all the data available for these objects,
we were able to derive crucial AGN and host galaxy quantities (i.e., the AGN bolometric luminosity, the AGN fraction in different wavelength ranges, the SFR)
and greatly reduce the uncertainties and degeneracies with respect to similar SED-decomposition works 
performed at high-z, where the very few photometric points available in the mid-IR/far-IR range (crucial to constrain 
and separate the AGN and the starburst contributions) lead to very uncertain results.
The relations found between the line luminosity and the SF or AGN or total IR luminosities, reflecting a connection between the black hole accretion and the SFR in the local
AGN, are used to estimate numbers, redshifts and luminosities of sources detectable in the different mid-/far-IR lines. 

The main results of this work can be summarised as follows:
\begin{itemize}
\item The observed broad-band SEDs of the 12MGS have been decomposed into three different components, peaking in three different regimes: evolved stars 
(optical/near-IR), dust-heated by SF (far-IR/sub-mm) and AGN dusty torus emission (mid-IR). Crucial physical quantities, as the SFR, the AGN bolometric luminosity, the total IR luminosity
due either to SF or AGN, the AGN fraction in different wavelength ranges, are accurately derived, thanks to the wealth of data available, including {\em IRS} spectra in the mid-IR. 
\item The 12MGS sources show all the range of AGN dominance in the mid-IR (e.g., 5--40 $\mu$m), from about 90\% to no AGN at all.
\item The AGN bolometric luminosity derived through the SED decomposition is in good agreement with previous derivations for the same objects from X-rays and high excitation 
mid-IR lines (i.e., [Ne~V], [O~IV]).
\item The SF luminosity is found to correlate nicely with the AGN bolometric luminosity, with the dispersion in the relation caused by the different fraction of AGN contribution to the 
IR luminosity: the normalisation of the relation changes with $f_{AGN}$, but the slope keeps similar to that found by Netzer et al. (2007, 2009)  
(e.g., sources with higher AGN fraction show lower SF luminosity at a given AGN bolometric luminosity). 
\item The total IR luminosity, the SF luminosity and the intrinsic AGN bolometric luminosity are found to correlate with the IR line luminosity. 
Variations of these relations with different AGN fractions are investigated resulting in higher AGN-line luminosities for sources with larger AGN fractions at given total IR luminosity, 
and no variation with the AGN contribution for lines produced by SF. On the contrary, the normalisation of the relation between the SF produced lines and the AGN luminosity
increases with decreasing AGN fraction, while it shows no variation with the amount of AGN relative contribution for AGN signature lines.
\item These local line versus AGN/SF luminosity relations are then used, together with the recent {\em Herschel} galaxy LF and
evolution results, to obtain IR line luminosity functions up to high redshifts, in order to estimate for how many SF galaxies and AGN we expect to detect and measure the mid-/far-IR lines 
at different redshifts and luminosities with the forthcoming/planned future IR space missions (e.g., JWST, SPICA).
\end{itemize}
 
Since the nuclear properties are likely to be more correlated than those averaged over the whole galaxy size,
the investigation of eventual relation between the nuclear AGN activity and the circum-nuclear SF (rather than the global SF)
would shed more light on the AGN/host galaxy connection. Spatially resolved observations of the nuclei of these
Seyfert galaxies in the near-/mid-IR would be a natural extension of this work. 
 
\section*{Acknowledgments}
We thank an anonymous referee for helpful comments that improved the
paper significantly.
We would also like to thank Gianni Zamorani for his valuable help on the statistical tests and results. 
CG acknowledges financial contribution from the contracts 
ASI-INAF I\/005\/07/1 and I\/005\/11\/0.

\bibliography{mybibliography}
\bibliographystyle{mnras}

\appendix
\section{Photometric Data}
\label{appendix}
The photometric data for our 12MGS subsample of 76 sources with {\em Spitzer-IRS} spectra have been collected from NED, Vizier and Simbad public databases in
fixed bands. Here, for the first time a ``homogenised'' photometric catalogue is put together for this sample, with derived total fluxes (corrected for aperture and magnitude zero point).
In Table~\ref{tab_phot_mir}  and \ref{tab_phot_fir} data from UV to {\em Spitzer-IRAC} bands and from 12 $\mu$m to 3 mm are presented respectively.
\begin{table*}
 \caption{12 MGS optical to mid-IR photometry}
 \rotatebox{90}{
\begin{tabular}{lrrrrrrrrrrrrrrrrr}
\hline \hline
         name &      \multicolumn{17}{c}{Flux (in mJy) at $\lambda$ (in $\mu$m):} \\ 
                     &  0.35 &      0.37 &    0.45 &   0.47 &      0.55 &      0.62 &      0.69 &      0.77 &      0.88 &      0.94 &      1.25 &      1.64 &      2.17 &      3.6 &      4.5 &      5.8 &      8.0   \\ \hline
          3C120 &      ... &      ... &      2.2 &      ... &      3.6 &      ... &      8.1 &      ... &      ... &      ... &     32.4 &     44.7 &     44.7 &      ... &      ... &      ... &      ... \\
          3C234 &      0.1 &      ... &      0.2 &      0.2 &      0.5 &      0.7 &      ... &      0.8 &      ... &      ... &      2.0 &      3.5 &      5.3 &      ... &     16.0 &      ... &      ... \\
          3C273 &     10.4 &      ... &     25.4 &     23.1 &     27.4 &     25.8 &      ... &     32.2 &      ... &      ... &     42.7 &     42.7 &    100.0 &      ... &      ... &      ... &      ... \\
          3C445 &      ... &      ... &      0.7 &      ... &      ... &      ... &      1.3 &      ... &      ... &      ... &      5.0 &      8.9 &     19.1 &      ... &      ... &      ... &      ... \\
    CGCG381-051 &      ... &      ... &      3.2 &      ... &      2.6 &      ... &      ... &      ... &      ... &      ... &     12.3 &     15.9 &     14.5 &      9.8 &      8.4 &     15.6 &    110.4 \\
    ESO012-G021 &      ... &      ... &      6.5 &      ... &      6.5 &      ... &     14.5 &      ... &      ... &      ... &     22.6 &     32.4 &     37.4 &     39.5 &     39.7 &     54.5 &    208.5 \\
    ESO033-G002 &      ... &      ... &      6.0 &      ... &      5.5 &      ... &     15.2 &      ... &      ... &      ... &     48.3 &     69.4 &     65.0 &     48.8 &     60.1 &     78.8 &    140.3 \\
    ESO141-G055 &      ... &      ... &      7.6 &      ... &     13.2 &      ... &     16.8 &      ... &      ... &      ... &     39.8 &     41.7 &     50.1 &      ... &      ... &      ... &      ... \\
    ESO362-G018 &      ... &      ... &     12.6 &      ... &     17.0 &      ... &     27.9 &      ... &      ... &      ... &     57.9 &     71.4 &     65.2 &     35.7 &     31.8 &     57.1 &    140.3 \\
        IC4329A &      ... &      ... &     12.6 &      ... &     13.0 &      ... &     46.3 &      ... &      ... &      ... &    109.6 &    158.5 &    177.8 &    313.7 &    382.4 &    483.8 &    684.0 \\
         IC5063 &      ... &      9.0 &     28.7 &      ... &     67.5 &      ... &     92.7 &      ... &      ... &      ... &    209.0 &    252.0 &    211.0 &     85.6 &    124.3 &    219.1 &    563.7 \\
IRASF01475-0740 &      ... &      ... &      0.8 &      ... &      2.1 &      ... &      1.9 &      ... &      ... &      ... &      7.0 &     10.1 &      9.7 &     13.1 &     19.9 &     42.1 &     87.7 \\
IRASF03450+0055 &      ... &      ... &      3.9 &      ... &      5.3 &      ... &      ... &      ... &      ... &      ... &     12.0 &     19.3 &     31.4 &     63.8 &     74.2 &     92.2 &    145.6 \\
IRASF04385-0828 &      ... &      ... &      2.2 &      ... &      6.0 &      ... &      ... &      ... &      ... &      ... &     16.2 &     26.3 &     37.2 &     64.9 &    109.3 &    175.7 &    355.7 \\
IRASF05189-2524 &      ... &      ... &      2.7 &      ... &      4.8 &      ... &      ... &      ... &      ... &      ... &     14.8 &     30.9 &     57.5 &      ... &      ... &      ... &      ... \\
IRASF07599+6508 &      5.4 &      ... &      8.0 &      6.4 &      ... &      6.7 &      7.9 &      8.0 &      ... &      7.0 &     10.0 &     20.9 &     32.4 &      ... &      ... &      ... &      ... \\
IRASF08572+3915 &      ... &      ... &      ... &      0.4 &      0.8 &      0.6 &      ... &      0.9 &      ... &      1.1 &      1.7 &      3.0 &      3.9 &      ... &      ... &      ... &      ... \\
IRASF13349+2438 &      1.8 &      ... &      ... &      2.2 &      3.8 &      6.3 &      ... &      5.3 &      ... &      7.7 &     14.1 &     26.9 &     52.5 &      ... &      ... &      ... &      ... \\
IRASF15480-0344 &      ... &      ... &      1.4 &      ... &      3.4 &      ... &      4.0 &      ... &      ... &      ... &     20.3 &     25.0 &     26.1 &     19.8 &     24.6 &     36.4 &     98.9 \\
         Izw001 &      ... &      ... &      7.3 &      ... &      9.2 &      ... &      ... &      ... &      ... &      ... &     34.7 &     58.9 &     89.1 &      ... &      ... &      ... &      ... \\
  MCG-02-33-034 &      ... &      ... &      5.4 &      ... &      9.2 &      ... &     25.7 &      ... &      ... &      ... &     70.0 &     85.7 &     76.6 &     25.2 &     20.3 &     27.3 &     71.0 \\
  MCG-03-34-064 &      ... &      ... &     10.6 &     12.6 &      ... &      ... &     18.8 &      ... &      ... &      ... &     69.7 &     87.3 &     76.6 &      ... &      ... &      ... &      ... \\
  MCG-03-58-007 &      ... &      ... &      2.4 &      ... &      4.6 &      ... &      ... &      ... &      ... &      ... &     20.0 &     29.5 &     36.3 &     37.3 &     45.1 &     58.0 &    123.0 \\
  MCG-06-30-015 &      ... &      ... &     11.7 &      ... &     13.6 &      ... &      ... &      ... &      ... &      ... &     71.8 &     94.8 &     98.0 &    106.8 &    127.8 &    157.3 &    239.4 \\
  MCG+00-29-023 &      ... &      ... &      4.2 &      ... &      ... &      ... &      ... &      ... &      ... &      ... &     39.8 &     57.5 &     49.0 &     30.0 &     24.8 &     55.0 &    372.4 \\
        MRK0006 &      ... &      1.7 &      4.2 &      ... &     14.2 &      ... &      ... &      ... &      ... &      ... &     59.4 &     83.9 &    100.0 &     92.2 &    101.5 &    104.9 &    142.9 \\
        MRK0009 &      ... &      ... &      5.2 &      ... &      6.8 &      ... &      ... &      ... &      ... &      ... &     21.9 &     33.1 &     46.8 &      ... &      ... &      ... &      ... \\
        MRK0079 &      1.4 &      ... &      5.4 &      5.7 &      7.4 &     16.1 &      ... &     38.0 &      ... &     50.6 &     39.8 &     57.5 &     64.6 &     90.5 &    108.3 &    146.1 &    243.8 \\
        MRK0231 &      3.0 &      ... &      5.7 &      5.6 &     11.0 &     11.9 &      ... &     23.2 &      ... &     21.7 &     49.0 &    107.2 &    186.2 &      ... &      ... &      ... &      ... \\
        MRK0273 &      1.1 &      ... &      2.3 &      4.1 &      4.1 &      7.3 &      ... &     10.7 &      ... &     14.1 &     19.6 &     60.3 &     61.7 &      ... &      ... &      ... &      ... \\
        MRK0335 &      ... &      2.1 &      8.9 &      ... &     10.9 &      ... &     16.5 &      ... &     15.4 &      ... &     33.1 &     55.0 &     95.5 &     92.6 &    110.8 &    127.7 &    176.0 \\
        MRK0463 &      ... &      2.2 &      3.0 &      ... &      7.8 &      ... &     10.3 &      ... &      ... &      ... &     21.7 &     35.1 &     62.0 &    115.0 &    156.0 &      ... &    412.0 \\
        MRK0509 &      ... &      ... &     19.3 &      ... &     21.4 &      ... &      ... &      ... &      ... &      ... &     91.2 &    107.2 &    128.8 &      ... &      ... &      ... &      ... \\
        MRK0704 &      1.9 &      ... &      5.0 &      4.6 &      7.9 &      8.9 &      ... &      8.3 &      ... &     15.5 &     20.9 &     32.4 &     41.7 &     81.8 &    107.3 &    132.0 &    237.2 \\
        MRK0897 &      ... &      ... &      2.9 &      ... &      4.7 &      ... &      ... &      ... &      ... &      ... &     50.1 &     72.4 &     60.3 &     27.6 &     25.0 &     26.6 &     78.5 \\
        MRK1239 &      1.7 &      ... &      4.3 &      4.6 &      6.6 &      9.0 &      ... &     10.8 &      ... &     14.7 &     22.9 &     44.7 &     85.1 &    233.6 &    277.0 &    344.1 &    500.1 \\
        NGC0034 &      ... &      ... &      9.3 &      ... &     14.8 &      ... &     20.6 &      ... &     42.9 &      ... &     51.1 &     67.1 &     62.3 &     42.5 &     36.2 &     92.2 &    590.3 \\
        NGC0262 &      ... &      4.4 &     11.7 &     19.0 &     16.9 &     28.7 &      ... &      ... &      ... &      ... &     89.1 &    104.7 &    100.0 &     55.0 &     74.2 &     98.3 &    179.9 \\
        NGC0424 &      ... &      ... &     13.2 &      ... &     44.2 &      ... &     38.5 &      ... &     84.8 &      ... &    100.7 &    131.0 &    148.7 &    207.3 &    301.0 &    413.7 &    671.5 \\
        NGC0513 &      3.0 &      ... &     18.4 &     11.1 &     16.5 &     22.4 &      ... &     31.8 &      ... &     41.3 &     66.2 &     83.3 &     72.2 &     37.7 &     29.8 &     37.4 &    190.1 \\
       NGC0526A &      ... &      ... &      5.3 &      ... &      6.8 &      ... &     14.3 &      ... &     19.9 &      ... &     28.2 &     34.7 &     33.1 &     70.6 &     83.7 &     99.2 &    137.7 \\
        NGC0931 &      ... &      ... &      2.2 &      ... &      4.8 &      ... &      ... &      ... &      ... &      ... &     57.5 &     81.3 &     87.1 &    124.9 &    142.7 &    205.4 &    312.7 \\
       NGC1056 &      ... &      ... &     20.0 &      ... &     41.8 &      ... &     48.9 &      ... &    152.0 &      ... &    140.0 &    181.0 &    150.0 &     55.5 &     33.6 &     58.7 &    404.6 \\
\hline
\end{tabular}
}
\end{table*}

\begin{table*}
\setcounter{table}{0}
\caption{$-$ Continue}
 \rotatebox{90}{
\begin{tabular}{lrrrrrrrrrrrrrrrrr}
\hline \hline
         name &      \multicolumn{17}{c}{Flux (in mJy) at $\lambda$ (in $\mu$m):} \\ 
                     &  0.35 &      0.37 &    0.45 &   0.47 &      0.55 &      0.62 &      0.69 &      0.77 &      0.88 &      0.94 &      1.25 &      1.64 &      2.17 &      3.6 &      4.5 &      5.8 &      8.0   \\ \hline
        NGC1125 &      ... &      ... &     10.8 &      ... &     21.2 &      ... &     35.2 &      ... &      ... &      ... &      ... &      ... &     32.8 &     26.8 &     24.4 &     47.9 &    159.6 \\
        NGC1194 &      1.4 &      3.5 &      6.1 &      9.8 &      ... &     25.3 &      ... &     40.4 &      ... &     57.6 &     68.2 &     79.7 &     73.6 &     54.0 &     79.9 &    137.0 &    232.8 \\
        NGC1320 &      ... &      7.0 &     19.8 &      ... &     39.6 &      ... &      ... &      ... &      ... &      ... &    100.0 &    125.0 &    112.0 &     64.4 &     73.5 &    117.1 &    264.9 \\
        NGC1365 &      ... &    116.0 &    345.0 &      ... &      ... &    512.0 &      ... &      ... &      ... &      ... &    507.0 &    546.0 &    516.0 &    360.2 &    327.0 &    603.5 &   3214.1 \\
        NGC1566 &      ... &    145.9 &    354.2 &      ... &    484.8 &      ... &    714.5 &      ... &    904.5 &      ... &   1264.7 &   1339.1 &   1173.8 &    574.0 &    395.4 &    460.0 &   1350.0 \\
        NGC2992 &      2.0 &      4.0 &      9.2 &      ... &     12.7 &      ... &     15.8 &     13.9 &      ... &     16.6 &     22.4 &     28.6 &     23.8 &    116.0 &    106.3 &    144.8 &    464.6 \\
        NGC3079 &      ... &     44.9 &    102.1 &      ... &    171.2 &      ... &      ... &      ... &      ... &      ... &    389.0 &    512.9 &    407.4 &    269.0 &    177.0 &    444.0 &   3350.0 \\
        NGC3516 &      ... &      ... &     23.8 &      ... &     41.5 &      ... &      ... &      ... &      ... &      ... &    246.6 &    274.7 &    262.5 &    120.4 &    117.6 &    142.1 &    220.3 \\
       NGC4051 &      5.3 &      ... &     15.5 &     26.2 &     25.7 &      ... &      ... &      ... &      ... &      ... &     74.1 &    109.6 &    107.2 &    108.8 &    119.8 &    166.2 &    382.9 \\
       NGC4151 &     13.8 &      ... &     56.6 &     44.3 &     96.7 &     82.4 &      ... &    134.2 &      ... &    240.3 &    245.5 &    263.0 &    239.9 &    302.4 &    386.0 &    520.8 &    997.8 \\
        NGC4253 &      4.0 &      ... &      7.7 &     13.8 &     14.1 &     24.4 &      ... &     31.1 &      ... &     40.6 &     26.9 &     49.0 &     63.1 &     66.8 &     84.4 &    104.9 &    257.7 \\
       NGC4388 &      9.1 &      ... &      ... &      ... &      ... &      ... &      ... &      ... &     34.6 &      ... &     40.9 &     51.1 &     41.9 &     85.6 &     82.1 &    164.7 &    408.4 \\
        NGC4593 &      ... &      ... &     11.1 &      ... &     20.8 &      ... &      ... &      ... &      ... &      ... &    169.8 &    195.0 &    158.5 &    124.9 &    133.8 &    163.2 &    290.4 \\
        NGC4602 &      ... &      ... &     43.4 &      ... &      ... &      ... &     41.1 &      ... &     52.0 &      ... &     58.9 &     75.9 &     61.7 &    136.2 &     84.6 &    122.4 &    433.8 \\
       NGC5135 &      ... &      ... &     39.2 &      ... &     76.1 &      ... &     84.3 &      ... &      ... &      ... &    195.0 &    295.1 &    257.0 &     73.9 &     65.2 &    132.0 &    756.9 \\
        NGC5256 &      ... &      ... &      8.9 &      ... &     16.2 &      ... &      ... &      ... &      ... &      ... &     44.2 &     56.3 &     52.5 &     23.6 &     17.8 &     43.3 &    346.0 \\
        NGC5347 &      4.2 &      ... &     17.4 &     19.9 &     31.4 &     38.7 &      ... &     55.9 &      ... &     71.4 &     89.9 &    105.0 &     92.5 &     31.7 &     37.2 &     61.5 &    152.4 \\
        NGC5506 &      5.0 &      ... &      ... &     25.2 &      ... &     57.0 &      ... &     76.8 &      ... &    115.2 &    208.0 &    299.0 &    354.0 &    437.1 &    568.3 &    732.3 &   1104.2 \\
        NGC5548 &      4.3 &      ... &      8.0 &      ... &     12.2 &     32.5 &      ... &     42.7 &      ... &     54.6 &     76.0 &     98.9 &    106.0 &     57.5 &     66.3 &     96.4 &    210.1 \\
        NGC5953 &      9.6 &      ... &     21.5 &     36.5 &     39.6 &     61.0 &      ... &     83.2 &      ... &    109.0 &    131.0 &    161.0 &    135.0 &     62.6 &     36.2 &     95.7 &    736.3 \\
        NGC5995 &      ... &      ... &      5.8 &      ... &     12.6 &      ... &     18.6 &      ... &      ... &      ... &     43.7 &     69.2 &     72.4 &    110.8 &    129.0 &    163.2 &    375.9 \\
        NGC6810 &      ... &      ... &     48.9 &      ... &    122.9 &      ... &    179.5 &      ... &      ... &      ... &     83.0 &     79.3 &     89.4 &    155.8 &     97.0 &    148.8 &   1025.8 \\
        NGC6860 &      ... &      ... &     14.0 &      ... &     14.6 &      ... &     48.1 &      ... &      ... &      ... &     92.5 &    118.8 &    116.9 &     66.8 &     75.6 &    107.8 &    183.3 \\
        NGC6890 &      ... &     12.5 &     31.7 &      ... &     51.7 &      ... &     72.4 &      ... &      ... &      ... &    147.0 &    177.0 &    151.0 &     44.5 &     37.2 &     52.6 &    185.0 \\
       NGC7130 &      ... &      ... &     31.5 &      ... &      ... &      ... &     72.7 &      ... &      ... &      ... &    115.0 &    141.0 &    125.0 &     47.5 &     39.0 &     78.1 &    558.5 \\
       NGC7213  &      ... &     51.7 &    168.0 &      ... &    326.0 &      ... &    482.0 &      ... &      ... &      ... &    973.0 &   1120.0 &    942.0 &    340.0 &    270.0 &    270.0 &    340.0 \\
       NGC7469 &      ... &     17.4 &     26.6 &      ... &     43.8 &      ... &      ... &      ... &      ... &      ... &    117.5 &    166.0 &    154.9 &    150.2 &    150.9 &    270.8 &   1188.7 \\
        NGC7496 &      ... &      ... &     72.0 &      ... &      ... &      ... &    143.9 &      ... &    172.3 &      ... &    169.0 &    196.0 &    156.0 &     28.3 &     21.0 &     47.5 &    272.3 \\
        NGC7603 &      3.8 &      ... &      5.4 &     13.9 &      9.4 &     24.2 &      ... &     28.0 &      ... &     51.4 &    354.8 &    501.2 &    416.9 &    143.4 &    156.5 &    178.9 &    298.6 \\
        NGC7674 &      ... &      5.5 &     11.4 &      ... &     19.3 &      ... &      ... &      ... &      ... &      ... &     51.2 &     64.6 &     68.4 &     71.2 &     94.3 &    147.5 &    443.7 \\
 TOLOLO1238-364 &      ... &      ... &     26.4 &      ... &      ... &      ... &     61.7 &      ... &      ... &      ... &     64.6 &    100.0 &    109.6 &     50.6 &     50.4 &     89.7 &    416.0 \\
      UGC05101 &      0.5 &      ... &      3.5 &      ... &      ... &      5.8 &      ... &      8.5 &      ... &     12.0 &     16.4 &     25.1 &     33.0 &     80.0 &    160.0 &    206.0 &    340.0 \\
       UGC07064 &      2.7 &      ... &      5.0 &      9.6 &      9.4 &     18.0 &      ... &     24.9 &      ... &     32.3 &     50.5 &     60.2 &     55.8 &     28.6 &     24.4 &     32.3 &    144.2 \\
\hline \hline
\end{tabular}
}
\label{tab_phot_mir}
\end{table*}

\begin{table*}
 \caption{12 MGS mid-IR to mm photometry}
 \rotatebox{90}{
\begin{tabular}{lrrrrrrrrrrrrrrrrrrrr}
\hline \hline
         name &      \multicolumn{20}{c}{Flux (in Jy) at $\lambda$ (in $\mu$m):} \\ 
               &       12 &       25 &       60 &      100 &      120 &      150 &      170 &      180 &      200 &      250 &      350 &      450 &      500 &      800 &      850 &      870 &     1100 &     1300 &     2000 &     3000 \\ \hline
           3C120 &     0.43 &     0.67 &     1.55 &     4.82 &     4.00 &     5.10 &     9.90 &      ... &      ... &      ... &      ... &      ... &      ... &     1.00 &     1.82 &      ... &     1.50 &      ... &     1.40 &     1.80\\
          3C234 &     0.10 &     0.26 &     0.31 &     0.26 &      ... &      ... &      ... &     0.80 &     1.09 &      ... &      ... &      ... &      ... &      ... &     2.27 &      ... &    12.00 &      ... &    15.00 &    16.50\\
          3C273 &     0.55 &     0.89 &     2.06 &     2.89 &     1.49 &     1.09 &     1.10 &     0.80 &     0.79 &      ... &      ... &      ... &      ... &     7.00 &     8.00 &      ... &    10.00 &      ... &    15.00 &    16.50\\
          3C445 &     0.21 &     0.26 &     0.31 &     0.60 &      ... &     0.56 &      ... &      ... &     0.49 &      ... &      ... &      ... &      ... &      ... &      ... &      ... &      ... &      ... &      ... &      ... \\
    CGCG381-051 &     0.16 &     0.51 &     1.75 &     2.76 &      ... &      ... &     1.82 &      ... &      ... &      ... &      ... &      ... &      ... &      ... &      ... &      ... &      ... &      ... &      ... &      ... \\
    ESO012-G021 &     0.22 &     0.19 &     1.51 &     3.22 &      ... &     2.21 &     1.81 &     1.14 &     0.92 &      ... &      ... &      ... &      ... &      ... &      ... &      ... &      ... &      ... &      ... &      ... \\
    ESO033-G002 &     0.24 &     0.47 &     0.82 &     1.84 &      ... &      ... &      ... &      ... &      ... &      ... &      ... &      ... &      ... &      ... &      ... &      ... &      ... &      ... &      ... &      ... \\
    ESO141-G055 &     0.28 &     0.46 &     0.47 &     1.48 &      ... &      ... &      ... &      ... &      ... &      ... &      ... &      ... &      ... &      ... &      ... &      ... &      ... &      ... &      ... &      ... \\
    ESO362-G018 &     0.23 &     0.57 &     1.28 &     2.34 &     2.54 &     1.86 &      ... &     0.86 &     0.55 &      ... &      ... &      ... &      ... &      ... &      ... &      ... &      ... &      ... &      ... &      ... \\
        IC4329A &     1.11 &     2.26 &     2.15 &     2.31 &      ... &     1.87 &      ... &      ... &     1.20 &      ... &      ... &      ... &      ... &      ... &      ... &      ... &      ... &      ... &      ... &      ... \\
         IC5063 &     1.15 &     3.95 &     5.79 &     4.41 &     4.97 &     4.13 &      ... &     3.10 &     2.26 &      ... &      ... &      ... &      ... &      ... &      ... &      ... &      ... &      ... &      ... &      ... \\
IRASF01475-0740 &     0.32 &     0.84 &     1.10 &     1.05 &      ... &      ... &      ... &      ... &      ... &      ... &      ... &      ... &      ... &      ... &      ... &      ... &      ... &      ... &      ... &      ... \\
IRASF03450+0055 &     0.29 &     0.39 &     0.87 &      ... &      ... &     1.03 &      ... &      ... &     0.68 &      ... &      ... &      ... &      ... &      ... &      ... &      ... &      ... &      ... &      ... &      ... \\
IRASF04385-0828 &     0.59 &     1.70 &     2.91 &     3.55 &      ... &      ... &      ... &      ... &      ... &      ... &      ... &      ... &      ... &      ... &      ... &      ... &      ... &      ... &      ... &      ... \\
IRASF05189-2524 &     0.71 &     3.41 &    13.27 &    11.90 &    10.60 &     7.00 &      ... &     3.20 &     2.90 &     1.94 &     0.72 &      ... &     0.23 &      ... &      ... &      ... &      ... &      ... &      ... &      ... \\
IRASF07599+6508 &     0.33 &     0.54 &     1.75 &     1.47 &      ... &      ... &      ... &      ... &      ... &      ... &      ... &      ... &      ... &      ... &      ... &      ... &      ... &      ... &      ... &      ... \\
IRASF08572+3915 &     0.32 &     1.87 &     7.33 &     4.98 &      ... &      ... &      ... &      ... &      ... &      ... &      ... &      ... &      ... &      ... &     0.02 &      ... &      ... &      ... &      ... &      ... \\
IRASF13349+2438 &     0.61 &     0.72 &     0.85 &     0.90 &      ... &      ... &     0.35 &      ... &      ... &      ... &      ... &      ... &      ... &      ... &      ... &      ... &      ... &      ... &      ... &      ... \\
IRASF15480-0344 &     0.24 &     0.72 &     1.09 &     4.05 &      ... &      ... &      ... &      ... &      ... &      ... &      ... &      ... &      ... &      ... &      ... &      ... &      ... &      ... &      ... &      ... \\
         Izw001 &     0.47 &     1.17 &     2.24 &     2.87 &     2.57 &     1.89 &     1.60 &     1.16 &     0.92 &      ... &      ... &      ... &      ... &      ... &      ... &      ... &      ... &      ... &      ... &      ... \\
  MCG-02-33-034 &     0.17 &     0.37 &     1.16 &     2.22 &      ... &      ... &      ... &      ... &      ... &      ... &      ... &      ... &      ... &      ... &      ... &      ... &      ... &      ... &      ... &      ... \\
  MCG-03-34-064 &     0.95 &     2.88 &     6.22 &     6.37 &     6.40 &     4.40 &      ... &     3.60 &     2.30 &      ... &      ... &      ... &      ... &      ... &      ... &      ... &      ... &      ... &      ... &      ... \\
  MCG-03-58-007 &     0.28 &     0.89 &     2.60 &     3.66 &      ... &      ... &      ... &      ... &      ... &      ... &      ... &      ... &      ... &      ... &      ... &      ... &      ... &      ... &      ... &      ... \\
  MCG-06-30-015 &     0.33 &     0.97 &     1.09 &     1.10 &      ... &     1.06 &      ... &      ... &     0.40 &      ... &      ... &      ... &      ... &      ... &      ... &      ... &      ... &      ... &      ... &      ... \\
  MCG+00-29-023 &     0.33 &     0.73 &     5.22 &     9.65 &      ... &      ... &      ... &      ... &      ... &      ... &      ... &     0.57 &      ... &      ... &     0.08 &      ... &      ... &      ... &      ... &      ... \\
        MRK0006 &     0.26 &     0.73 &     3.50 &      ... &      ... &     5.00 &      ... &      ... &      ... &      ... &      ... &      ... &      ... &      ... &      ... &      ... &      ... &      ... &      ... &      ... \\
        MRK0009 &     0.23 &     0.39 &     0.76 &     0.98 &      ... &     0.74 &      ... &     0.36 &     0.23 &      ... &      ... &      ... &      ... &      ... &      ... &      ... &      ... &      ... &      ... &      ... \\
        MRK0079 &     0.36 &     0.73 &     1.55 &     2.35 &     2.74 &     2.04 &     1.88 &     1.12 &     0.85 &      ... &      ... &      ... &      ... &      ... &      ... &      ... &      ... &      ... &      ... &      ... \\
        MRK0231 &     1.87 &     8.66 &    32.00 &    30.30 &    24.30 &    16.10 &    20.80 &     9.75 &     6.88 &      ... &      ... &      ... &      ... &      ... &      ... &      ... &      ... &      ... &      ... &      ... \\
        MRK0273 &     0.22 &     2.30 &    22.80 &    22.20 &    23.60 &    16.95 &     8.30 &     7.48 &     5.75 &      ... &     1.34 &     0.71 &      ... &      ... &     0.06 &      ... &      ... &     0.01 &      ... &      ... \\
        MRK0335 &     0.27 &     0.45 &     0.35 &     0.57 &      ... &      ... &      ... &      ... &      ... &      ... &      ... &      ... &      ... &      ... &      ... &      ... &      ... &      ... &      ... &      ... \\
        MRK0463 &     0.47 &     1.49 &     2.21 &     1.99 &     2.26 &     0.99 &      ... &      ... &     0.45 &      ... &      ... &     0.11 &      ... &      ... &      ... &      ... &      ... &      ... &      ... &      ... \\
        MRK0509 &     0.30 &     0.73 &     1.39 &     1.36 &      ... &      ... &     1.00 &      ... &      ... &      ... &      ... &      ... &      ... &      ... &      ... &      ... &      ... &      ... &      ... &      ... \\
        MRK0704 &     0.42 &     0.60 &     0.36 &     0.45 &      ... &      ... &      ... &      ... &      ... &      ... &      ... &      ... &      ... &      ... &      ... &      ... &      ... &      ... &      ... &      ... \\
        MRK0897 &     0.37 &     0.86 &     2.97 &     5.59 &      ... &      ... &      ... &      ... &      ... &      ... &      ... &      ... &      ... &      ... &      ... &      ... &      ... &      ... &      ... &      ... \\
        MRK1239 &     0.76 &     1.21 &     1.68 &     2.42 &      ... &      ... &      ... &      ... &      ... &      ... &      ... &      ... &      ... &      ... &      ... &      ... &      ... &      ... &      ... &      ... \\
        NGC0034 &     0.40 &     2.51 &    16.84 &    17.61 &    17.20 &    10.60 &     8.50 &     5.40 &      ... &     3.28 &     1.15 &      ... &     0.35 &      ... &      ... &      ... &      ... &      ... &      ... &      ... \\
        NGC0262 &     0.49 &     1.02 &     1.43 &     1.43 &     2.24 &     2.28 &     1.89 &     1.29 &     1.14 &      ... &      ... &      ... &      ... &      ... &      ... &      ... &      ... &      ... &      ... &     0.36\\
        NGC0424 &     1.22 &     1.76 &     2.00 &     1.74 &      ... &      ... &     1.16 &      ... &      ... &      ... &      ... &      ... &      ... &      ... &      ... &      ... &      ... &      ... &      ... &      ... \\
        NGC0513 &     0.17 &     0.28 &     1.93 &     4.05 &      ... &      ... &      ... &      ... &      ... &      ... &      ... &      ... &      ... &      ... &      ... &      ... &      ... &      ... &      ... &      ... \\
       NGC0526A &     0.23 &     0.48 &     2.31 &     4.08 &      ... &      ... &      ... &      ... &      ... &      ... &      ... &      ... &      ... &      ... &      ... &      ... &      ... &      ... &      ... &      ... \\
        NGC0931 &     0.62 &     1.42 &     2.80 &     5.66 &      ... &     5.54 &     5.25 &     3.72 &     2.88 &      ... &      ... &      ... &      ... &      ... &      ... &      ... &      ... &      ... &      ... &      ... \\
       NGC1056 &     0.36 &     0.48 &     5.49 &    10.22 &      ... &      ... &      ... &      ... &      ... &     2.15 &     1.09 &      ... &     0.24 &      ... &      ... &      ... &      ... &      ... &      ... &      ... \\
\hline
\end{tabular}
}
\end{table*}

\begin{table*}
\setcounter{table}{1}
\caption{$-$ Continue}
 \rotatebox{90}{
\begin{tabular}{lrrrrrrrrrrrrrrrrrrrr}
\hline \hline
         name &      \multicolumn{20}{c}{Flux (in Jy) at $\lambda$ (in $\mu$m):} \\ 
               &       12 &       25 &       60 &      100 &      120 &      150 &      170 &      180 &      200 &      250 &      350 &      450 &      500 &      800 &      850 &      870 &     1100 &     1300 &     2000 &     3000 \\     \hline   
        NGC1125 &     0.32 &     1.00 &     3.71 &     4.04 &      ... &      ... &      ... &      ... &      ... &      ... &      ... &      ... &      ... &      ... &      ... &      ... &      ... &      ... &      ... &      ... \\
        NGC1194 &     0.28 &     0.85 &     0.92 &     0.71 &      ... &      ... &      ... &      ... &      ... &      ... &      ... &      ... &      ... &      ... &      ... &      ... &      ... &      ... &      ... &      ... \\
        NGC1320 &     0.31 &     1.32 &     2.21 &     2.82 &     3.70 &     2.58 &      ... &     1.44 &     1.06 &      ... &      ... &      ... &      ... &      ... &      ... &      ... &      ... &      ... &      ... &      ... \\
        NGC1365 &     4.42 &    13.07 &    84.20 &   185.40 &   217.00 &   194.00 &   167.00 &   103.00 &    85.20 &   145.80 &    62.30 &    24.70 &      ... &      ... &      ... &      ... &      ... &      ... &      ... &      ... \\
        NGC1566 &     1.91 &     3.02 &    22.50 &    58.10 &      ... &    95.26 &      ... &      ... &      ... &    51.50 &    22.20 &     9.10 &      ... &      ... &      ... &      ... &      ... &      ... &      ... &      ... \\
        NGC2992 &     0.51 &     1.58 &    10.78 &    17.05 &      ... &      ... &      ... &      ... &      ... &      ... &      ... &      ... &      ... &      ... &      ... &      ... &      ... &      ... &      ... &      ... \\
        NGC3079 &     2.60 &     3.65 &    50.70 &   105.17 &      ... &      ... &    91.30 &      ... &      ... &      ... &    10.70 &     3.70 &      ... &     0.80 &      ... &      ... &      ... &     0.50 &      ... &      ... \\
        NGC3516 &     0.39 &     0.96 &     2.09 &     2.73 &      ... &      ... &      ... &      ... &      ... &      ... &      ... &      ... &      ... &      ... &      ... &      ... &      ... &      ... &      ... &      ... \\
       NGC4051 &     0.86 &     2.20 &    10.53 &    24.93 &      ... &      ... &      ... &      ... &      ... &     1.49 &     0.78 &      ... &     0.34 &      ... &      ... &      ... &      ... &      ... &      ... &      ... \\
       NGC4151 &     1.95 &     4.87 &     6.46 &     8.50 &     4.40 &      ... &     3.00 &      ... &      ... &     0.59 &     0.31 &      ... &     0.13 &      ... &      ... &      ... &      ... &      ... &      ... &      ... \\
        NGC4253 &     0.35 &     1.47 &     3.89 &     4.20 &      ... &      ... &      ... &      ... &      ... &      ... &      ... &      ... &      ... &      ... &      ... &      ... &      ... &      ... &      ... &      ... \\
       NGC4388 &     1.03 &     3.72 &    10.27 &    17.15 &      ... &      ... &    19.51 &      ... &      ... &     2.39 &     1.33 &      ... &     0.61 &      ... &     0.29 &      ... &      ... &     0.01 &      ... &      ... \\
        NGC4593 &     0.47 &     0.96 &     3.43 &     6.26 &      ... &     8.10 &      ... &      ... &     4.70 &      ... &      ... &      ... &      ... &      ... &      ... &      ... &      ... &      ... &      ... &      ... \\
        NGC4602 &     0.54 &     0.56 &     4.75 &    12.60 &      ... &      ... &      ... &      ... &      ... &      ... &      ... &      ... &      ... &      ... &      ... &      ... &      ... &      ... &      ... &      ... \\
       NGC5135 &     0.58 &     2.39 &    16.60 &    31.18 &      ... &      ... &      ... &      ... &      ... &     9.20 &     3.85 &      ... &     1.37 &      ... &      ... &      ... &      ... &      ... &      ... &      ... \\
        NGC5256 &     0.30 &     1.13 &     7.27 &    10.07 &      ... &      ... &     8.73 &      ... &      ... &      ... &      ... &      ... &      ... &      ... &     0.08 &      ... &      ... &      ... &      ... &      ... \\
        NGC5347 &     0.30 &     0.96 &     1.43 &     2.64 &      ... &      ... &      ... &      ... &      ... &      ... &      ... &      ... &      ... &      ... &      ... &      ... &      ... &      ... &      ... &      ... \\
        NGC5506 &     1.25 &     4.24 &     8.44 &     9.24 &      ... &      ... &      ... &      ... &      ... &      ... &      ... &      ... &      ... &      ... &      ... &      ... &      ... &      ... &      ... &      ... \\
        NGC5548 &     0.43 &     0.81 &     1.07 &     1.61 &      ... &      ... &      ... &      ... &      ... &      ... &      ... &      ... &      ... &      ... &      ... &      ... &      ... &      ... &      ... &      ... \\
        NGC5953 &     0.82 &     1.67 &    11.85 &    20.47 &    17.00 &    17.20 &    15.50 &      ... &      ... &      ... &      ... &      ... &      ... &      ... &     0.18 &      ... &      ... &      ... &      ... &      ... \\
        NGC5995 &     0.41 &     1.45 &     4.09 &     7.06 &      ... &      ... &      ... &      ... &      ... &      ... &      ... &      ... &      ... &      ... &      ... &      ... &      ... &      ... &      ... &      ... \\
        NGC6810 &     1.33 &     3.61 &    18.90 &    33.30 &      ... &    29.80 &      ... &    16.90 &    14.00 &      ... &      ... &      ... &      ... &      ... &      ... &      ... &      ... &      ... &      ... &      ... \\
        NGC6860 &     0.25 &     0.31 &     0.96 &     2.19 &      ... &      ... &      ... &      ... &      ... &      ... &      ... &      ... &      ... &      ... &      ... &      ... &      ... &      ... &      ... &      ... \\
        NGC6890 &     0.36 &     0.80 &     4.01 &     8.26 &    11.70 &     8.80 &      ... &     5.10 &     4.36 &      ... &      ... &      ... &      ... &      ... &      ... &      ... &      ... &      ... &      ... &      ... \\
       NGC7130 &     0.64 &     2.15 &    16.85 &    26.96 &    28.90 &    21.90 &    16.90 &    10.80 &     7.29 &     6.49 &     2.92 &      ... &     1.04 &      ... &      ... &      ... &      ... &      ... &      ... &      ... \\
       NGC7213 &     0.65 &     0.81 &     2.70 &     9.15 &      ... &    13.62 &      ... &      ... &     6.60 &      ... &      ... &      ... &      ... &      ... &      ... &      ... &      ... &      ... &      ... &      ... \\
       NGC7469 &     1.58 &     6.04 &    28.57 &    35.83 &    33.00 &    26.00 &    29.00 &      ... &      ... &     7.82 &     3.07 &      ... &     0.99 &      ... &     0.19 &      ... &      ... &      ... &      ... &      ... \\
        NGC7496 &     0.62 &     2.00 &    10.21 &    16.59 &    20.80 &    16.20 &    13.90 &     8.89 &     6.25 &      ... &      ... &      ... &      ... &      ... &      ... &      ... &      ... &      ... &      ... &      ... \\
        NGC7603 &     0.40 &     0.24 &     1.25 &     2.00 &      ... &      ... &      ... &      ... &      ... &      ... &      ... &      ... &      ... &      ... &      ... &      ... &      ... &      ... &      ... &      ... \\
        NGC7674 &     0.64 &     1.79 &     5.64 &     8.46 &     9.38 &     7.65 &     6.02 &     3.67 &     2.78 &      ... &      ... &      ... &      ... &      ... &     0.11 &      ... &      ... &      ... &      ... &      ... \\
 TOLOLO1238-364 &     0.66 &     2.63 &     9.08 &    14.03 &      ... &      ... &     7.42 &     4.60 &     3.62 &      ... &      ... &      ... &      ... &      ... &      ... &      ... &      ... &      ... &      ... &      ... \\
      UGC05101 &     0.35 &     1.11 &    11.64 &    20.52 &      ... &      ... &      ... &      ... &      ... &     5.83 &     2.35 &      ... &     0.79 &     0.14 &      ... &     0.04 &     0.07 &     0.01 &      ... &      ... \\
       UGC07064 &     0.17 &     0.38 &     2.75 &     5.57 &      ... &      ... &      ... &      ... &      ... &      ... &      ... &      ... &      ... &      ... &      ... &      ... &      ... &      ... &      ... &      ... \\
\\
\hline \hline
\end{tabular}
}
\label{tab_phot_fir}
\end{table*}

\label{lastpage}

\end{document}